\documentclass[varenna]{cimento}

\usepackage{graphicx}
\usepackage{dcolumn} \usepackage{bm} \usepackage{amsmath}
\usepackage{amsfonts} \usepackage{amssymb}

\begin{document}
\title{Strongly Interacting Fermi Gases}

\author{Wilhelm Zwerger}
\institute{Physik-Department,
Technische Universit\"at M\"unchen, D-85748 Garching, Germany}

\begin{abstract}
The experimental realization of stable, ultracold Fermi gases near a Feshbach resonance
allows to study gases with attractive interactions of essentially arbitrary strength. They
extend the classic paradigm of BCS into a regime which has never been accessible before.
We review the theoretical concepts which have been developed in this context, including the
Tan relations and the notion of fixed points at zero density, which are at the origin of universality.
We discuss in detail the universal thermodynamics of the unitary Fermi gas which allows a fit free
comparison between theory and experiment for this strongly interacting system.  In addition,
we adress the consequences of scale invariance at infinite scattering length and the 
subtle violation of scale invariance in two dimensions. Finally we discuss the
Fermionic excitation spectrum accessible in momentum resolved RF-spectroscopy  
and the origin of universal lower bounds for the shear viscosity and the spin diffusion 
constant.     
\end{abstract}

\date{July 2014}


\maketitle
\tableofcontents
\newpage

\section{FESHBACH RESONANCES}

The very notion of a strongly interacting gas seems to be a
contradiction in itself, so we start by discussing how strongly interacting gases
can be realized and why this is possible only with Fermions 
not, however - at least in equilibrium or without an optical lattice - with Bosons.

A gas requires densities $n$ which are low enough that the average interparticle
spacing $n^{-1/3}\gg r_e$ is much larger than the range $r_e$
of interactions.  In order to have nontrivial correlations in a gaseous state,
the temperature has to be small enough that the thermal wavelength $\lambda_T$ 
is of the order of or larger than the interparticle spacing. At these temperatures, the wave nature of the 
particles becomes relevant. In contrast to the non-degenerate limit, 
interactions can then no longer be described by point like collisions of
particles which approach each other on distances of order $r_e$. Instead,
the relevant scale which determines the strength of the interactions is
the scattering length $a$, which may be much larger than the interaction range $r_e$.
Now, as will be shown below in Eqs.~(\ref{eq:effrange}) and~(\ref{eq:FB-range}), the effective range
of the interactions at low energies turns out to be essentially the van der Waals length $l_{\rm vdw}$. 
Irrespective of their interaction strength, ultracold gases are thus characterized by the following hierarchy of length scales
\begin{equation}
\label{eq:hierarchy}
l_\text{vdw}\ll n^{-1/3}\ll \lambda_T \, .
\end{equation}
Since $l_{\rm vdw}\ll\lambda_T$, the thermal energy is
necessarily much smaller than the centrifugal barrier, which is of order $E_{\rm vdw}=\hbar^2/ml_{\rm vdw}^2$
for interactions with a van der Waals tail as in Eq.~(\ref{eq:vdwpot}) below.
As a result, the two-body interactions are due to s-wave scattering only.
The distinction between weak and strong interactions now depends on whether the 
associated scattering length $a$ is much smaller or  larger than the average interparticle 
spacing $n^{-1/3}$. The former limit is in fact the standard situation because, as 
follows from Eq.~(\ref{eq:s-length}), generic values of the scattering length 
are of the order of the van der Waals length $ l_{\rm vdw}$, which also determines the 
effective range $r_e$. Using Feshbach resonances, however, the scattering length 
can be increased to values 
far beyond typical interparticle distances, which are about $0.5\,\mu$m, still 
keeping the effective range of order $ l_{\rm vdw}$ i.e. in the few nm range.
This allows to realize ultracold gases with strong interactions $n^{1/3}a\gg 1$.

\subsection{Two-body scattering}

We start by recalling some elementary facts about 
two-body scattering at low energies, using a simple toy model~\cite{grib93scatt}, where the van der Waals attraction 
at large distances is cutoff by a hard core at some distance $r_c$ on the order of an atomic 
dimension. The resulting spherically symmetric potential
\begin{equation}
\label{eq:vdwpot}
V(r)=\begin{cases} -C_6/r^6 & \text{if} \qquad r>r_c\\
\infty & \text{if} \qquad r\leq r_c
\end{cases}
\end{equation}
involves two quite different  characteristic lengths, namely $r_c$ and 
the van der Waals length
\begin{equation}
\label{eq:vdwlength}
l_\text{vdw}=\frac{1}{2}\,\left(\frac{mC_6}{\hbar^2}\right)^{1/4}
\end{equation}
which is determined by the strength $C_6$ of the attractive interaction at large distances.  
For alkali atoms, which are strongly polarizable, this length is typically on the order of several 
nano-meters, much larger than the atomic scale $r_c$. As a result, the potential~(\ref{eq:vdwpot})
supports many bound states. Moreover, the low energy scattering properties in the limit $l_{\rm vdw}\gg r_c$
are essentially determined by the van der Waals length $l_{\rm vdw}$. 

To see this, we recall that the scattering length $a$ and effective range $r_e$ are 
defined by the low energy expansion
\begin{equation}
\label{eq:s-ampl}
f(k)=\frac{1}{k\cot{\delta_0(k)}-ik}\,\to\,
\frac{1}{-1/a+r_ek^2/2+\dots -ik}
\end{equation}
of the s-wave scattering amplitude.
For the simple model potential~(\ref{eq:vdwpot}), the exact expression~\cite{grib93scatt} 
 \begin{equation}
\label{eq:s-length}
a=\bar{a}\left[ 1-\tan\left(\Phi-3\pi/8\right)\right]
\end{equation}
for the  scattering length shows that its
characteristic magnitude is set by the mean scattering length $
\bar{a}=0.956\, l_\text{vdw}$, which is basically identical with 
the van der Waals length. The short range part of the interaction,
which is sensitive to the hard core scale $r_c$,
only enters via the WKB-phase 
\begin{equation}
\label{eq:vdwphase}
\Phi=\int_{r_c}^{\infty}dr\sqrt{m\vert V(r)\vert}/\hbar
=2 l^2_\text{vdw}/r_c^2\gg 1
\end{equation}
at zero energy. Scattering lengths with a magnitude much larger than $\bar{a}$ thus only
arise near a zero energy resonance, where the phase $\Phi$ happens to be close to a value where the $\tan{}$ 
in Eq.~(\ref{eq:s-length}) diverges and an additional s-wave bound state is pulled in from the continuum.
Concerning the effective range $r_e$, 
the exact result for the toy-model potential~(\ref{eq:vdwpot}) is~\cite{flam99scatt}
\begin{equation}
\label{eq:effrange}
r_e=2.92\,\bar{a}\left( 1-2\frac{\bar{a}}{a}+2\left(\frac{\bar{a}}{a}\right)^2\right)\, .
\end{equation}
It is positive, with a typical magnitude which is again set by the 
van der Waals length, unless $a\ll\bar{a}$. In contrast to what  
might have been expected naively, the effective range in low energy scattering 
is thus much larger than the short range scale $r_c$. The property $r_e\gg r_c$ is in fact a generic result for low 
energy scattering in long range potentials, provided
the number $N_b\simeq  (l_\text{vdw}/r_c)^2$ of bound states is much larger than 
one~\cite{flam99scatt}. 

\subsection{Feshbach Resonances} 

The regime of strong interactions in dilute, ultracold gases 
can be reached by exploiting Feshbach resonances, 
which allow to increase the scattering length in a systematic manner to values 
far beyond the average interparticle spacing. In the following, we will
focus on the case of magnetically tunable Feshbach resonances and 
their minimal description in terms of a two-channel model. For a detailed
exposition of the subject, see the review by Chin et.al.~\cite{chin10feshbach}.

\begin{figure}[t]
\begin{center}
\includegraphics[width=3.5in]{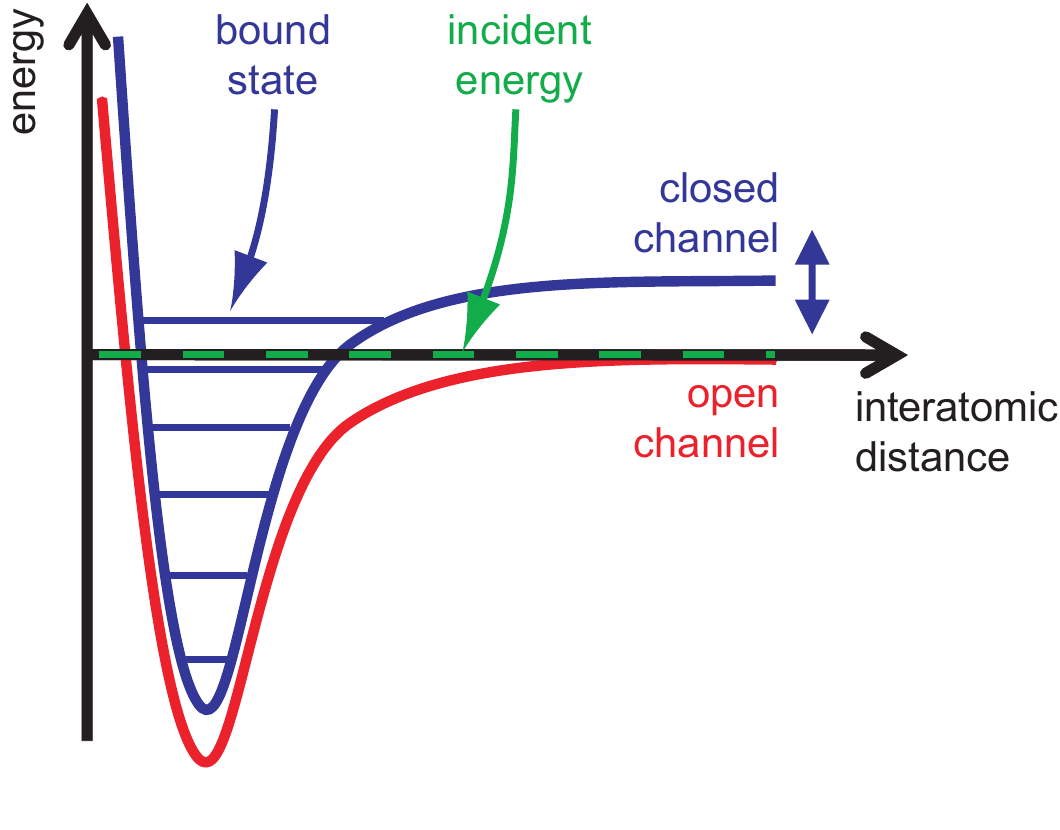}
\caption[The two-channel model for a Feshbach resonance.]{Atoms prepared in the open channel undergo a collision at low incident energy. 
The coupling to a bound state in the closed channel near zero energy leads to a scattering resonance. The position of the closed channel can 
be tuned with respect to the open one by varying the magnetic field $B$ (from Ref.~\cite{bloc08review}).}
\label{fig:Feshbachmodel}
\end{center}
\end{figure}

Quite generally, a Feshbach resonance in a two-particle collision
appears whenever a bound state in a closed channel is coupled
resonantly with the scattering continuum of an open channel, as 
shown schematically in Fig.~\ref{fig:Feshbachmodel}.  
Taking the specific example of fermionic $^6$Li atoms,
which have electronic spin $S=1/2$ and nuclear spin $I=1$,
for typical magnetic fields above $500\rm\,G$, the electron spin is almost
fully polarized by the magnetic field, and aligned in the same direction for
the three lowest hyperfine states. Low energy scattering 
of two lithium atoms is thus essentially determined 
by the triplet potential.
At any finite field, however, the initial configuration is not a pure
triplet. The spin dependent part $\hat{\mathbf S}_1\cdot\hat{\mathbf S}_2\;(V_t(r)-V_s(r))$
of the full two-body interaction 
thus couples the initial state to other
scattering channels, provided only that the $z$ projection of the total spin is conserved. 
The closed channel consists of states in the singlet potential $V_s(r)$ which
has the same strength of van der Waals attraction than $V_t(r)$ but differs
considerably at short distances, with a much deeper attractive well for the singlet. 
When the atoms are far apart, the
Zeeman+hyperfine energy of the available closed channel states exceeds the initial
kinetic energy of the pair of atoms
by an energy on the order of the hyperfine energy. Since 
the thermal energy is much smaller than that for ultracold collisions,
the channel is closed and the atoms always
emerge from the collision in the open channel state.
Yet, as will be derived below,
the coupling to the closed channel gives rise to a resonant contribution
to the effective open channel interaction and thus allows to reach 
scattering lengths much larger than their characteristic values of order $\bar{a}$.

What makes Feshbach resonances in the scattering of cold atoms
particularly useful, is the ability to tune the scattering length simply by 
changing the magnetic field. This
tunability relies on the finite difference $\Delta\mu$ in the magnetic moments of the
closed and open channels, which allows to change the position of
closed channel bound states relative to the open channel threshold
by an external magnetic field. A standard parametrization for the 
magnetic field dependent scattering length near 
a particular Feshbach resonance at $B=B_0$ is given by 
\begin{equation}
\label{eq:FBs-length}
a(B)=a_{\rm bg}\left(1-\frac{\Delta B}{B-B_0}\right)\to -\frac{\hbar^2}{mr^{\star}\nu(B)}+\ldots\, .
\end{equation}
Here, $a_{\rm bg}$ is the off-resonant background scattering length
in the absence of the coupling to the closed channel, while 
$\Delta B$ describes the width of the resonance
expressed in magnetic field units. More generally, as indicated 
in the second form of Eq.~(\ref{eq:FBs-length}), it is often sufficient to 
focus on just the resonant contribution to the scattering length, 
which is inversely proportional to  the detuning $\nu(B)=\Delta\mu (B-B_0)$ away from the resonance. 
This dependence defines a characteristic length $r^{\star}>0$
\footnote{Formally, the length $r^{\star}$ may be defined by $a_{bg}\, \Delta\mu\Delta B=\hbar^2/mr^{\star}$. 
The proper definition, as given in Eq.~(\ref{eq:rstar}) below is, however, independent of a specific parametrization of $a(B)$ as in~(\ref{eq:FBs-length}).},  
whose inverse turns out to be a measure of how strongly the open and closed channels are coupled.

A microscopic description of the resonant contribution 
to the scattering length can be obtained within a two-channel model:
two atoms in an open channel are converted into a closed channel bound state 
and back with an amplitude which is determined by the strength of
the off-diagonal coupling $W(r)$ induced by the spin-dependent interaction
proportional to $V_t(r)-V_s(r)$, as discussed above. 
For a two-component Fermi gas, the effective Hamiltonian is 
\begin{eqnarray}
\label{eq:BFM}
&\hat{H}=\int d^3x \Bigg[ \sum_{\sigma}\hat{\psi}_{\sigma}^{\dagger}
\big(-\frac{\hbar^2}{2m}\nabla^2\big)\hat{\psi}_{\sigma}\; +\hat{\Phi}^{\dagger}\big(-\frac{\hbar^2}{4m}\nabla^2+\nu_{c}(B)\big)
\hat{\Phi} &\nonumber \\
&+ \tilde{g}\,\int d^3x'\,  \chi(|\mathbf{x}-\mathbf{x'}|) \Big( \hat{\Phi}^{\dagger}(\frac{\mathbf{x}+\mathbf{x'}}{2})\,
 \hat{\psi}_{\uparrow}(\mathbf{x}) \hat{\psi}_{\downarrow}(\mathbf{x'})+\rm{h.c.}\Big)\Bigg]\, .&
\end{eqnarray}

Here, the fermionic field operators $\hat{\psi}_{\sigma}(\mathbf{x})$ 
describe atoms in the open channel, with a formal 
spin variable $\sigma=\uparrow,\downarrow$ distinguishing 
two different hyperfine states. The bound
state in the closed channel is denoted by the bosonic operator $\hat{\Phi}$,
which is often called the dimer field.
Its energy $\nu_{c}(B)$ measures the detuning of the bare closed channel 
bound state with respect to two atoms at zero energy. 
The coupling is characterized by a strength $\tilde{g}$ and a cutoff function 
$\chi(\mathbf{x})$, which only depends on the magnitude $r=|\mathbf{x}-\mathbf{x'}|$ of the 
distance between two atoms in the open channel, consistent with the pure s-wave nature of scattering.
The function $\chi(\mathbf{x})$ is normalized by $\int_{\mathbf{x}}\chi(\mathbf{x})=1$.
Its Fourier transform $\chi(\mathbf{q})$, which depends on $q=|\mathbf{q}|$ only, thus obeys $\chi(q\to 0)=1$. 
As will be shown below, the characteristic range of the cutoff function is essentially 
the mean scattering length. The conversion between the open channel scattering states 
and the closed channel bound state is therefore spread out over a separation of the order $l_{\rm vdw}$
despite the fact that the coupling itself becomes strong only near the short distance scale $r_c$.  
The absence of a term quartic in the
fermionic fields in~(\ref{eq:BFM}) implies that  background scattering 
between Fermions is neglected. This is justified close enough to resonance $|B-B_0|\ll
|\Delta B|$, where the scattering
length is dominated by its resonant contribution $a_{\rm res}\sim -(r^{\star}\nu)^{-1}$. 

For just two atoms, the model in~(\ref{eq:BFM}) is equivalent to an an off-diagonal coupling~\cite{cast07} 
\begin{equation}
\hat{W}\, |\phi_{\rm res}\rangle =\tilde{g}\,\sum_k \chi(\mathbf{k})\, ||\mathbf{k}\rangle  \quad \quad {\rm and} \quad\quad
\hat{W}\, ||\mathbf{k}\rangle = \tilde{g}\,\chi(\mathbf{k})\,   |\phi_{\rm res}\rangle
\label{eq:2channel}
\end{equation}
which transfers a single bound state $|\phi_{\rm res}\rangle$ in the closed-channel  
into a pair of atoms with opposite momenta $\mathbf{k}, -\mathbf{k}$ in an open channel state $||\mathbf{k}\rangle$
and vice versa. The associated two-body problem can be reduced to a
coupled eigenvalue equation in momentum space
\begin{align}
\frac{\hbar^2k^2}{m}\,\alpha(\mathbf{k})+\tilde{g}\,\chi(\mathbf{k})\,\sqrt{Z}=& E\,\alpha(\mathbf{k}) \nonumber \\
\nu_{c}(B)\,\sqrt{Z}+\tilde{g}\,\sum_k\, \chi(\mathbf{k}) \alpha(\mathbf{k})= & E\, \sqrt{Z}\, .
\label{eq:coupled}
\end{align}
by decomposing an eigenstate with energy $E=\hbar^2\mathbf{k}_0^2/m$ in the form 
$\sqrt{Z}\,|\phi_{\rm res}\rangle+\sum_k\, \alpha(\mathbf{k})\, ||\mathbf{k}\rangle$,
with $Z$ as a measure of the closed channel admixture. The set of Eqs.~(\ref{eq:coupled}) can be solved easily, giving
\begin{equation}
  \alpha(\mathbf{k})=(2\pi)^3\delta(\mathbf{k}-\mathbf{k}_0)+
  \frac{\tilde{g}^2\gamma\chi(\mathbf{k})/(E-\nu_c)}{E-\hbar^2\mathbf{k}^2/m+i\,0}\, .
  \label{eq:Aimplicit}
\end{equation}
This is an implicit equation only, however, since 
$\gamma=\sum_k\, \chi(\mathbf{k}) \alpha(\mathbf{k})$ still depends on $\alpha(\mathbf{k})$. 
The solution may be made explicit
by multiplying~(\ref{eq:Aimplicit}) by $\chi(\mathbf{k})$ 
and summing over $\mathbf{k}$. As a result one obtains
\begin{equation}
 \frac{\gamma \chi(\mathbf{k})}{E-\nu_c}= \frac{ \chi(\mathbf{k}) \chi(\mathbf{k}_0)}{E-\nu_c -\tilde{g}^2Y}\;\;
 {\rm with}\;\;   Y=\int_q\,\frac{\chi^2(\mathbf{q})}{E-\hbar^2\mathbf{q}^2/m+i\,0}
 \label{eq:XYZ}
\end{equation}
Now general scattering theory implies that Eq.~(\ref{eq:Aimplicit}) describes an
outgoing scattering state with
\begin{equation}
\tilde{g}^2\gamma\chi(\mathbf{k})/(E-\nu_c)=\langle\mathbf{k}|T(E+i\,0)|\mathbf{k}_0\rangle
  \label{eq:asympTmatrix}
\end{equation}
the exact $T$ matrix for scattering $\mathbf{k}_0\to\mathbf{k}$. This allows to directly read off 
the resulting open channel scattering amplitude
\begin{equation}
  f(k)=-\frac{m}{4\pi\hbar^2}\, \langle\mathbf{k}|T(E+i\,0)|\mathbf{k}_0\rangle=
  \frac{m}{4\pi\hbar^2}\frac{\tilde{g}^2\chi^2(k)}{\nu_{c}(B)-\frac{\hbar^2k^2}{m}
  +\frac{m\tilde{g}^2}{\hbar^2}\,\int_q\frac{\chi^2(q)}{k^2-q^2 +i\,0}}\, .
  \label{eq:scattampcastin}
\end{equation}

Expanding this at low energies, the resulting expressions for the scattering length and the effective range 
defined in~(\ref{eq:s-ampl}) are~\cite{schm12efimov}
\begin{equation}
       \frac{1}{a}=-\frac{mr^\star}{\hbar^2}\nu_{c}(B)+\frac{1}{2\sigma}
    \quad {\rm and} \quad
    r_e =-2r^\star+3\sigma\left( 1-\frac{4\sigma}{3a} \right)\, .
      \label{eq:FB-range}
\end{equation}
Here,  we have introduced $r^\star=4\pi\hbar^4/(m^2\tilde{g}^2)$ as an 
intrinsic length scale. As anticipated above, its inverse is a direct measure of the strength $\tilde{g}^2$ of the Feshbach coupling.
The explicit result~(\ref{eq:FB-range}) is obtained by using a Lorentzian cutoff 
$\chi(\mathbf{k})=1/\left(1+(k\sigma)^2\right)$ in momentum space.  
This choice is convenient since the resulting effective range  
$r_e\to 3\,\sigma=3\, \bar a$ (see below) of two-body scattering for an open-channel dominated Feshbach resonance
with $\bar{a}\gg r^\star$  
is very close to the value $r_e\to 2.92\, \bar a$ of Eq.~(\ref{eq:effrange}) for a single-channel potential with a $1/r^6$ tail.
By contrast, the more standard Gaussian cutoff~\cite{wern09closed} yields $r_e=8\bar{a}/\pi=2.54\,\bar{a}$ in this limit.
Expanding the detuning $\nu_{c}(B)=\Delta\mu(B-B_{c})$
of the closed channel molecular state to linear order around a bare resonance position $B_{c}$
leads to a scattering length which is indeed of the form given in~(\ref{eq:FBs-length}) with $a_{\rm bg}=0$.
The resonance position is, however, shifted from its bare value by
\footnote{A generalization of this result to a finite value $r_{\rm bg}=a_{\rm bg}/\bar{a}$ of the
dimensionless background scattering length is given in Eqs.~(37) or (42) of the review by Chin et.al.~\cite{chin10feshbach}.}
\begin{equation}
\Delta\mu\left( B_0-B_{c}\right)=\frac{\hbar^2}{2mr^\star\sigma}\, ,
\label{eq:shift}
\end{equation} 
which is a consequence of the level repulsion due to the off-diagonal coupling $\tilde{g}$.
The magnetic field $B_0$ where the scattering length diverges thus
differs from the value $B_c$ at which the energy of a bare molecule crosses zero. 
 This resonance shift has been calculated within a microscopic description of the
Feshbach coupling based on interaction potentials with a van der Waals tail~\cite{gora04shift}. 
Comparison with this result yields the identification 
$\sigma=\bar a$~\cite{schm12efimov}, thus fixing the effective range $\sigma$ of the Feshbach 
coupling to be equal to the mean scattering length $\bar a$.
 \begin{figure}
\begin{center}
\includegraphics[width=0.8\linewidth]{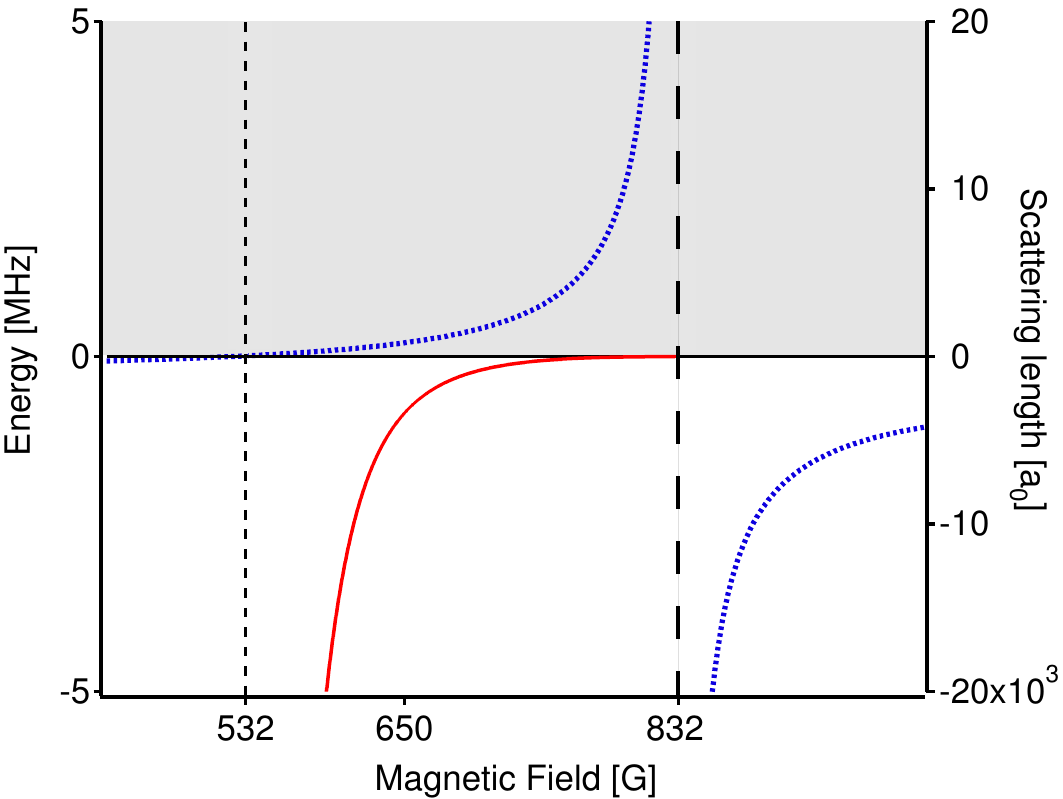}
\caption{Magnetic field dependence of the scattering length (dotted curve)
between the two lowest hyperfine levels of $^6$Li with a Feshbach
resonance near $B_0\simeq 832\,$G~\cite{zuer13Li-resonance} and a zero crossing at
$B_0+\Delta B\simeq 532\,$G. The background scattering length is $a_{\rm bg}=-1405\, a_B$ with $a_B$ the Bohr radius.
The energy of the bound state is shown as a full line (adapted from Ref.~\cite{kett08varenna}).
\label{fig:FeshbachLi6}}
\end{center}
\end{figure}
The ratio
\begin{equation}
s_{\rm res}=\frac{\bar{a}}{r^\star}
\label{eq:sres}
\end{equation}
between the two intrinsic microscopic lengths which characterize the scattering length and range in potentials with a van der Waals tail 
and the strength of the coupling to the closed channel is called the resonance strength~\cite{chin10feshbach}.  
It allows  to classify Feshbach resonances into two limiting cases: when $s_{\rm res}\gg 1$, the resonance is called an open channel dominated
one because the closed channel fraction $Z$ remains small compared to one over the whole range of detunings $|B-B_0|<| \Delta B|$.
Indeed, consider the regime where the scattering length
is dominated by its resonant contribution $a_{\rm res}\sim -1/\nu r^{\star}\gg\bar{a}$.
The two-body bound state energy at $a>0$ then
follows the universal behavior $\epsilon_b=\hbar^2/(ma^2)=\nu^2/\epsilon^{\star}$ 
determined by the scattering length only.  Its quadratic dependence on  
the detuning $\nu=\Delta\mu(B-B_0)$ with a characteristic energy
$\epsilon^{\star}=\hbar^2/m(r^{\star})^2$ leads to a linearly vanishing
closed channel admixture $Z$ near resonance~\cite{chin10feshbach}
 \begin{equation}
Z=-\frac{\partial\epsilon_b}{\partial\nu}=2\frac{|\nu|}{\epsilon^*}=2\frac{r^*}{a}\simeq 2\frac{r^*}{|a_{\rm bg}|}\; \frac{|B-B_0|}{|\Delta B|} \, .
\label{eq:Z}
 \end{equation}
Since $\vert a_{\rm bg}\vert $ is typically of the order of the mean scattering length $\bar{a}$, 
open channel dominated resonances also have $r^{\star}\ll\vert a_{\rm bg}\vert $
\footnote{The Feshbach resonance of $^6$Li at $B_0\simeq 832$ G shown in Fig.~\ref{fig:FeshbachLi6}
has an exceptionally large background scattering length $|a_{\rm bg}|\gg\bar{a}$.
The closed channel admixture thus remains negligible in an even larger regime on the 
positive scattering length side of the resonance.}. 
As a result, the closed channel admixture is much smaller than one 
over the full magnetic field range $|B-B_0|\lesssim |\Delta B|$.
Moreover, as emphasized above, in the relevant regime where $a\gg\bar{a}$, the effective range $r_e\to 3\,\bar a$ of 
open channel dominated resonances obtained from Eq.~(\ref{eq:FB-range}) 
is essentially identical to the corresponding result~(\ref{eq:effrange}) for a single-channel potential with a $1/r^6$ tail. 
Resonances with $s_{\rm res}\ll 1$, in turn,  are called closed channel dominated. Here, the near-threshold scattering and bound states 
have an open channel character only very close to resonance. Indeed, Eq.~(\ref{eq:Z})  shows that $Z$ reaches values of order one
already at a detuning $\vert\nu\vert\simeq\epsilon^{\star}$ which is now much less than the van der Waals
energy $\hbar^2/ml_{\rm vdw}^2$. In addition, the effective range $r_e\to -2r^\star$ is 
negative and large compared to the characteristic scale set by $\bar{a}\simeq l_{\rm vdw}$.

\subsection{Three-body losses}

While (s-wave) Feshbach resonances appear
for both Bose or two component Fermi gases, the strong interaction limit
$a\gg n^{-1/3}\gg\bar{a}$ is in practice only accessible for
Fermions. This is a result of the fact that for Fermions the lifetime due to three-body
collisions is large near a Feshbach resonance, quite in 
contrast to Bosons, where it goes to zero. 
The basic physics which underlies the stability of Fermions
near a resonance of the scattering length is the fact that relaxation into 
deep bound states is strongly suppressed by the Pauli-principle. 
Indeed, by energy and momentum conservation,
a relaxation into one of the deeply bound states requires that at least
three Fermions are close together, at a distance of order $l_{\rm vdw}\ll a$. In a two-component gas
two of them are necessarily equal
\footnote{Note that this is no longer the case for Fermi gases with three or more components, 
which therefore do not exhbit an enhanced stability for large scattering lengths.}. 
As was shown by Petrov et.al.~\cite{petr04dimers}, the dependence of the three-body 
loss rate on the scattering length can be
inferred from the behavior of the corresponding wave function at short distances.
Quite generally, for a system of $N$ Fermions with effectively zero range interactions, 
one may define an exponent $\gamma(N_{\uparrow},N_{\downarrow})$ by the 
behavior~\cite{tan04}
\begin{equation}
 \Psi(\mathbf{r}_1, \sigma_1\, \mathbf{r}_{2}, \sigma_2 \ldots \mathbf{r}_N, \sigma_N)\;\rightarrow\;
 r^{\gamma(N_{\uparrow},N_{\downarrow})}\;\; {\rm as} \;\; r\to 0
\label{eq:short-range}
\end{equation}
of the many-body wave function as $N_{\uparrow}$ up-spin Fermions and $N_{\downarrow}$
down-spin Fermions are within a small radius $r\ll a$. 
All other particles remain at a finite distance.  
In the case of two Fermions with opposite spin and scattering length $a$,  
the standard expression  $\psi_0(r)=1/r\, -1/a$ for the two-body wave function in a zero range approximation
shows that
$\gamma(1,1)=-1$. For three Fermions, the solution of the three-body Schr\"odinger 
equation with a zero range interaction~\cite{petr04dimers} yields $\gamma(2,1)=-0.2273\ldots$. 
The wave function is thus less  singular than for two particles, reflecting the fact that two identical Fermions can get close
only in a relative p-wave configuration. The physical origin of the non-integer power law is an effective 
$1/r^2$-potential which appears in the three-body Schr\"odinger equation expressed in terms
of the hyperradius $r=\sqrt{r_{12}^2+r_{13}^2+r_{23}^2}$~\cite{petr04dimers}. 
By dimensional analysis, the probability 
that three Fermions get close depends on $a$ via the prefactor $A(a)\sim a^{-3/2 -\gamma}$ 
of the three-body wave function $\Psi(r\to 0)=a^{-3/2}\, (r/a)^{\gamma}\, F(\Omega)=A(a)\, r^{\gamma}\, F(\Omega)$.
Here,  $F(\Omega)$  is a function which depends on the remaining angular degrees of freedom. 
The relaxation rate $\alpha_3$ into deep bound states 
will be proportional to $|A(a)|^2$. Expressed in physical units  cm$^3$/sec, it thus follows a power law~\cite{petr04dimers} 
\begin{equation}
\label{eq:relaxation}
\alpha_{3}(a)={\rm const}\,\frac{\hbar\, l_{\rm vdw}}{m}\cdot\Bigl(\frac{l_{\rm vdw}}{a}\Bigr)^s
\end{equation}
with a positive exponent $s=3+2\gamma\simeq 2.55$.
The dimensionless prefactor depends on short 
range physics below the scale $l_{\rm vdw}$ and thus cannot 
be calculated within the zero range approximation. 
Experimental results for the lifetime of Fermionic $^{40}$K or $^{6}$Li 
near their respective Feshbach resonances at $B_0\simeq 202\,$G~\cite{rega04lifetime} and
$B_0\simeq 832\,$G~\cite{bour04coll} are consistent with the dependence predicted by Eq.~(\ref{eq:relaxation}).
They do not allow, however, to determine the exponent $s$ with the precision necessary to 
extract a reliable value for the anomalous dimension $\gamma$. 
In a system with finite density $n\sim k_F^3$, the power 
law dependence on $a$ is cut off at values beyond $k_Fa=\mathcal{O}(1)$. 
The ratio of the rate $\Gamma_3=-\dot{N}_3/N$ for three-body losses due to decay into deeply bound states 
and the rate $\Gamma_2$ for equilibration due to two-body collisions is therefore 
expected to be~\cite{petr04dimers} 
\begin{equation}
\frac{\Gamma_3}{\Gamma_2}\simeq\frac{\hbar\Gamma_3}{\varepsilon_F}\simeq \hbar\, n\, \alpha_{3}(1/a\to k_F)/\varepsilon_F 
\simeq (k_F\, l_{\rm vdw})^{s+1}\ll 1\, .
\label{eq:stability}
\end{equation}
Indeed,  the cross-section  for two-body scattering in a deeply degenerate Fermi gas at $a=\pm\infty$ is 
$\sigma\simeq 1/k_F^2$. The associated equilibration rate $\Gamma_2\simeq n\sigma v_F\simeq \varepsilon_F/\hbar$ 
is thus essentially set by the Fermi energy. The fact that $\Gamma_3\ll \Gamma_2$ 
in the experimentally relevant limit  $k_F\, l_{\rm vdw}\ll 1$ is essential for the stability of a 
degenerate gas of Fermions at unitarity: the time scale for equilibration via two-body scattering is
much faster than the decay associated with three-body losses. For concrete numbers, consider a balanced gas of
 $^6$Li atoms in their two lowest Zeeman split hyperfine levels. A typical Fermi energy 
 of around $1\,\mu$K then corresponds to $k_F\simeq 1/(3800\, a_B)$. With $l_{\rm vdw}\simeq 31\, a_B$ for  $^6$Li this
 gives $k_Fl_{\rm vdw}$ values of order $10^{-2}$ or smaller and lifetimes of a degenerate unitary gas 
 of up to a minute~\cite{kett08varenna}.  For a gas of Bosons, the situation is, unfortunately, completely different.
 Indeed, as will be shown below, for Bosons at $a=\pm\infty$, the requirement 
 $\Gamma_3\ll\Gamma_2$ is valid only in the non-degenerate limit $n\lambda_T^3\ll 1$.\\
 
The rate $\Gamma_3$ of three-body losses is expected to scale like the square of 
the density. The associated loss rate coefficient $L_3$ defined by $\Gamma_3=L_3\cdot n^2$
should thus be density independent. The result~(\ref{eq:stability}) above, however, 
shows that this is not the case. Indeed, $L_3\sim (k_F\, l_{\rm vdw})^{2\gamma}$ 
exhibits a dependence on density $n\sim k_F^3$ which directly reflects the presence of 
a nontrivial scaling exponent  $\gamma$. From a field-theoretic point of view, the 
unexpected density dependence of $L_3$ 
can be understood as a result of the appearance of 
an anomalous dimension for operators whose matrix elements first appear at 
the three-body level. Specifically, the value $\gamma=-0.227..$ appears in the anomalous dimension
\begin{equation}
 \Delta_{\mathcal{O}}=\Delta_{\phi}+\Delta_{\psi\uparrow}+1+\gamma= 2+3/2 +1 +( -0.227\ldots) =4.272\ldots
 \label{eq:a-dimension}
 \end{equation}
 of the operator
 \begin{equation}
  \mathcal{O}= \mathcal{O}_{\uparrow\uparrow\downarrow}^{(l=1)}(\mathbf{x}) = Z^{-1}(\Lambda)
  \left[2\phi\,\partial_i\psi_\uparrow -(\partial_i\phi)\psi_\uparrow \right](\mathbf{x})\, .
   \label{eq:three-body-operator}
\end{equation}
It contains the gradient ($l=1$) of a renormalized diatom operator $\hat{\phi}$ 
introduced in Eq.~(\ref{eq:dimer-operator2}) below, combined
with one additional up-spin Fermion. Here $i=x,y,z$ and $Z\sim\Lambda^{-\gamma}$ is the 
renormalization factor which is necessary for giving  finite matrix elements of the operator
in the zero range limit $\Lambda\to\infty$. The value $ \Delta_{\mathcal{O}}$ also determines
the energy $E_0= \Delta_{\mathcal{O}}\hbar\omega$ of the ground state of three Fermions in a
harmonic trap with frequency $\omega$ precisely at infinite scattering length, which has $l=1$~\cite{wern06traplevels}.
For a detailed discussion of these connections and an explicit calculation of the anomalous dimension $\gamma$, 
see the review by Nishida and Son~\cite{nish12book}.  \\

The issue of inelastic collisions has an additional aspect, which
is crucial for the eventual stability of a many-body system of Fermions for
arbitrary large scattering lengths. On the three-body level, this is related 
to the  repulsive nature of elastic atom-dimer scattering, which is 
described by a positive scattering length $a_{ad}\simeq1.18\ a$ in the 
regime $a>0$ where two Fermions form a bound state
with wave function $\varphi_0(r)\sim\exp{(-r/a)}$~\cite{petr05dimer}.
The underlying statistical repulsion due to the Pauli principle also shows up in the 
four-body problem.  In quantitative terms, it can be derived from
an exact solution of the four-particle Schr\"odinger equation with
zero range interactions in the limit where 
the distance $R$ between the 
centers of mass of two bosonic dimers is much larger than the 
dimer size $a$ and at collision energies much 
smaller than their respective binding energies $\hbar^2/ma^2$.
The wave function has the asymptotic form~\cite{petr04dimers}
\begin{equation}
\label{eq:a_dd}
\Psi(\mathbf{x_1},\mathbf{x_2},\mathbf{R})=
\varphi_0(r_1)\varphi_0(r_2)\bigl( 1-a_{dd}/R\bigr)\;\;\;\text{with}\;\;\;
a_{dd}=0.6\, a\, .
\end{equation}
Here, $\varphi_0(r)$ is the bound state wave function 
of an individual dimer and $\mathbf{x}_{1,2}$ are the respective
interparticle distances between the two distinguishable Fermions
which they are composed of.  Eq.~(\ref{eq:a_dd}) implies 
that the effective dimer-dimer interaction at low energies is 
characterized by a positive scattering length 
proportional to the original scattering length between its 
fermionic constituents. The fact that $a_{dd}>0$ guarantees
the stability of molecular condensates and implies that, at least 
for short range interactions, there are no four-particle bound states.
More generally, the stability of a Fermi gas
at the many-body level for arbitrary strong attractive interactions
relies crucially on the assumption that the range of the
interactions is negligible. In fact, it is easy to show that 
the Pauli principle alone is unable to stabilize a Fermi gas
with purely attractive interactions if they 
have a finite range~\cite{cast12book}.  \\

\subsection{Unitary Bosons and the Efimov effect} A completely different behavior appears for Bosons near a 
Feshbach resonance. Indeed, it turns out that for scattering lengths which exceed about ten times
its characteristic value $\bar{a}$, they form an unstable system with a rather short lifetime. 
The instability is connected with the fact  that the three boson scattering amplitude has an attractive rather than a repulsive 
character. As shown by Efimov in 1970~\cite{efim70}, this effective
attraction gives rise to an infinite sequence of three-body bound states.
They appear already for negative scattering lengths, where there is no two-body bound state.  
For $a<0$, therefore, Efimov trimers behave like 
Borromean rings: three atoms are bound together but cutting one of the bonds 
makes the whole system fly apart. Now, as will be shown below, the 
first Efimov state appears at a scattering length $a_{-}\simeq -9\, \bar{a}$.  This is about one order of
magnitude larger than the characteristic values of the scattering length in the absence of a resonance. 
Bose gases are therefore stable unless tuned to an interaction rather close to the unitary point, where $|a|\gg\bar{a}$.  
In quantitative terms, the rate of three-body losses can be written in the form~\cite{fedi96reco, niel99boserecomb}
\begin{equation}
\label{eq:Bose-lifetime}
\Gamma_3=-\dot{N}_3 /N\, =L_3\cdot n^2 = 3\, C(a)\, \frac{\hbar}{m}\cdot \left( n a^2\right)^2 \, .
\end{equation}
Since $\Gamma_3$ is expected to scale like $n^2$, 
this result is in fact fixed by dimensional analysis provided that only the scattering length $a$ enters 
and not a microscopic length like $l_{\rm vdw}$. Clearly, this scale has to eventually appear in the
dimensionless factor $C(a)$, which accounts for the detailed structure due the Efimov effect (see below). 
Neglecting the variation of $\Gamma_3$ due to this factor, the three-body loss rate 
diverges $\sim a^4$ with increasing scattering length.
This is in stark contrast to the result~(\ref{eq:relaxation}) for Fermions, where the rate approaches zero 
as $a\to\pm\infty$.  Bosons near unitarity can therefore not be realized in an equilibrium configuration
unless one enters the non-degenerate limit. Indeed, at finite temperature
the divergence of $L_3\sim a^4$ with increasing scattering length is cutoff by 
the thermal wavelength $\lambda_T$. Provided $\lambda_T\ll |a|$, the three-body loss rate  
$\Gamma_3=L_3(T)n^2\simeq \hbar \cdot n^2\lambda_T^4/m\sim 1/T^2$ thus exhibits a power law
dependence on temperature, which has been verified experimentally~\cite{flet13, rem13}. 
The rate $\Gamma_2\simeq n\sigma v_T$ of equilibration 
due to two-body scattering in a thermal gas at unitarity, in turn, is of order 
$\Gamma_2\simeq \hbar n\lambda_T /m$ since $\sigma\simeq\lambda_T^2$ and $v_T=\hbar/m\lambda_T$. 
For Bosons at infinite scattering length, the condition $\Gamma_3\ll\Gamma_2$ for a thermodynamically 
stable gas is therefore obeyed only in the 
non-degenerate regime, where $\Gamma_3/\Gamma_2\simeq n\lambda_T^3\ll 1$.\\

The result $L_3\simeq \hbar a^4/m$ at low temperatures, where $\lambda_T\gg |a|$, only describes the dependence on $a$ {\it on average}.
In fact, as realized by Esry et.al.~\cite{esry99recomb} and by Bedaque et.al.~\cite{beda00boserecomb}, the Efimov effect gives rise
to a nontrivial, log-periodic structure in the pre-\- factor $C(a)$. It leads to pronounced 
maxima in the three-body  loss coefficient on top of the $a^4$-law at scattering lengths $a_{-}^{(n)}<0$ 
where the Efimov bound states detach from the two-particle continuum.
It is this feature on which the first experimental observation of the Efimov effect 
by Kraemer et.al.~\cite{krae06efimov} is based on, via
an enhanced three-body recombination rate at the scattering length 
$a_{-}\! =a_{-}^{(0)}$ where the first trimer state appears. 
In turn, there are minima of $C(a)$ at a set of positive $a_{*}^{(n)}$, where 
the $n$-th trimer state crosses the two-body bound state energy $\epsilon_b(a)$.

\begin{figure}
\begin{center}
\includegraphics[width=1.0\linewidth]{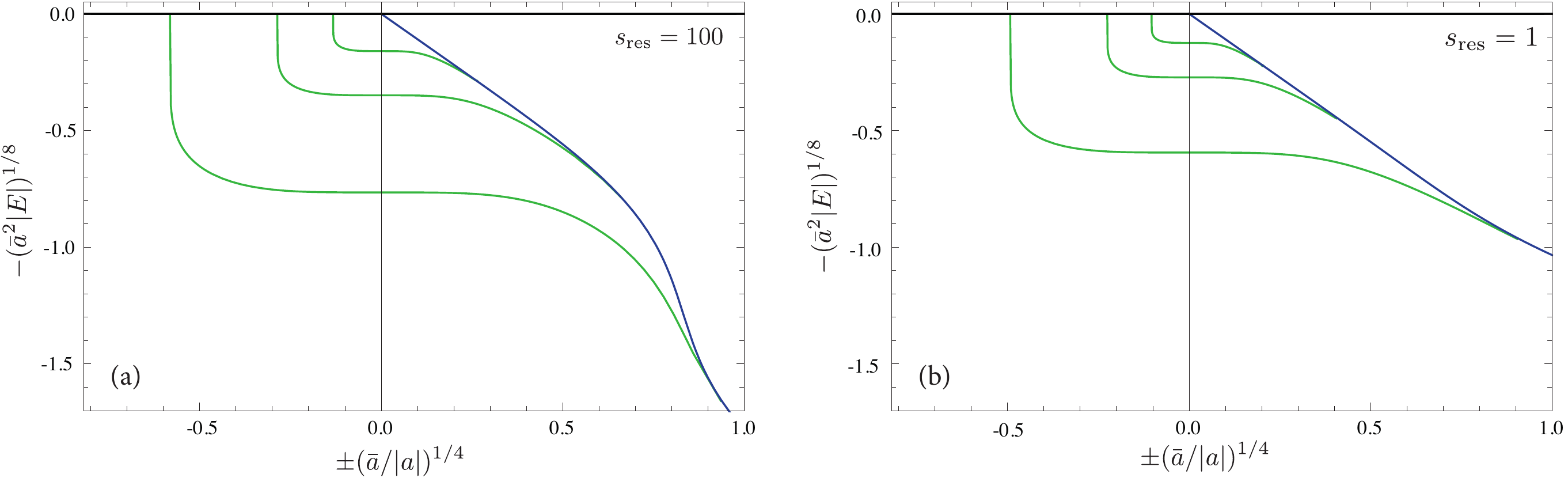}
\caption{The energies of the three lowest Efimov states as a function of the inverse scattering length, both in dimensionless units,
 for an open-channel dominated Feshbach resonance of strength $s_\text{res}=100$
and a resonance of intermediate strength $s_\text{res}=1$. The binding energy of the two-body bound state (dimer) is shown in addition. 
\label{fig:Efimov}}
\end{center}
\end{figure}

The spectrum of the Efimov trimers can be calculated fully within the  
microscopic two-channel model~(\ref{eq:BFM}), where the two-component Fermions $\psi_{\sigma}\to\psi$ 
are replaced by  a single component Bose field~\cite{schm12efimov}. 
Formally, the infinite sequence of trimer energies $E_T^{(n)}<0$ is obtained from 
the poles of a three-body vertex $\lambda_3(q_1,q_2;E)$ which describes the 
scattering of a single atom and the dimer as a function of the total energy $E$ in the center of mass frame.
For a two-component Fermi gas, this amplitude has a well defined low energy 
limit $\lambda_3(0)\sim a_{ad}=1.18 a$ in the regime where the scattering length is
positive. It describes the effective short range repulsion between a single Fermion
and a two-particle bound state~\cite{dieh08}. For Bosons, the vertex $\lambda_3$ 
develops a nontrivial dependence on both energy and the momenta $q_1$ and $q_2$ of the in- and outgoing atoms. 
In particular, it exhibits poles associated with three-body bound states. Close to these poles, it can be parametrized by \cite{braa06}
\begin{equation}\label{ef3_polstruc}
\lambda_3^{(n)}(q_1,q_2;E)\approx\frac{\mathcal{B}(q_1,q_2)}{E-E_T^{(n)}}\, .
\end{equation}
In Fig.~\ref{fig:Efimov}, the resulting energies $E_T^{(n)}(\bar{a}/a)$ of the three lowest Efimov states are depicted in a dimensionless form,
with the scattering length and energy measured in units of $\bar{a}$ and $\bar{E}=\hbar^2/2m\bar{a}^2$, respectively. 
The rescaling by a power $1/4$ for $\bar{a}/a$ and $1/8$ for the dimensionless energy is choosen for convenience,
to make several Efimov states visible. The appearance of $\bar{a}\simeq l_{\rm vdw}$ as the natural unit for
the scattering length and the fact that the van der Waals energy sets the characteristic scale
for the binding energies of the Efimov trimers is a consequence of the simple two-channel  model~(\ref{eq:BFM}),
where the  finite range $\sigma=\bar{a}$ of the Feshbach coupling described by the function $\chi(\mathbf{x})$ 
provides the characteristic length and energy scale.  Within this model, the spectrum of Efimov trimers  
follows a universal set of curves which only depend    
on the resonance strength parameter $s_{\rm res}=\bar{a}/r^{\star}$~\cite{schm12efimov}. 
The overall appearance of the spectrum remains similar as the strength of the resonance is varied. 
In the limit $s_{\text{res}}\ll 1$, it gets pushed towards the 
unitarity point $E=1/a=0$, while for open-channel dominated resonances it reaches its maximal extent in the $(\bar{a}/a,E)$ plane.  
Specifically, for open channel dominated resonances with $s_{\rm res}\gg 1$, the first trimer state within this model 
detaches from the continuum at $a_{-}=-8.3\, l_\text{vdw}$.  
As discussed by Schmidt et.al.~\cite{schm12efimov}
\footnote{For different approaches towards an understanding of the  'universal' value of the three-body 
parameter, see~\cite{wang12,naid14}.}, 
this result provides an explanation for the
surprising observation of an apparently universal value of the scattering
length $a_{-}$ where the first Efimov trimer state appears:
for many different resonances, the measured values for $a_{-}$ clustered around an average value 
$\langle a_{-}\rangle \approx -9.45\, l_{\text{vdw}}$~\cite{bern11,wild12}.

As was predicted already by Efimov, the infinite sequence of scattering lengths $a_{-}^{(n)}$ 
where the three-body bound states detach from the two-particle continuum obeys an asymptotic scaling law:  
the ratio of consecutive values of $a^{(n)}_{-}$ approaches $a_{-}^{(n+1)}/a_{-}^{(n)}\!\to\! e^{\pi/s_0}\simeq 22.6942$
for $n\gg 1$, with a universal number  $s_0\approx1.00624$
\footnote{For a derivation of these results see e.g. the review by Petrov~\cite{petr13}. An effective field theory approach 
to the Efimov effect and its connection 
to a limit cycle in a renormalization group flow of the three-body scattering amplitude is discussed in the review 
by Braaten and Hammer~\cite{braa06}.}. 
In practice, an observation of higher order trimer states 
is rather difficult since their size eventually becomes larger  
than typical trap sizes and the lifetime of a degenerate gas of Bosons 
approaches zero near unitarity. Recently, it has been possible to observe the second
trimer state near a Feshbach resonance in $^{133}$Cs 
at $B_0=786\,$G with $s_{\rm res}\simeq 1500$
at a scattering length $a_{-}^{(1)}\simeq 21.0(1.3)\, a_{-}$~\cite{huan14efimov}. 
The fact that the ratio $a_{-}^{(1)}/a_{-}$ for the experimentally accessible lowest Efimov states 
is smaller than the asymptotic value $22.69\ldots$ expected for $n\gg 1$
is consistent with the result obtained from the two-channel model~(\ref{eq:BFM})~\cite{schm12efimov}.  
Its precise value depends, however, on the detailed form of the cutoff function $\chi(\mathbf{x})$. More importantly, 
it is affected by genuine three-body forces which may, in fact, also explain the observed variation of 
the ratio  $a_{-}/l_{\rm vdw}$ between around $-8$ and $-10$~\cite{lang15}.

Remarkably, for Bosons, many-body bound states
exist also for particle numbers beyond $N=3$. This has been studied in detail 
for $N=4$, where one finds an infinite sequence of
two tetramer states per Efimov trimer~\cite{hamm07,stec09tetramers,schm10tetramers,delt12}. 
The lowest one at $a_{-(4)}\simeq 0.43\, a_{-(3)}$ has been seen by Ferlaino et.al.~\cite{ferl09}. 
More recently, even signatures of a five-body bound state have been observed~\cite{zene13}. 
Theoretically, many-body bound states have been found by van Stecher et.al~\cite{stec10N-body}
from numerical solutions of the Schr\"odinger equation up to $N=13$.
It is an open question whether the family of universal bound states for Bosons
persists for arbitrary $N$. A theorem due to 
Seiringer~\cite{seir12} which states that any
pairwise interaction potential with negative scattering 
length $a$ has an $N$-body bound state for {\it some} value 
of $N$, no matter how small $|a|$ may be, suggests that 
the sequence indeed continues up to $N=\infty$.   

\pagebreak

\section{TAN-RELATIONS}
\label{sec:tan}

This chapter provides an introduction to a series of exact relations due to Shina Tan,
which hold for Fermions
\footnote{For an extension of the Tan relations to strongly interacting Bose gases, where the 
presence of  the Efimov effect has to be accounted for, see~\cite{braa11bosons,wern12bosons}.}
with short range interactions~\cite{tan08energy,tan08momentum,tan08virial}. The Tan relations connect the short 
distance behavior of one- and two-particle correlations with thermodynamic properties.
They can be extended to time dependent correlations, giving rise to sum rules and power law 
tails at high frequency of RF-spectra~\cite{punk07rf,schn10shortrange,braa10rf} or of response
functions like the dynamic shear viscosity~\cite{tayl10visc,hofm11response,enss11viscosity}.  For a  
detailed discussion of the subject see the review by Braaten~\cite{braa12book}.

The study of a non-relativistic system of Fermions with spin-independent two-body interactions appears as a 
generic many-body problem in different areas of physics. Except for the particular case of one dimension,
 there are, unfortunately, very few exact results on this problem 
beyond the perturbative regime. It is therefore of considerable
interest to derive relations for the many-body problem that hold independent of the interaction strength. 
As realized by Tan and - independently - by Zhang and Leggett~\cite{zhan09universal},  a whole new class 
of exact relations may be derived in the context 
of strongly interacting ultracold gases, where the range of the interactions can effectively be set to zero. 
In this special case, it turns out that the momentum distribution $n_{\sigma}(k)$ exhibits a universal $C/k^4$ 
decay in the regime where $k$ is larger than other characteristic momentum scales in the problem. 
The constant $C$ is independent of the spin orientation $\sigma=\pm 1$  
and is called the contact, because it is a measure of the probability that two Fermions with opposite spin 
are close together.  A crucial feature of the Tan relations is the fact that 
they  apply to {\it any} state of the system, e.g. both to a Fermi-liquid or a superfluid state, at zero or at finite temperature 
and also in a few-body situation.  The only change is the value of the contact. The origin of this universality was elucidated 
by Braaten and Platter~\cite{braa08contact} who have shown that the Tan relations are a consequence of operator 
identities that follow from a Wilson operator product expansion. 

\subsection{Thermodynamic relations}
It is convenient to start by defining the concept of a contact first in a purely thermodynamic fashion. 
The equilibrium thermodynamics of any system in a micro-canonical situation
is  determined by its entropy $S(U,V,N)$ as a function of the 
conserved variables energy $U$, volume $V$
and total particle number $N$.   The condition $S(\lambda U, \lambda V,\lambda N)=\lambda\, S(U,V,N)$
of an extensive system implies the Gibbs-Duhem relation $G=\mu N$ for the 
free enthalpy $G=U-TS+pV$, or - equivalently - the relation $dp=n\, d\mu +s\, dT$, where $s=S/V$ is the entropy density.
These relations are completely general, however a concrete result for the equation of state
requires of course to explicitely calculate the entropy $S(U,V,N)$ from the associated microcanonical partition function
for a given form of the interaction between the particles. Usually, the microscopic interaction potentials are 
complicated functions of the interparticle distances which are neither known precisely nor can they be changed externally.  
In the context of ultracold  gases, however, a new situation arises because at energies below $E_{\rm vdw}$ 
\begin{itemize}
\item the whole interaction is embodied in a {\it single} parameter, namely the scattering length
\item the interaction can be changed externally via Feshbach resonances.
\end{itemize}   

It thus makes sense to consider the entropy of the gas not only as a function of the conserved and extensive 
variables $U,V,N$ but also of the - for later convenience - inverse scattering length $1/a$. The associated 
complete differential      
 \begin{equation}
 dS(U,V,N, 1/a)=\frac{1}{T}\, dU + \frac{p}{T}\, dV - \frac{\mu}{T}\, dN -\frac{X_{1/a}}{T}\, d\left(1/a\right)
\label{eq:entropy}
\end{equation}
then defines a new 'generalized force' $X_{1/a}$ \cite{bali91}. Its thermodynamic meaning becomes
clear by rewriting (\ref{eq:entropy}) as the differential change in free energy
 \begin{equation}
 dF(T,V,N, 1/a)=-S\, dT -p\, dV +\mu \, dN + X_{1/a}\, d\left(1/a\right)\, .
\label{eq:free-energy}
\end{equation}  
Thus $X_{1/a}\, d\left(1/a\right)$ is the work done {\it on} the gas in an infinitesimal  change $d(1/a)$ 
of the scattering length, keeping $T,V$ and $N$ fixed.  Consider, for instance, a situation
where a gas with a strongly repulsive interaction $l_{\rm vdw}/a\ll 1$ is turned into a weakly interacting gas
$l_{\rm vdw}/a\simeq 1$. Similar to the case of an expansion $dV>0$ at fixed interaction strength, the gas will perform work on its 
environment. As a result, $X_{1/a}=-\hbar^2\, C/(4\pi m)<0$ defines an extensive and positive quantity $C$ which has
dimensions of an inverse length. For reasons that will become clear below, $C$ is called the contact. 
Due to the extensive nature of the entropy,  (\ref{eq:entropy}) leads to a generalized form of the Gibbs-Duhem relation 
\begin{equation}
 d\, p(\mu, T, 1/a)=n\, d\mu +s\, dT + \frac{\hbar^2}{4\pi m}\,\mathcal{C}\, d\left(1/a\right)\, ,
\label{eq:pressure}
\end{equation} 
where $\mathcal{C}=C/V$ is an intensive contact density. 
In the case of a trapped gas with non-uniform particle density $n(\mathbf{R})$, the contact density $\mathcal{C}(\mathbf{R})$
is also varying in space and the full contact is $C=\int_{\mathbf{R}} \mathcal{C}(\mathbf{R})$. 
At fixed temperature, particle number and volume or  - in the case of  trapped gas - at a given confining 
potential,  the thermodynamic relation (\ref{eq:free-energy}) implies that 
 \begin{equation}
\frac{\partial F(T)}{\partial (1/a)}=\frac{\partial U(S)}{\partial (1/a)}=-\frac{\hbar^2}{4\pi m}\cdot\int_{\mathbf{R}}\,\mathcal{C}(\mathbf{R})\, 
\label{eq:Tan-adiabatic}
\end{equation} 
which is called the Tan adiabatic theorem~\cite{tan08energy}. 
The full contact is therefore just the  derivative of the total internal energy $U$ or the free energy $F$ 
with respect to the inverse scattering length 
at fixed values of the entropy $S$ or temperature $T$, respectively.
As a result, knowledge of the contact $C(1/a)$ as a function 
of the inverse scattering length determines the free energy of the 
interacting gas by an integration which starts with the non-interacting system at $a=0$.

An important exact relation for the total energy in an inhomogeneous 
situation is provided by the Tan virial theorem~\cite{tan08virial}
 \begin{equation}
U=\langle\hat{H}_{\text{kin}}+\hat{H}_{\text{int}}+\hat{H}_{\text{ext}}\rangle=2\, \int_{\mathbf{R}}\,V_{\text{ext}}(\mathbf{R})\, n(\mathbf{R})\, 
-\frac{\hbar^2}{8\pi ma}\,\int_R\,\mathcal{C}(\mathbf{R})\, ,
\label{eq:virial}
\end{equation}
which holds for harmonic trap potentials $V_{\text{ext}}(\mathbf{R})$ even if they are anisotropic. 
For a unitary gas, in particular,  the last term vanishes since the contact density is finite at infinite scattering length (see below).
Its total energy may thus be determined directly from in situ measurements of the density profile $n(\mathbf{R})$~\cite{thom05virial,thom08virial}
\footnote{This may be viewed as a trivial example of density functional theory, where the non-trivial part 
of the functional $E[n]$ related to the kinetic and interaction energy is simply $\int_{\mathbf{R}}\,V_{\rm ext}\, n(\mathbf{R})$!}.  
The relation (\ref{eq:virial}) is a simple consequence of dimensional analysis combined with the Tan adiabatic theorem.
Consider, for simplicity,  an isotropic harmonic trap potential $V_{\rm ext}(\mathbf{R})=m\omega^2\mathbf{R}^2/2$. At a fixed number of particles,
the free energy
\begin{equation}
\label{eq:freeenergy-dimlessfct}
 F(T,\omega, 1/a) = \hbar\omega\, \tilde{F} \left(\frac{k_B T}{\hbar\omega}, \frac{\omega}{\hbar/ma^2}\right)
\end{equation}
can be expressed in terms of a dimensionless function $\tilde{F}$, which depends only on 
dimensionless ratios. From (\ref{eq:freeenergy-dimlessfct}), one can deduce the simple scaling law
$ F(\lambda T, \lambda \omega, \sqrt{\lambda}/a) = \lambda\, F(T,\omega, 1/a)$. Its derivative
with respect to $\lambda$ at $\lambda = 1$ yields
\begin{equation}
 \label{eq:freeenergy-de-viral}
\left( T \frac{\partial}{\partial T} + \omega \frac{\partial}{\partial \omega} + \frac{1}{2a} \frac{\partial}{\partial (1/a)} \right) F = F\, ,
\end{equation}
where all the partial derivatives are to be understood as leaving all other system variables constant. Since the free energy is just the Legendre transform of the energy, its partial derivatives at constant temperature $T$ with respect to $\omega$ and $1/a$ are equal to those of the energy at the 
associated value of the entropy. Therefore, using $\partial F / \partial T = -S$, the energy turns out to obey the differential equation
\begin{equation}
\label{eq:energy-diffop}
 \left( \omega \frac{\partial}{\partial \omega} + \frac{1}{2a} \frac{\partial}{\partial (1/a)} \right)U=U.
\end{equation}
 This leads immediately to the relation (\ref{eq:virial}) by using the Tan adiabatic theorem  (\ref{eq:Tan-adiabatic}) and
 $\omega\, \partial E/\partial\omega= 2\langle V_{\rm ext}\rangle$. 
  
  For a uniform gas, a further exact relation is the Tan pressure relation
  \begin{equation}
p=\frac{2}{3}\,\epsilon+\frac{\hbar^2}{12\pi ma}\,\mathcal{C}\,  ,
\label{eq:Tan-pressure}
\end{equation}
which relates pressure $p$ and energy density $\epsilon$. Similar to the argument above, its proof 
relies on dimensional analysis. Indeed, again at a fixed number of particles,  the entropy 
$S(U,V,1/a)=\tilde{S}(u/(\hbar^2/ma^2), v/a^3)$ is only a function of dimensionless ratios that can be 
formed from the energy $u$ and volume $v$ per particle and the scattering length $a$.  Taking the
derivative of this relation with respect to $1/a$, the definition of the contact via (\ref{eq:entropy})
implies that 
\begin{equation}
\frac{\hbar^2}{4\pi m}\, C= a\left[ -2U+3pV\right]
\end{equation}
from which the pressure relation immediately follows. Anticipating again that $\mathcal{C}$ is finite  
at $a=\pm\infty$, this  implies that pressure and energy density are related simply by $p=2\epsilon/3$ for the unitary gas. 
The relation is identical to the one which holds in the non-interacting case and is valid irrespective of whether
the particles obey Fermi or Bose statistics .  
The deep underlying reason for this remarkable 
result is that at infinite scattering length, the gas is scale invariant, a property
that will be discussed in more detail in section 3.4 below.

\subsection{Quantitative results for the contact}
The Tan adiabatic relation
\begin{equation}
\mathcal C = \frac{4\pi m a^2}{\hbar^2}~\frac {\partial \epsilon(S)}{\partial a}
\label{adiabatic-C}
\end{equation}
for a homogeneous gas implies that the contact density $\mathcal C$ can be determined from the knowledge of
the energy density at fixed entropy as a function of the scattering length $a$. In the following, we
consider the case of a balanced gas, in which the two 
spin states are equally populated. Its ground state is 
superfluid for arbitrary values of the dimensionless 
interaction variable $v=1/k_F a$, where  $k_F = (3 \pi^2 n)^{1/3}$
is the  Fermi wave vector associated with a given total density $n$.
Upon changing $1/k_Fa$ over the range from 
$- \infty$ to $+\infty$, the nature of the pairing changes from 
a weak coupling BCS-type to a BEC of tightly bound dimers. 
Now, in the regime of small negative scattering lengths, 
the effects of pairing are exponentially small. As shown by Diener et.al.~\cite{dien08},
the ground state energy density of the superfluid has an expansion 
\begin{equation}
\epsilon_0 =
\frac{\hbar^2 k_F^5}{10 \pi^2 m} 
\left(1 + \frac{10}{9 \pi} k_F a + \frac{4(11-2\ln{2})}{21\pi^2}\, (k_Fa)^2+\ldots \right) 
\end{equation}
in powers of $k_F a$ which is identical to the one
obtained for a {\it repulsive} and normal Fermi liquid with $a>0$.
Using~(\ref{adiabatic-C}),  this gives rise to a contact density 
\begin{equation}
\mathcal C(k_Fa\to 0^-\!, T\!=0) = (2\pi n a)^2+\ldots=
k_F^4 \left(\frac{2k_Fa}{3\pi}\right)^2\! \left(1 + \frac{12(11-2\ln{2})}{35\pi}\, k_Fa+\ldots \right)\, .
\label{C-BCS}
\end{equation}
Its leading order contribution vanishes like $a^2$ 
and is independent of the sign of the interaction.
The BCS pairing, which leads to a finite energy gap
$\Delta\sim\exp{(- \pi/2k_F|a|)}$ for $a<0$ in the weak coupling limit $k_F|a|\ll 1$, only gives an
exponentially small reduction of the energy of order $\Delta^2/\varepsilon_F$.  At 
the level of the contact this is reflected in a corresponding
enhancement of the contact density by $\delta\mathcal{C}=k_F^4\,(\Delta/2\varepsilon_F)^2$~\cite{haus09rf}.
Within a BCS description, this follows from the standard expression for the condensation energy 
via Eq.~(\ref{adiabatic-C}) or from the fact that the associated ground state momentum distribution 
$n_{\sigma}(q)=v_q^2\to\delta\mathcal{C}/q^4$
exhibits a power law decay or $q\gg k_F$. 
As will be shown in Eq.~(\ref{eq:Tan-momentum}) below, such a tail is a universal feature of Fermions with 
zero range interactions and allows to read off the contact density from $n_{\sigma}(q)\to\mathcal{C}/q^4$.
More generally, it turns out that  the transition to a superfluid with gap $\Delta$
is always accompanied with an anomalous contribution $m^2\Delta^2/\hbar^4$ to the contact density $\mathcal{C}$.
This holds for arbitrary coupling and is 
a consequence of the connection between the contact density and 
the short distance limit of the vertex function, see Eq.~(\ref{eq:contact-Delta}) below. 

\begin{figure}[t]
\centerline{\includegraphics*[width=3.in,angle=0]{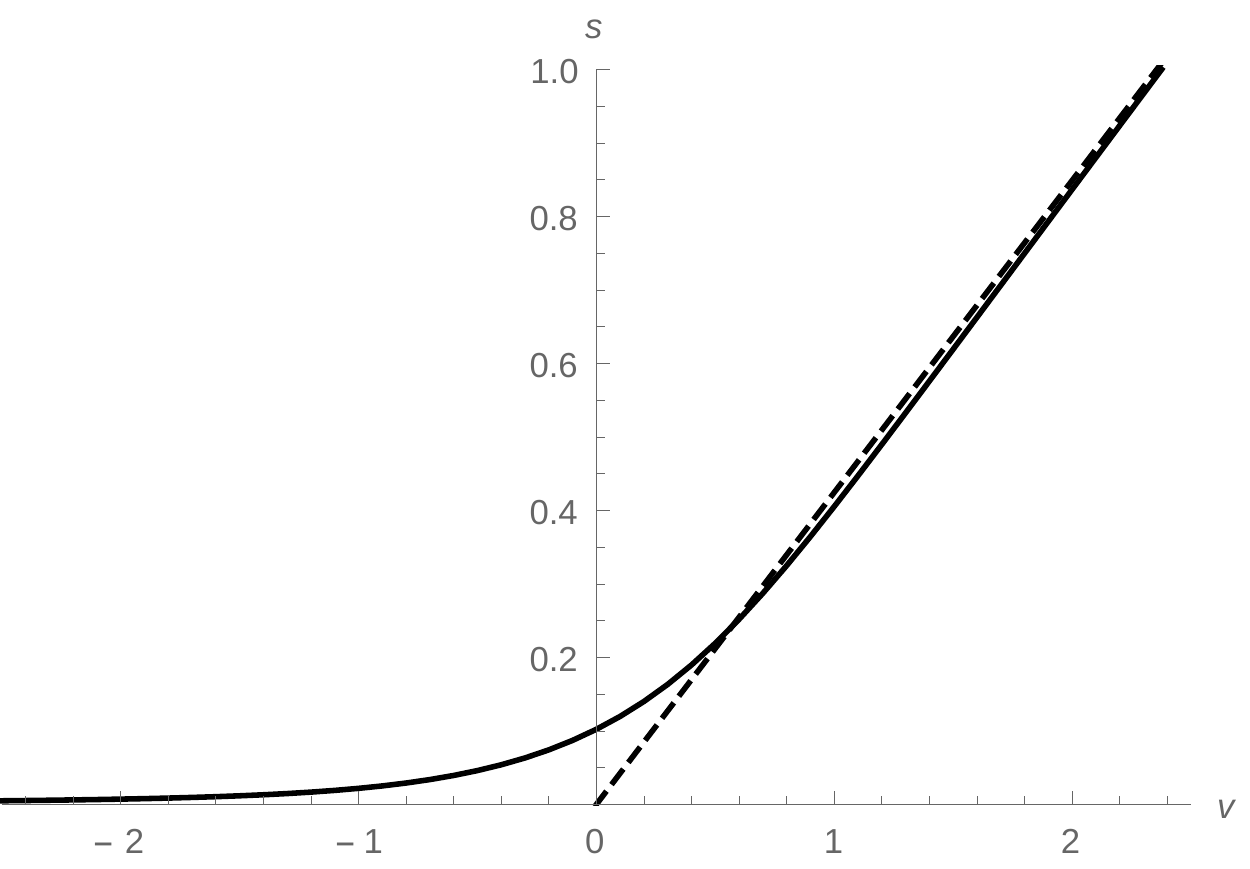}}
\caption{
The dimensionless contact density $s = {\cal C}/k_F^4$ 
for the balanced Fermi gas as a function of 
the coupling strength $v=1/k_F a$ from the Luttinger-Ward approach~\cite{haus09rf}.
In the dilute Fermi gas regime $v<-2$ the result is in perfect agreement with the expansion~(\ref{C-BCS}).
The right dashed line describes the leading two-body contribution from Eq.~(\ref{C-BEC}). 
The correction due to the dimer-dimer repulsion should be positive, but even at $v=1$ it is less than $0.01$.
}
\label{fig:contact_2}
\end{figure}

In the opposite limit of a molecular condensate $k_Fa \to 0^+$, 
the ground state energy density can be expanded in the form
\begin{equation}
\epsilon_0=  - \frac{\hbar^2 n}{2 m a^2} 
+ \frac{\pi \hbar^2 n^2 a_{dd}}{4m} 
\left( 1 + \frac{128}{15} \sqrt{n a_{dd}^3/2 \pi} + \ldots \right),
\end{equation}
where $a_{dd}=0.6~a$ is the 
dimer-dimer scattering length introduced in Eq.~(\ref{eq:a_dd}).
The leading term is the total binding energy density
for dimers with number density $n/2$, while the second term is the energy of the 
corresponding molecular BEC, including the well known 
Lee-Huang-Yang corrections of an interacting, dilute Bose gas
\footnote{These corrections have been observed experimentally
both from measuring the frequency of the radial compression mode of a trapped, elongated gas on the BEC-side
of the BCS-BEC crossover~\cite{altm07precision} and also, more directly, from the equation of state~\cite{navo10thermo}.
For a discussion of how these corrections arise from collective modes in the superfluid state, see~\cite{dien08}.}.  
Using Eq.~(\ref{adiabatic-C}), we find that 
the contact density in the BEC limit is
\begin{equation}
\mathcal C(k_Fa\to 0^+, T=0)= \frac{4 \pi n}{a}+0.6\, (\pi na)^2+\ldots =k_F^4\left[ \frac{4}{3\pi k_Fa} + 0.6\left(\frac{k_Fa}{3\pi}\right)^2+\ldots\right]
\label{C-BEC}
\end{equation}
%
Quantitative results for the dimensionles contact parameter
$s(v)=\mathcal{C}/k_F^4$ of the balanced Fermi gas at zero temperature for arbitrary values
of the coupling constant $v=1/k_Fa$ have been given by Haussmann et.al.~\cite{haus09rf},
using a Luttinger-Ward description of the BCS-BEC crossover problem. 
As shown in Fig.\ \ref{fig:contact_2}, it interpolates smoothly 
between the BCS and the BEC limit, with a finite value  $s(0)=0.102$ at unitarity.   
On the BEC side, the extraction of the contact from the tail of the momentum distribution
is apparently not precise enough to capture the repulsive interaction
between dimers, which from Eq.~(\ref{C-BEC}) gives rise to a contact density which lies
{\it above} the two-body contribution. Near unitarity, however, the predicted result for the 
parameter $c_0=3\pi^2 s=3.02$ is close to the experimental value 
$c_0^{\rm exp}=3.17\pm 0.09$ obtained from extrapolating precise 
measurements of the contact via the high momentum scaling 
$S_{\uparrow\downarrow}(q)\to\mathcal{C}/(8n_{\downarrow}q)$ of the static structure factor
(see Eq.~(\ref{eq:S12}) below) down to zero temperature~\cite{hoin13contact}.  

The zero temperature value of the contact density of the unitary gas 
is directly related to the slope of the ground state energy as a function of the coupling constant. 
Indeed, by dimensional analysis, the energy density $\epsilon_0(a=\pm\infty)=\xi_s\,\epsilon_0^{(0)}$ 
at unitarity must be some constant $\xi_s<1$ times the corresponding value of the non-interacting
Fermi gas. The universal number  $\xi_s$ is known as the Bertsch parameter and will be discussed 
in section 3.2 below. Expanding in powers of  $1/k_F a$ around unitarity, one has 
\begin{equation}
\epsilon_0(k_F|a|\gg 1) = \frac{\hbar^2 k_F^5}{10 \pi^2 m} 
\left( \xi_s  - \frac{\zeta}{k_F a} + \ldots\right) ,
\label{eq:zeta}
\end{equation}
with a positive numerical constant $\zeta$.
Using Eq.~(\ref{adiabatic-C}), the constant is directly related to the dimensionless  
contact at unitarity by $\zeta=5\pi s(0)/2\simeq 0.84$, with the numerical result based on the experimental value of $s(0)$.  
Quantitative results for the contact density of the unitary gas at finite temperature 
have been obtained in Refs~\cite{enss11viscosity} and~\cite{vanh13contact} in the normal phase above $T_c$.  
There, $\mathcal{C}$ is a monotonically decreasing function of temperature, approaching $\mathcal{C}\to 4\pi (\lambda_T n)^2\sim 1/T$
at high temperatures, consistent with the result obtained from a viral expansion~\cite{yu09correlations}. 
 In practice,  the virial expansion for $\mathcal{C}(T)$ is applicable at temperatures  $\theta = T/T_F\gtrsim 2$, 
 where the degeneracy parameter $n\lambda_T^3 = 8/(3\sqrt\pi\,\theta^{3/2})$ is smaller than one. Note that 
in the regime $k_Fl_{\rm vdw}\ll 1$ of dilute gases, the non-degenerate limit is still compatible with the 
condition $E\simeq k_BT\ll E_{\rm vdw}$ which is necessary to describe the interactions completely in terms of s-wave   
scattering. The theoretical results for $\mathcal{C}(T)$ agree well
with the experimental data by Kuhnle et.al.~\cite{kuhn11contact}, again based on the measurement of the 
static structure factor~(\ref{eq:S12}) via Bragg spectroscopy.

\subsection{Closed channel fraction}
The first example where Tan's concept of the contact turned out to be relevant for 
understanding strongly interacting Fermi gases was given by Punk and Zwerger~\cite{punk07rf}, who showed 
that the average clock shift observed in RF-spectroscopy is a direct measure of the contact. This 
will be discussed in the section 4.1. As noted by Werner et.al.~\cite{wern09closed} and by Zhang and Leggett~\cite{zhan09universal},
an observable which allows to extract the contact in a rather direct manner is the number 
 \begin{equation}
N_b=\int d^3R\;\langle\hat{\Phi}^{\dagger}(\mathbf{R})\hat{\Phi}(\mathbf{R})\rangle
\label{eq:Nb}
\end{equation} 
of closed channel molecules near a  Feshbach resonance. 
The connection between $N_b$ and the contact is a simple consequence of the observation above that the contact 
determines the work done on a gas upon changing the (inverse) scattering length.     
For a magnetically tunable Feshbach resonance, this is directly related
to the work needed to change the magnetic field by an infinitesimal amount $dB$
in the presence of a finite magnetization $M=N_b \Delta\mu$,  where $\Delta\mu$ is the difference in the
 magnetic moment between the molecule and the open-channel atoms. Indeed, using
 the two-channel description of Eq.~(\ref{eq:BFM}),  the only term in the Hamiltonian 
which depends on the magnetic field $B$ is the bare detuning $\nu_{c}(B)=\Delta\mu(B-B_{c})$ 
of the closed channel bound state. Using $dF=M dB$ and the Tan adiabatic theorem (\ref{eq:Tan-adiabatic})
in the form 
\begin{equation}
N_b \Delta\mu=\frac{\partial F}{\partial B}=\frac{\partial F}{\partial(1/a)}\cdot\frac{d(1/a)}{dB}=-
\frac{\hbar^2}{4\pi m}\cdot\int_{\mathbf{R}}\,\mathcal{C}(\mathbf{R})\,
\cdot\frac{d(1/a)}{dB}
\label{eq:Nb2}
\end{equation} 
shows that a knowledge of the dependence $a(B)$ of the two-body scattering length on the external field $B$ 
allows to determine the contact 
from the number of closed channel molecules~\cite{wern09closed}.
This general relation simplifies in the vicinity of a Feshbach resonance, where  the inverse 
scattering length 
\begin{equation}
\frac{1}{a(B)}=-\frac{\Delta\mu(B-B_0)\cdot mr^{\star}}{\hbar^2}+\ldots
\label{eq:rstar}
\end{equation} 
can be expanded to leading order in the renormalized detuning $\Delta\mu(B-B_0)$ around its zero 
crossing at $B=B_0$. 
Using~(\ref{eq:Nb2}), the number of closed channel molecules near resonance 
\begin{equation}
N_b(B\approx B_0)=\frac{r^{\star}}{4\pi}\,\int_{\mathbf{R}}\,\mathcal{C}(\mathbf{R})
\label{eq:Nb3}
\end{equation} 
is therefore a direct measure of the many-body contact $C=\int_{\mathbf{R}}\,\mathcal{C}(\mathbf{R})$
multiplied with the two-body parameter $r^{\star}$ which characterizes the intrinsic width
of the Feshbach resonance. An experiment based on this connection has 
been performed by Partridge et.al.~\cite{part05} in a  two component gas of $^6$Li near the 
Feshbach resonance at $B_0\simeq 832\,$G even before 
the relation with the Tan contact was realized. Specifically, they have determined 
the loss of atoms which results from exciting the closed channel molecules to a short lived molecular state. 
The resulting loss rate $\Gamma_{\rm loss}=\tilde{Z}\cdot\Omega^2/\gamma$
defined by $\dot{N}=-2 N_b(t)\,\Omega^2/\gamma=-\Gamma_{\rm loss}\, N$
\footnote{For a gas near infinite scattering length, the decay in the trap is actually not exponential and thus
both $\Gamma_{\rm loss}$ and $\tilde{Z}$ will be time-dependent~\cite{wern09closed}. In practice, an initial decay rate 
is measured.} 
depends on the effective Rabi frequency $\Omega$ of the transition, the spontaneous emission rate $\gamma$ of the molecular state
and the closed channel fraction $\tilde{Z}=N_b/(N/2)$ (we use the tilde to distinguish this from the closed channel
admixture $Z$ at the two-body level, as defined in Eq.~(\ref{eq:coupled})). Now, as noted above, the contact density
 $\mathcal{C}(\mathbf{R})=s(0) \cdot  k_F^4(\mathbf{R})$ of the unitary gas at $T=0$ 
scales with the fourth power of the local Fermi wavevector $k_F(\mathbf{R})$, with $s(0)\simeq 0.1$.
Neglecting the trap inhomogeneity, this gives a closed channel fraction 
\begin{equation}
\tilde{Z}(B\approx B_0)\simeq k_Fr^{\star}/2
\label{eq:ztilde}
\end{equation} 
near resonance. Specifically, for the open channel dominated resonance in $^6$Li near $832\,$G with $s_{\rm res}=\bar{a}/r^{\star}\simeq 59$
and $\bar{a}\simeq 30\, a_B$~\cite{chin10feshbach}, the finite density closed channel fraction 
$\tilde{Z}\simeq k_F\bar{a}/(2s_{\rm res})\simeq10^{-4}$ near $B_0$ is very small since both $k_F\bar{a}$ and $1/s_{\rm res}$ are 
much less than one.
More generally, as  shown by Werner et.al.~\cite{wern09closed}, the theoretical estimate for $\tilde{Z}$ based on its connection with the Tan contact 
and its dependence on $1/k_Fa$ near unitarity is in fair agreement with the values observed earlier by Partridge et.al.~\cite{part05}.
Note that $\tilde{Z}$ has a finite value precisely {\it at} the Feshbach resonance, 
in contrast to the closed channel admixture at the two-body level, which obeys $Z(B=B_0)\equiv 0$, see~(\ref{eq:Z}). 
In fact, the latter result is obtained as a trivial limit of the expression $N_b=r^{\star}C/4\pi$ from Eq.~(\ref{eq:Nb3})
by noting that the contact in the two-body limit $N=2$ is simply $C_{2-\rm body}=8\pi/a$ for $a>0$ and
zero otherwise. This follows easily from the Tan adiabatic theorem by noting that the energy 
of the two-body problem depends on the scattering length via $E_{2-\rm body}=-\epsilon_b(a)=-\hbar^2/ma^2$.   
The resulting value $\tilde{Z}\to N_b=2r^{\star}/a$ then agrees with the expression given in Eq.~(\ref{eq:Z}).

\subsection{Single channel model and zero range limit}
\begin{figure}[b]
\begin{minipage}[t]{\textwidth}
\centering
\setlength{\unitlength}{\linewidth}
  \includegraphics[width=0.2\linewidth]{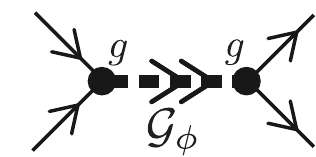}	
  \end{minipage}
  \caption{Tree level diagram yielding the effective atom-atom scattering amplitude Eq.~\eqref{eq:fkexact}.}
  \label{ef3_fig_tree}
\end{figure}

For a derivation of the connection between the thermodynamically defined contact
and microscopic correlation functions at short distances or time scales, it is convenient to replace 
the two-channel model of Eq.~(\ref{eq:BFM}) by an effective single channel model whose
interaction potential is adjusted to give the correct scattering length. 
On a formal level, the reduction 
to an effective single channel description can 
be obtained by integrating out  the bosonic field $\Phi$ in an effective 
action version of Eq.~(\ref{eq:BFM}). The scattering of two atoms in the open channel is 
mediated by the exchange of the dimer field $\Phi$. It gives rise to s-wave scattering only
because $\chi(\mathbf{x})$ is rotation invariant. By construction, therefore, the two-channel model
only describes an effective interaction between Fermions of opposite spin, i.e. in the 
'pairing channel'.  Evaluating the diagram shown in Fig.~\ref{ef3_fig_tree}, where the dashed (solid) line denotes the propagation of a \textit{full} dimer (atom) and the dot represents the Yukawa coupling, the effective scattering amplitude 
of two atoms with momenta $\pm\mathbf{k}$ in their center of mass frame is given by
\begin{equation}\label{eq:fkexact}
f(k)=\frac{m}{4\pi\hbar^2}\,\tilde{g}^2\chi^2(k)\, \mathcal{G}_\Phi(E=\frac{\hbar^2k^2}{m},\mathbf{Q}=0)\, .
\end{equation}
It has a nontrivial energy dependence which is determined by the full inverse propagator 
\begin{equation}
\label{ef3_FullG}
\mathcal{G}_\Phi^{-1}(E,\mathbf{Q}=0) = -E+\nu_c(B)+\frac{m\tilde{g}^2}{\hbar^2}\,\int_q\frac{\chi^2(q)}{k^2-q^2 +i0}
\end{equation}
of the dimer field at vanishing total momentum $\mathbf{Q}=0$ and energy $E=\hbar^2k^2/m$.
As expected, the expression~(\ref{eq:fkexact}) is identical to the one derived in 
Eq.~(\ref{eq:scattampcastin}) at the two-body level. The resulting scattering length and the effective 
range are therefore  given by Eq.~(\ref{eq:FB-range}). Now, in order to justify the replacement of
the interaction associated with the two-channel model by a single channel description 
involving only the scattering length for a Fermi gas at finite density,
it is necessary that the effective two-body scattering amplitude~(\ref{eq:fkexact})  
is of the idealized form $f(k)=-a/(1+ika)$ of a contact interaction 
at all relevant wave vectors up to $k_F$. This requires
the effective range $r_e$ of the interactions to be negligible, which is generically true for a dilute and degenerate gas in the
\begin{itemize}
\item {\bf zero  range limit}  $k_F|r_e|\to 0$.
\end{itemize}
(Note that this is different from the 'scaling limit' $r_e/a\to 0$ at the two- or few-body level~\cite{braa06},
where e.g. the two-body bound state energy $\epsilon_b=\hbar^2/ma^2$ at positive $a$ has a universal form depending only on the value of $a$.)
Now, for open-channel dominated Feshbach resonances, the effective range $r_e=3\bar{a}$ on resonance
is  of the order of the van der Waals length.  Since $k_F l_{\rm vdw}\ll 1$ is a necessary condition for a dilute gas 
(see Eq.~(\ref{eq:hierarchy})), the interaction between two 
opposite spin Fermions is therefore well descibed by an effective single channel potential with zero range.
It is important to emphasize that the criterion $k_F l_{\rm vdw}\ll 1$ is more restrictive than the
widely used condition $k_Fr^{\star}\ll 1$ (see e.g.~\cite{bruu04eff}) for the irrelevance of the closed channel in describing the physics near a 'broad'
Feshbach resonance, where the finite density closed channel fraction $\tilde{Z}(B=B_0)$ is negligible. The 
difference is relevant for the question whether the Bertsch parameter $\xi_s$ is indeed a universal 
number. For open channel dominated resonances, where $r^{\star}\ll l_{\rm vdw}$, this is 
true only in the limit where $k_F l_{\rm vdw}$ is taken to zero.     
Considering closed channel dominated resonances, their effective range $r_e\to -2r^{\star}$ is much larger
in magnitude than the van der Waals length.  Moreover - as pointed out in section 1.2  - the closed channel admixture 
even at the two-body level is of order one at detunings much less than the van der Waals energy.  In this limit, a 
single channel description is not possible.  In particular, the effective range may be such that $k_Fr^{\star}$ is
large compared to one, a situation which defines the regime of narrow Feshbach 
resonances. An example is the resonance between the two lowest hyperfine states of $^6$Li at 
$B_0\simeq 543\,$G, which has $s_{\rm res}\simeq 10^{-3}$~\cite{chin10feshbach}, implying $k_Fr^{\star}\simeq 10$
for typical values of the density.    
In the following, we will only consider open channel dominated resonances and approximate the
effective two-body scattering amplitude from Eq.~(\ref{eq:fkexact}) by that of an idealized contact interaction.
It is important to emphasize, that there are a number of physically relevant observables which 
{\it cannot} be described within this approximation, for example three-body losses, 
the Efimov effect and also - of course -  the closed channel fraction. They involve the additional microscopic lengths $l_{\rm vdw}$
and $r^{\star}$, thus violating the simple notion of universality which assumes 
that the interaction is completely described by the scattering length only. Within a description based 
on zero-range interactions, the breakdown of this assumption shows up through the appearance of an 
anomalous dimension. An example, discussed above, is  the three-body loss coefficient $L_3\sim (k_F\, l_{\rm vdw})^{2\gamma}$, 
which exhibits a dependence on the density characterized by the anomalous exponent $\gamma$.\\

For the single parameter which characterizes the interaction $V_{\uparrow\downarrow}(\mathbf{x})=V(\mathbf{x})$
between opposite spin Fermions, it is convenient to take its 
integrated strength $\bar{g}(\Lambda)=\int_{\mathbf{x}} V(\mathbf{x})$ or - equivalently - the associated scattering length 
$a_B(\Lambda)=m\bar{g}(\Lambda)/(4\pi\hbar^2)$ in Born-approximation.
The parameter $\Lambda$ is a high momentum cutoff. It accounts for a finite effective range,
with $\Lambda\to\infty$ the zero range limit. Formally, it is convenient to replace the 
interaction $V(\mathbf{x})\to\bar{g}(\Lambda)\,\delta(\mathbf{x})$ by a delta function. 
For any finite $\bar{g}$, this leads to a vanishing scattering amplitude.   
The cutoff dependent coupling constant
$\bar{g}(\Lambda)=4\pi\hbar^2a_B(\Lambda)/m$ must therefore be adjusted properly to 
give rise to a non-vanishing scattering length $a$.  Quite generally, the relation 
between the bare coupling constant $\bar{g}$ and the resulting 
value $g=4\pi\hbar^2 a/m$ of the low energy scattering amplitude may be determined from the solution of 
the Lippman-Schwinger equation 
 \begin{equation}
\tilde{f}(\mathbf{k}\to\mathbf{k}')=v(\mathbf{k}' - \mathbf{k}) +\int_q\,\frac{ v(\mathbf{k}' - \mathbf{q}) \tilde{f}(\mathbf{k}\to\mathbf{q})}
{k^2-q^2+i0}\, ,
\label{eq:LS-equ}
\end{equation}
for the scattering amplitude $\tilde{f}=-4\pi f$, where $v(\mathbf{k})$ is the Fourier transform of the microscopic  two-body potential 
$mV(\mathbf{x})/\hbar^2$. 
Replacing the latter by its expression $v(\mathbf{k})\to 4\pi a_B(\Lambda)$ within the zero range approximation 
and taking the limit $\mathbf{k}\to 0$, where  $\tilde{f}\to gm/\hbar^2$, gives
 \begin{equation}
g=\bar{g}(\Lambda)-g\,\bar{g}(\Lambda) \,\int_q\frac{1}{2\epsilon_q}\;\;\;\;\; {\rm or} \;\;\;\;\;
\frac{1}{\bar{g}}=\frac{1}{g}-\int_q\frac{1}{2\epsilon_q}\, ,
\label{eq:gbar}
\end{equation}
where $\epsilon_q=\hbar^2q^2/2m$ is the energy of a free particle. The divergent integral in~(\ref{eq:gbar})
is now made finite by using a sharp cutoff $\Lambda$ in momentum space. Within this specific regularization, 
the relation between the bare and the physical value of the scattering length is given by
\begin{equation}
a_B(\Lambda)=\frac{a}{1-2a\Lambda/\pi} \;\;\;\;\; {\rm or} \;\;\;\;\; a=\frac{a_B(\Lambda)}{1+2a_B(\Lambda)\Lambda/\pi}\, .
\label{eq:abar}
\end{equation}
Starting with an interaction which is repulsive at the microscopic scale, i.e. $a_B(\Lambda)> 0$, the second form of
Eq.~(\ref{eq:abar}) shows that in the zero-range limit $\Lambda\to\infty$, the scattering length $a$ approaches zero. 
This is a reflection of the well known fact that purely repulsive potentials can feature a scattering length at most as 
large as their range~\cite{land77qm}. 
For cold atoms, the microscopic interactions are, however, attractive, i.e. $a_B(\Lambda)<0$ is negative.
As a result,  any desired scattering length $a$ can be achieved by fine-tuning $a_B(\Lambda)$ for a given finite value 
$|r_e|\simeq 1/\Lambda$ of the potential range. 
In particular, it is also possible to take the zero-range limit $\Lambda\to\infty$, while still retaining an arbitrary finite value of $a$.  
Thus, by adjusting  $a_B(\Lambda)$ according to Eq.~(\ref{eq:abar}) and afterwards sending 
$\Lambda$ to infinity, all information about the short-range details is hidden in the single parameter $g=4\pi\hbar^2 a/m$.
The somewhat counterintuitive result that 
finite values of $a$ require the strength $a_B(\Lambda)$  of the attractive delta function potential to 
vanish inversely with the cutoff can be understood by considering a simple example: 
consider two particles with mass $m$ which interact via an attractive 
square well potential with range $b$ and depth $V_0=\hbar^2k_0^2/m$.
In order to obtain a non-vanishing scattering amplitude 
in the zero range
limit $b\sim 1/\Lambda \to 0$, it is necessary that the depth parameter $k_0\sim\Lambda$ diverges 
linearly with $1/b$. As a result, the integrated strength $\int V(\mathbf{x})=\bar{g}(\Lambda)\sim -V_0b^3$
vanishes like $1/\Lambda$.

\subsection{Short distance correlations}
In order to derive the Tan relations within the zero range model, we consider the 
formal expression for the operator of the interaction energy density 
\begin{equation}
\hat{\epsilon}_{\text{int}}(\mathbf{R})=\bar{g}(\Lambda)\,
\hat{\psi}_{\uparrow}^{\dagger}(\mathbf{R})
\hat{\psi}_{\downarrow}^{\dagger}(\mathbf{R})
 \hat{\psi}_{\downarrow}(\mathbf{R})\hat{\psi}_{\uparrow}(\mathbf{R})\equiv \bar{g}(\Lambda)\,\mathcal{\hat{O}}_c(\mathbf{R})
 \label{eq:Hshortrange2}
\end{equation}
which contains the bare coupling constant $\bar{g}$. 
Its expectation value $\epsilon_{\text{int}} =\bar{g}(\Lambda)\,\langle\mathcal{\hat{O}}_c\rangle\sim\Lambda$ 
diverges linearly with the cutoff because, as follows from Eq.~(\ref{eq:contact-def}) below, it is the 
product $\bar{g}^2(\Lambda)\,\langle\mathcal{\hat{O}}_c\rangle$ which is finite in the zero range limit $\Lambda\to\infty$.   
In physical terms, the result $\epsilon_{\text{int}}\sim\Lambda$ means that the interaction energy density is
linearly sensitive to the range of interactions. Both $\epsilon_{\text{int}}$ and the product $\mathcal{\hat{O}}_c$ of the 
four field operators at the same point in space, which scales like $\Lambda^2$, are therefore ill-defined in the zero range limit.
By contrast, the expectation value of the {\it total} energy density $\epsilon=\epsilon_{\text{kin}}+\epsilon_{\text{int}}$
should be finite as $\Lambda\to\infty$. Using the Hellman-Feynman theorem and the first Equation in~(\ref{eq:abar}), its 
dependence on the scattering length $a$ is determined by
 \begin{equation}
\frac{\partial\epsilon}{\partial a}=\frac{\partial\bar{g}}{\partial a}\cdot\langle\mathcal{\hat{O}}_c\rangle=
\frac{\bar{g}^2}{ga}\cdot\langle\mathcal{\hat{O}}_c\rangle\, .
\label{eq:Hellman}
\end{equation} 
The requirement of a finite energy density which depends in a continuous manner on $a$,
therefore implies that the combination
 \begin{equation}
\lim_{\Lambda\to\infty}\,\bar{g}^2(\Lambda)\cdot\langle\mathcal{\hat{O}}_c(\mathbf{R})\rangle=
\frac{\hbar^4}{m^2}\cdot\mathcal{C}(\mathbf{R})
\label{eq:contact-def}
\end{equation} 
remains finite as the cutoff is taken to infinity. The relation~(\ref{eq:contact-def}) defines the 
contact density $\mathcal{C}$ in the zero range limit.
It leads immediately to the local form of the Tan adiabatic theorem
 \begin{equation}
\frac{\partial\epsilon}{\partial a}(\mathbf{R})=\frac{\hbar^2}{4\pi ma^2}\cdot\mathcal{C}(\mathbf{R})\;\;\;\;\; {\rm or} \;\;\;\;\;
\frac{\partial\epsilon}{\partial (1/a)}(\mathbf{R})=-\frac{\hbar^2}{4\pi m}\cdot\mathcal{C}(\mathbf{R})\, ,
\label{eq:Tan-adiabatic-local}
\end{equation}
which may be viewed as a special case of the Hellman-Feynman theorem for systems with zero
range interactions.
In its integral form, this coincides with Eq.~(\ref{eq:Tan-adiabatic}) where the contact has been 
defined in a purely thermodynamic manner. 
In order to derive the connection between the contact density and the tail of the momentum distribution
mentioned above, we consider the total energy density in the translation invariant case
 \begin{equation}
\epsilon=\sum_{\sigma}\int_q\, \epsilon_q\, n_{\sigma}(q) +\frac{1}{\bar{g}}\cdot\frac{\hbar^4\mathcal{C}}{m^2}\, .
\label{eq:Tan-energy}
\end{equation} 
It is a sum of the kinetic and the interaction contribution which involve the momentum distribution $n_{\sigma}(q)$  
and the contact density, according to its definition in~(\ref{eq:contact-def}).
With a finite value for $\mathcal{C}$, the interaction term apparently diverges linearly with the cutoff, as noted 
above. This divergence is cancelled, however, by a divergence in the kinetic energy. To see this, 
the interaction term is rewritten 
by using~Eq.~(\ref{eq:gbar}). This leads to a sum of two finite contributions for the total energy density
 \begin{equation}
\epsilon=\sum_{\sigma}\int_q\,\epsilon_q\, \left[n_{\sigma}(q) -\frac{\mathcal{C}}{q^4}\right]+\frac{\hbar^2\mathcal{C}}{4\pi m a}\, ,
\label{eq:Tan-energy2}
\end{equation} 
which is the Tan energy theorem~\cite{tan08energy}. The finiteness of the momentum integral implies that the contact density
determines the weight of the tail
 \begin{equation}
\lim_{q\to\infty} n_{\sigma}(q)=\frac{\mathcal{C}}{q^4} \qquad {\rm for} \qquad q\gg k_F\, ,1/\lambda_T
\label{eq:Tan-momentum}
\end{equation}  
of $n_{\sigma}(q)$ at large momentum, which is identical for both spin polarizations $\sigma=\pm 1$, 
even if the gas is not balanced.  In practice, the power law behavior $\sim 1/q^4$,  
which was realized first by Haussmann~\cite{haus94}, applies 
for momenta larger than the characteristic scales $k_F\, ,1/\lambda_T$
\footnote{For $a>0$, one needs $q\gg 1/a$ in addition due to the presence of two-body bound states.}. 
It is therefore observed most easily in a deeply degenerate gas near unitarity, where 
it appears already at $q\gtrsim 2k_F$~\cite{haus94}. A remarkable consequence of the 
result~(\ref{eq:Tan-energy2}) is that the total energy of
Fermions with zero range interactions can be expressed completely in terms of the
momentum distribution, i.e. the Fourier transform of the {\it one}-particle density matrix
\footnote{For interactions which are not of zero range, the total energy is still determined by 
the {\it one}-particle Green function, however it requires knowledge of its full 
momentum and frequency dependence~\cite{fett71}.}. 
A direct proof of the asymptotic behavior~(\ref{eq:Tan-momentum}) of the momentum distribution
can be given by using the operator product expansion. Indeed, the singular operator $\mathcal{\hat{O}}_c$ 
which arises in the interaction energy also appears as a non-analytic term 
$\sim|\mathbf{x}|\,\bar{g}^2\mathcal{\hat{O}}_c$ in the short-distance expansion 
\begin{equation}
 \hat\Psi_{\sigma}^{\dagger} (\mathbf{R}+\frac{\mathbf{x}}{2}) \hat\Psi_{\sigma} (\mathbf{R}-\frac{\mathbf{x}}{2})\!=\hat{n}_{\sigma}(\mathbf{R})
 +i\mathbf{x}\cdot\hat{\mathbf{p}}_{\sigma}(\mathbf{R}) -\frac{\vert\mathbf{x}\vert}{8\pi}\,\bar{g}^2(\Lambda)
\,\hat\Psi_{\uparrow}^{\dagger}\hat\Psi_{\downarrow}^{\dagger}\hat\Psi_{\downarrow}\hat\Psi_{\uparrow}(\mathbf{R})+\ldots
\label{eq:OPE}
\end{equation}
of the one-particle density matrix  
as $|\mathbf{x}|\to 0$~\cite{braa08contact}. Taking the expectation value and noting that the Fourier
transform of $|\mathbf{x}|$ is $-8\pi/q^4$ immediately gives~(\ref{eq:Tan-momentum}).  

For a better understanding of the physical meaning of the contact density introduced
formally in Eq.~(\ref{eq:contact-def}), it is convenient to consider the density correlation function
\begin{equation}
\langle\hat{n}_{\uparrow}(\mathbf{R}+\mathbf{x}/2)\,\hat{n}_{\downarrow}(\mathbf{R}-\mathbf{x}/2)\rangle\!=\!
 n_{\uparrow}(\mathbf{R}) n_{\downarrow}(\mathbf{R}) 
 g_{\uparrow\downarrow}^{(2)}(\mathbf{x},\mathbf{R})
 \label{eq:g2}
\end{equation}
between two Fermions with opposite spin at short separation $|\mathbf{x}|$. 
In the limit $k_F l_{\rm vdw}\ll 1$ of a dilute gas, the probability $\sim (k_F l_{\rm vdw})^{s}$ (see section 1.3) 
that a third Fermion is also close by is negligible. The two 
Fermions thus only feel their two-body interaction.
Formally, the product
\begin{equation}
\lim_{r\to 0}\,\hat{\psi}_{\downarrow}(\mathbf{R}-\mathbf{x}/2)\,\hat{\psi}_{\uparrow}(\mathbf{R}+\mathbf{x}/2)\, =
\frac{\psi_0(r)}{4\pi}\,\hat{\phi}(\mathbf{R})
\label{eq:dimer-operator1}
\end{equation}
 of two operators at short distances is therefore proportional to the two-body wavefunction $\psi_0(r)$ 
 in vacuum and an operator 
 $\hat{\phi}(\mathbf{R})$, which is regular as $r\to 0$ and finite in the zero range limit.
To determine the form and normalization of $\psi_0(r)$ and the connnection between the 
operator $\hat{\phi}(\mathbf{R})$ and the contact density, we note that the Schr\"odinger equation 
for relative motion of two particles with zero angular momentum - which always dominate at short distances - reads
\begin{equation}
\label{eq:chi}
\left[ \frac{d^2}{dr^2}-\frac{mV(r)}{\hbar^2}\right]\,\chi(r; k)=k^2\,\chi(r; k)
\end{equation}
for $\chi(r;k)=r\cdot\psi(r;k)$.
The asymptotic behavior $\chi(r;k=0)\sim (1-r/a)$
of the associated zero energy solution at distances large compared to the range of the 
potential defines the exact scattering length. Within the zero range approximation, 
this behavior remains valid also at short distances, where the actual two-body potential 
$V(r)$ becomes infinitely repulsive and thus always dominates the kinetic energy $\sim k^2$. 
Choosing $\psi_0(r)=1/r-1/a$, it turns out that the operator introduced in~(\ref{eq:dimer-operator1}) is given by  
\begin{equation}
\hat{\phi}(\mathbf{R})=\lim_{\Lambda\to\infty}\, 4\pi a_B(\Lambda)\, \hat{\psi}_{\downarrow}(\mathbf{R})
\hat{\psi}_{\uparrow}(\mathbf{R})\, .
\label{eq:dimer-operator2}
\end{equation}
Indeed, in terms of this operator, the definition~(\ref{eq:contact-def}) of the contact density may be rewritten in the form
 \begin{equation}
\mathcal{C}(\mathbf{R})=\lim_{\Lambda\to\infty}\,\left(4\pi a_B(\Lambda)\right)^2\,\langle\mathcal{\hat{O}}_c(\mathbf{R})\rangle=
\langle\hat{\phi}^{\dagger}(\mathbf{R})\hat{\phi}(\mathbf{R})\rangle\, .
\label{eq:contact-dimer}
\end{equation}
A simple product of Eq.~(\ref{eq:dimer-operator1}) and its hermitean conjugate, which is legitimate to leading order, then leads to a singular behavior 
\begin{equation}
\label{eq:OPE}
 \langle\hat{n}_{\uparrow}(\mathbf{R}+\frac{\mathbf{x}}{2})\,\hat{n}_{\downarrow}(\mathbf{R}-\frac{\mathbf{x}}{2})\rangle
 =\frac{\mathcal{C}(\mathbf{R})}{16\pi^2}\,\left(\frac{1}{r^2}-\frac{2}{ar}+\ldots\right)
\end{equation}
of the dimensionless pair distribution function $g_{\uparrow\downarrow}^{(2)}(\mathbf{x},\mathbf{R})\sim\mathcal{C}(\mathbf{R})/|\mathbf{x}|^2$  
for opposite spins as $|\mathbf{x}|\to 0$, consistent with the linear divergence with cutoff 
of the interaction energy density $\epsilon_{\rm int}=\hbar^4\mathcal{C}/(m^2 \bar{g}(\Lambda))$.
The short distance behavior~(\ref{eq:OPE}) is valid for length scales $l_c$ smaller than the inverse 
characteristic momenta in Eq.~(\ref{eq:Tan-momentum}), i.e. for $|\mathbf{x}|<n^{-1/3}, \lambda_T$.  
For a homogeneous, balanced Fermi gas with density $n=n_{\uparrow}+n_{\downarrow}=2n_{\uparrow}$, the 
standard connection between $g^{(2)}(\mathbf{x})$ 
and the static structure factor implies that the singular behavior~(\ref{eq:OPE}) gives rise to a quite slowly decaying tail 
\begin{equation}
\label{eq:S12}
S_{\uparrow\downarrow}(q)
 =\frac{\mathcal{C}}{8n_{\downarrow}}\,\left(\frac{1}{q}-\frac{4}{\pi aq^2}+\ldots\right)
\end{equation}
in the structure factor for opposite spins at large momenta~\cite{comb06,hu10}. 
As mentioned above, this relation can be used to measure the contact density 
of the Fermi gas near unitarity via Bragg spectroscopy, which in fact gives access to the 
full dynamical structure factor $S(\mathbf{q},\omega)$~\cite{kuhn11contact,hoin13contact}.     
For a simple physical interpretation of the anomalous behavior~(\ref{eq:OPE}), 
it is useful to recall the standard definition of a pair distribution function and ask how many
$\uparrow$-Fermions will - on average - be found in a sphere of radius $b$ around a $\downarrow$-Fermion
fixed at some position $\mathbf{R}$. At large distances $b\gg n^{-1/3}$, the presence of a $\downarrow$-Fermion
at $\mathbf{R}$ becomes irrelevant and  $N_{\uparrow}(b,\mathbf{R})=n_{\uparrow}(\mathbf{R})\cdot 4\pi b^3/3$
scales linearly with the volume of the sphere. 
By contrast,  in the limit where $r_e< b\ll l_c$ this 
number can be calculated using the short distance behavior~(\ref{eq:OPE}), which gives  
  \begin{equation}
\label{eq:Nup}
N_{\uparrow}(b\to 0,\mathbf{R})=\int_{|\mathbf{x}|<b} d^3x\,\frac{\langle\hat{n}_{\uparrow}(\mathbf{R}+\mathbf{x})\,\hat{n}_{\downarrow}(\mathbf{R})\rangle}
{ n_{\downarrow}(\mathbf{R})} =\frac{\mathcal{C}(\mathbf{R})}{4\pi  n_{\downarrow}(\mathbf{R})}\cdot b\, .
\end{equation}
For a strongly interacting, degenerate gas where $l_c\simeq n^{-1/3}$, the number of $\uparrow$-Fermions 
in a sphere whose radius is below the average interparticle spacing therefore scales linearly with the {\it radius}
of the sphere instead of linearly with its volume! This anomalous behavior is a result of the $1/r^2$-dependence 
of the probability density $|\psi_0(r)|^2$ at short distances, which cancels the factor $4\pi r^2$ from
the volume element. The number of $\uparrow$-Fermions in the range  $r_e< b\ll n^{-1/3}$
is therefore linear in $b$, vanishing much more slowly than the naively expected $b^3$-law, 
which would apply if $g_{\uparrow\downarrow}^{(2)}(0,\mathbf{R})$
were finite. For distances below the effectice range $r_e$, 
the details of the short range repulsion matter, typically leading to an exponentially small
$N_{\uparrow}(b\ll r_e)\sim\exp{(-(r_e/b)^{\alpha})}$
\footnote{This is a result of the fact that the exact two-body wave function $\psi(r)$ 
vanishes very quickly at short distances. Taking a Lennard-Jones potential for instance, the
$1/r^{12}$ repulsion leads to an exponent $\alpha =5$.}.
In the limit $n_{\downarrow}\to 0$ of a strongly imbalanced gas, the contact density
has to vanish because there is no tail in the momentum distribution for a single-component,
non-interacting Fermi gas.  It turns out that in this limit $\mathcal{C}\sim \tilde{s}\, k_F n_{\downarrow}$ 
vanishes {\it linearly} with the minority density with a dimensionless prefactor $ \tilde{s}$ 
of order one and $k_F$ the Fermi wave vector of the majority component~\cite{punk09molaron}.
The ratio $\mathcal{C}/n_{\downarrow}$ in Eqs.~(\ref{eq:S12}) and~(\ref{eq:Nup}) thus stays finite in the limit 
of a single down-spin. Finally, we add three brief comments: \\

a) the derivation of the Tan relations in the zero range model
provides a simple example of an anomalous dimension. Indeed, the 
naive scaling dimension of the operator  $\hat{\phi}(\mathbf{R})$, which consists of a 
product of two Fermionic field operators $\hat{\psi}_{\downarrow}(\mathbf{R})
\hat{\psi}_{\uparrow}(\mathbf{R})$ is three. The short distance singularity $\psi_0(r)\sim(1/r-1a)$
of  the two-body wavefunction within the zero range model, however, 
makes the product of two field operators at the 
same point in space ill-defined. Eq.~(\ref{eq:dimer-operator2}) shows that a finite limit is only obtained 
by multiplying this product with $a_B(\Lambda)\sim 1/\Lambda$,
which gives an additional factor of length. As a result, the operator $\hat{\phi}(\mathbf{R})$ has scaling dimension 
$\Delta_{\phi}=2\Delta_{\psi}-1=2$, as used in Eq.~(\ref{eq:a-dimension}). This anomalous
dimension shows up in the linear dependence $\tilde{Z}\simeq k_Fr^{\star}/2$ of the closed channel
fraction of the unitary Fermi gas on $k_F$ discussed above.  Indeed, since the product  $\hat{\phi}^{\dagger}(\mathbf{R})\hat{\phi}(\mathbf{R})$
has dimension four, the integral $\int_{\mathbf{R}}\mathcal{C}(\mathbf{R})$ which appears
in the number of closed channel molecules~(\ref{eq:Nb3}) 
is not simply $\sim N$ but 
scales like $\sim k_F\cdot N$, as is necessary to make $\tilde{Z}$ dimensionless.

b) The Tan relations can also be derived for  
Fermions in either two or in one dimension with only slight changes, see Werner and Castin~\cite{wern12}
(who also provide a careful discussion of finite range 
corrections $\sim k_Fr_e$) or Barth and Zwerger~\cite{bart11}. In both cases,
zero range interactions give rise to a momentum distribution $n_{\sigma}(q)\to\mathcal{C}/q^4$
at large momenta. The associated contact density $\mathcal{C}$ again determines the dependence of
the free energy as a function of an appropriately defined scattering length. 

c) An analog of the Tan relations can also be proven for repulsive two-body potentials 
of the form $V(r\to 0)\sim 1/r^s$, in particular for  
Coulomb interactions, where $s=1$. In this case, the two-body wave function $\psi_0(r)= 1 + r/(2a_B) + \ldots$ is 
finite near the origin but non-analytic. This gives rise to a universal power law 
$n(q\to\infty)\sim g_{\uparrow\downarrow}^{(2)}(0)/q^8$ in the momentum distribution.
Its strength is determined by the pair distribution function $g_{\uparrow\downarrow}^{(2)}(0)$
at vanishing separation which is now finite, see Hofmann et.al.~\cite{hofm13jellium}.

 \newpage

\section{UNITARY FERMIONS: UNIVERSALITY AND SCALE INVARIANCE}

The ability to tune the strength of interactions in a two component 
Fermi gas via a Feshbach resonance allows to explore the
crossover from a BCS superfluid, when the attraction is weak and
pairs overlap strongly, to a molecular condensate 
of tightly bound pairs which may properly be viewed as Bosons
\footnote{For further information on this subject see the reviews~\cite{bloc08review,gior08review,rand14}
or the book~\cite{zwer12book}.} . 
Of particular interest in this context is the unitary regime $k_F|a|\gg 1$, 
where the scattering length is much larger than the interparticle spacing. 
As emphasized by Bertsch~\cite{bert00}, this limit is relevant not only 
for cold atoms near a Feshbach resonance but may also serve as an idealized model for 
understanding the equation of state of low density nuclear matter in neutron stars
(for a recent discussion of this subject see~\cite{geze14}).
Historically, the point at which the 
scattering length diverges was not expected to be of particular interest, because for
the many-body problem at finite density bound states are present at arbitrary coupling 
and the ground state is a superfluid on both sides of the unitary point. 

As realized by Nikolic and Sachdev~\cite{niko07renorm}, however, 
the unitary gas at zero density is a quantum multicritical point, which 
separates the onset transition from the vacuum to a finite density superfluid state into 
a regime where the flow is towards a weakly interacting gas of either Fermions or Bosons.
The thermodynamics of Fermions near unitarity is therefore
governed by a novel strong coupling fixed point and associated universal scaling functions. Moreover, as 
pointed out by Nishida and Son~\cite{nish07CFT}, the unitary gas realizes a non-relativistic 
field theory which is both scale and conformally invariant. The additional 
symmetries entail a number of exact results like the existence of 
a breathing mode at twice the trap frequency due to a hidden SO$\,(2,1)$ dynamical symmetry~\cite{wern06unitary}
or a vanishing bulk viscosity at arbitrary temperatures~\cite{son07bulk}. In addition, 
they allow for a controlled calculation of thermodynamic properties in a systematic expansion
around an upper and lower critical dimension $d=4$ and $d=2$, respectively~\cite{nish06epsilon,nish07eps}.

\subsection{Quantum critical point and universality}

In the context of cold atoms, the notion of universality is usually associated with the fact
that observables depend only on the scattering length $a$ while all other details
of the microscopic interaction are irrelevant. Universality in this sense applies e.g. 
to weakly interacting Bose-Einstein condensates, whose properties are fully 
characterized by the dimensionless parameter $na^3$. As discussed above, the origin of this
kind of universality is that - at low energies and densities - the range $|r_e|$ of the interactions
is negligible compared to the average interparticle spacing.  Now, 
in more general terms, universality appears for systems near a critical point, where
a correlation length diverges and therefore the details of the underlying 
microscopic Hamiltonian become irrelevant. As a result, there is a wide class
of Hamiltonians - the critical manifold - which share the same critical behavior.  
To understand the connection between this point of view and the simple
truncation of the full interaction to a zero range potential which has 
the scattering length as a single variable and - moreover - to see which are the underlying critical points  
responsible for universality in dilute, ultracold quantum gases, 
it is convenient to consider the onset transition
at zero temperature from the vacuum state with no particles whatsoever to the state with a finite density~\cite{sach11book}.
For Bosons with repulsive interaction $g_B\!=\! 4\pi\hbar^2\, a_{B}/m_{B}>0$, the onset transition is well described 
by a Gross-Pitaevskii (or Bogoliubov in next-to-leading order) theory. In particular, the density of Bosons 
$n_B(\mu, T=0)=\mu_B/g_B+\ldots$ vanishes linearly
for positive chemical potential while $n_B(\mu, T=0)\equiv 0$ if $\mu_B<0$.  For an arbitrary finite 
value $a_B>0$ of the interaction between Bosons, $\mu_B=T=0$ is
therefore a quantum critical point.  It separates the formally incompressible vacuum state,
 where the density is pinned at zero and the superfluid, 
where it starts to rise linearly with $\mu_B$. The fixed point is a weak coupling one because
at low densities the associated correlation length is much larger
than the average interparticle spacing. Indeed, approaching the fixed point by decreasing 
the chemical potential at zero temperature, the correlation length is just the well known healing
length $\xi_{\mu}=\hbar/\sqrt{2m_B|\mu_B|}=(8\pi n_Ba_B)^{-1/2}$. It is large 
compared to the  average interparticle spacing since $n_Ba_B^3\ll 1$ in the low density limit.  
Similarly one finds $\xi_T\simeq\sqrt{\lambda_T^3/a_B}\gg n_B^{-1/3}$
if the critical point is approached within the quantum critical regime, say at $\mu_B=0$. 
As discussed by Sachdev~\cite{sach11book}, the universality of both 
thermodynamics and correlation functions of dilute Bose gases can be understood by studying the  
relevant perturbations around this seemingly trivial fixed point. \\

For attractive, two-component Fermi gases the zero density limit turns out to 
be of a fundamentally different nature depending on whether the associated 
scattering length $a$ is positive or negative. For $a>0$, the existence of a two-body bound state 
with energy $\epsilon_b=\hbar^2/ma^2$ implies that a finite density of Fermions appears
already for a negative (Fermion) chemical potential $\mu>-\epsilon_b/2$. Since the effective interaction 
between two bound Fermion pairs is repulsive with $a_B=a_{\rm dd}=0.6\, a$ (see Eq.~(\ref{eq:a_dd})),
a dilute gas of dimers realizes a weakly interacting
BEC.  It is described by the theory of a dilute Bose gas above with $\mu_B=2\mu+\epsilon_b$.
In particular, a mean-field approach is adequate in $d=3$ since this is above the upper critical dimension
two~\cite{sach11book}.
For negative values of $a$, there is no bound state. A finite  density of Fermions thus only
appears for $\mu>0$ with $n(\mu, T=0)\sim\mu^{3/2}$ to leading order. The attractive 
interaction between the Fermions leads to the well known BCS instability, so the ground state
is again a superfluid. In the low density regime $\mu\ll\hbar^2/ma^2$ or - equivalently -
$k_F|a|\ll 1$ - pairing affects only a tiny range around the Fermi energy. 
A dilute Fermi gas at any finite, negative value of the scattering length is thus described by 
a different but again well understood quantum critical point at $\mu=T=0$  which separates the vacuum 
from a weak coupling superfluid at $\mu>0$ and $k_BT\ll\mu$~\cite{sach11book}.  A completely different situation arises
precisely at infinite scattering length. There, lowering the density or chemical potential from 
a finite value towards zero, one never reaches a dilute gas of either Bosons or Fermions.  Instead,
the system remains at strong coupling for arbitrary small values of the density. As shown by
Nikolic and Sachdev~\cite{niko07renorm}, the physics of Fermions near a Feshbach 
resonance is governed by a novel, non-perturbative fixed point whose properties are fundamentally
different from its weak coupling counterparts discussed above.

\begin{figure}[b]
\begin{center}
 \includegraphics[width=4in]{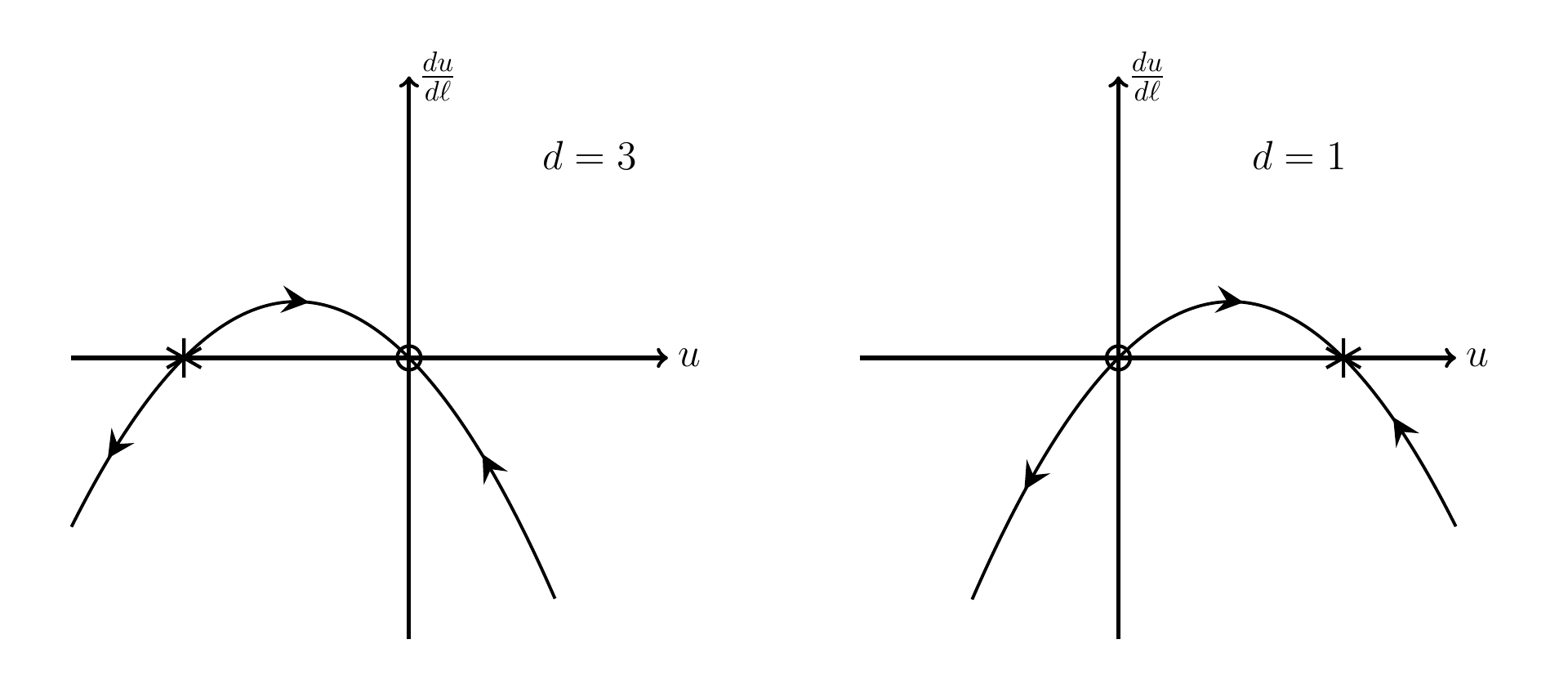}	
   \end{center}
  \caption{Flow of the dimensionless coupling $du/d\ell=\beta_d(u)$ in $d=3$ and $d=1$.
  The unstable attractive fixed point in $d>2$ turns into a stable, repulsive one in $d<2$. }
  \label{fig:flow}
\end{figure} 

To understand the crucial difference between the approach to zero density 
at either positive or negative values of $a$ and the properties of the fixed point at $\mu = T=1/a = 0$, 
it is sufficient to consider the dependence of the dimensionless coupling constant 
 \begin{equation}
u(\Lambda)=\frac{m\Lambda}{\pi^2\hbar^2}\, \bar{g}(\Lambda)
=\frac{4 a\Lambda/\pi}{1-2a\Lambda/\pi}\, .
\label{eq:ubar}
\end{equation}
on the momentum cutoff $\Lambda$. Since $\Lambda$ is 
inversely proportional to the effective range, 
$u(\Lambda)$ is essentially the ratio between the microscopic value $a_B(\Lambda)$ 
of the scattering length and the effective range. The explicit dependence on $\Lambda$ 
here follows from the first form of Eq.~(\ref{eq:abar}), i.e. it is based on a 
particular form of the cutoff procedure. This will affect the specific value $u^{\star}$
for the fixed point, but it does not affect the overall flow. In particular,
the result ${\rm dim} [\nu]=d-2$ below for the scaling dimension of the detuning away from
unitarity is universal.
Using Eq.~(\ref{eq:ubar}), it is easy to show that
 under an infinitesimal reduction $d\Lambda=-\Lambda d\ell$ of the cutoff 
$\Lambda(\ell)=\Lambda e^{-\ell}$, the coupling $u$ changes according to 
 \begin{equation}
\frac{du}{d\ell}=-u-\frac{u^2}{2}=\beta_3(u)\, ,
\label{eq:uflow}
\end{equation} 
which is exact to all orders in $u$. The flow has a stable fixed point at $u_0^{\star}=0$ which attracts either positive or small negative values of 
$u$.   In addition, there is an unstable fixed point at $u^{\star}=-2$, which is just the value of
$u$ attained in the limit $a\to\pm\infty$ at fixed $\Lambda$. Here the flow is 
towards $u=0$ if $u>u^{\star}$ and to more negative values if $u<u^{\star}$, see Fig.~\ref{fig:flow}.
A simple physical interpretation of this dependence is obtained by considering 
the relation between the scattering length and microscopic parameters for an 
attractive square well potential with range $b$ and depth parameter $k_0$,
 as discussed above. Its first bound state appears at a critical value $(k_0b)_c=\pi/2$. 
 In this specific example, the role of $u(\ell=0)$ is $-(k_0b)^2$, up to a factor of order one.
 The regime $k_0b<\pi/2$, where the scattering length is negative and no bound state exists, 
 thus corresponds to $u^{\star}<u(\ell=0)<0$. In this regime, lowering of the cutoff - which 
increases the scale for measuring lengths  - is associated with a flow of 
$u(\ell)$ towards the non-interacting limit.  For $k_0b>\pi/2$, in turn, there is a two body-bound state
and a positive scattering length for particles in the continuum. This is the regime $u(\ell=0)<-2$,
where a lowering of the cutoff leads to a flow towards more negative values of the dimensionless coupling
and thus to an increase of the two-body bound state energy.
The non-trivial fixed point at $u^{\star}=-2$, finally, 
describes the situation at infinite scattering length. For this fine tuned value of 
the microscopic interaction, a reduction of the cutoff leaves the 
system staying at an infinite value of $a$. Deviations away from this point 
grow under a lowering of the cutoff, i.e.
they are a relevant perturbation. To determine the dependence of 
this fixed point on dimensionality and in particular the associated lower critical dimension, 
it is useful to generalize     
the flow equation~(\ref{eq:uflow}) to dimensions $d\ne 3$, where 
the right hand side is replaced by $\beta_d(u)=(2-d)u -u^2/2$~\cite{ niko07renorm}.
The unstable fixed point corresponding to a  resonance in
the scattering amplitude is now at $u^{\star}=-2(d-2)$. It is
associated with attractive interactions only for $d>2$. By contrast, 
for dimensions less than two, $u^{\star}$ is positive and
the fixed point turns into a stable one, as shown in Fig.~\ref{fig:flow}.
This stable fixed point 
describes dilute gases of either Fermions or  Bosons with {\it repulsive} 
interactions in $d<2$, which again exhibit universal behavior in the 
limit of low density~\cite{sach11book}. For example, for 
Bosons in one dimension, the fixed point at $u^{\star}=+2$ corresponds 
to the well known Tonks-Girardeau limit 
which describes a repulsive 1D Bose gas at low densities~\cite{bloc08review}. \\

In $d>2$, the unstable fixed point which generalizes the physics at 
a Feshbach resonance to non-integer dimensions where the low-energy 
scattering amplitude at 'unitarity' scales like $f(q)\to -1/(iq)^{d-2}$,  
has three relevant perturbations: the first one is the dimensionless detuning $\nu=u-u^{\star}$
away from the resonance. In $d=3$, the associated microscopic 
parameter is $\nu\simeq -\bar{a}/a$
\footnote{This follows from expanding Eq.~(\ref{eq:ubar}) near the fixed point
$u^{\star}=-2$ at $a=\pm\infty$ and identifying the short distance cutoff  $\pi/\Lambda\simeq \bar{a}$ 
with the effective range.
 Note that the parameter $\nu$ characterizes the microscopic two-body interaction. It should not be
confused with the dimensionless coupling constant $-1/k_Fa$ of the gas at {\it finite} density.}.
The scaling dimension of this
perturbation is obtained by linearization of the flow equation around the 
fixed point, which gives $d\nu/d\ell=\beta_d'(u^{\star})\,\nu+\ldots$ with a positive slope
$\beta_d'(u^{\star})=\dim[\nu]=d-2$. The two other relevant perturbations are the chemical potential and 
a possible finite difference $h=(\mu_{\uparrow}-\mu_{\downarrow})/2$ of the 
chemical potentials for the two spin species, which have both scaling dimension
$\dim[\mu]=\dim[h]=2$, see the discussion below Eq.~(\ref{eq:Lagrange-mu}).  
\begin{figure}[t]
\begin{center}
\includegraphics[width=5in]{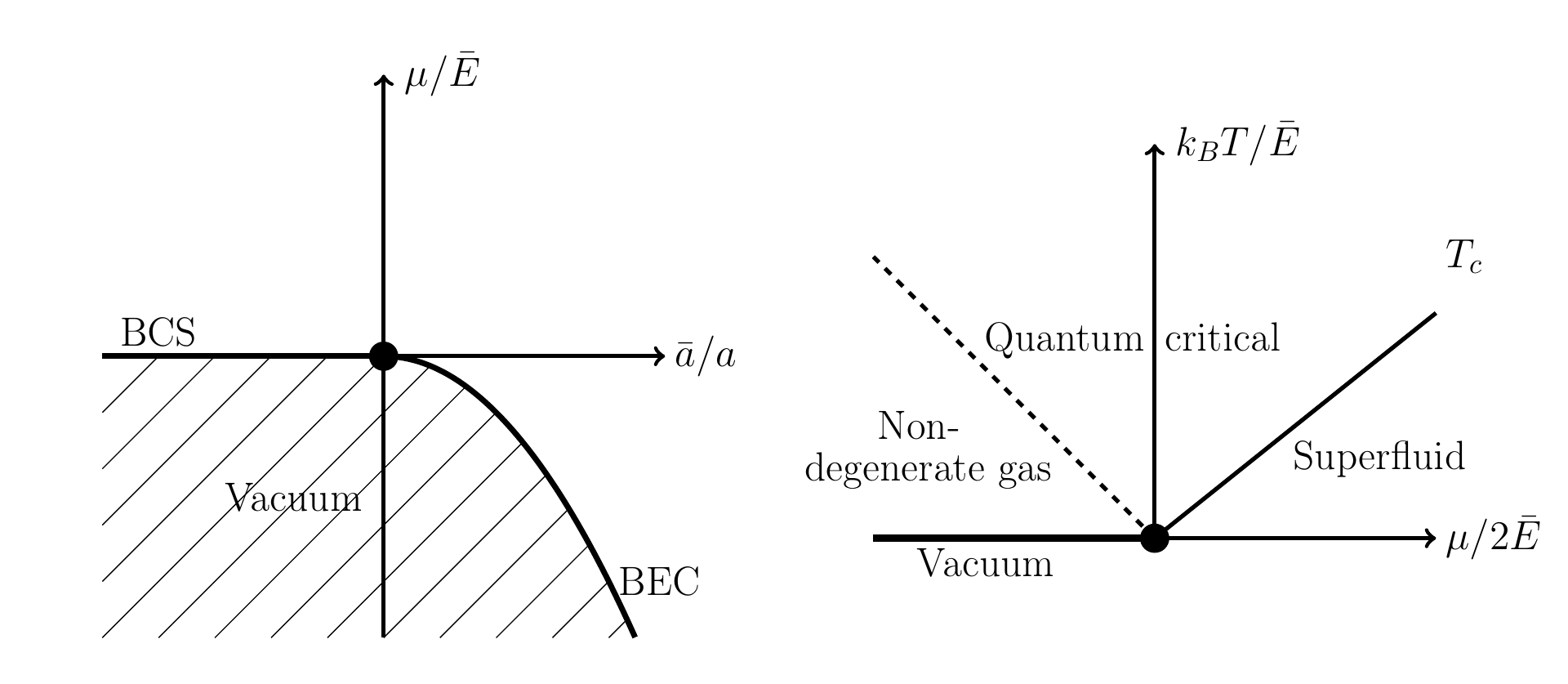}
\caption{Left: Zero temperature phase diagram of a dilute Fermi gas with
attractive interactions. The unstable fixed point at zero chemical potential 
and infinite scattering length $\bar{a}/a=0$ describes the physics near a Feshbach resonance (from~\cite{niko07renorm}). 
Right:  Phase diagram at finite temperature at unitarity $\bar{a}/a=0$. For $\mu>0$ there is a superfluid phase 
below $k_BT_c\simeq 0.4\mu$. In addition, there is a quantum critical regime above the fixed point for $|\mu|\ll k_BT$ (from~\cite{enss12QCP}).} 
\label{fig:phase}
\end{center}
\end{figure}
The zero temperature phase diagram in $d=3$ is shown in Fig.~\ref{fig:phase} for the case of a spin-balanced gas at $h=0$.
There are two lines of continuous quantum phase transitions:
$\mu=0,\,\bar{a}/a<0$ on the fermionic side and $\mu/\bar{E}=-(\bar{a}/a)^2,\,\bar{a}/a<0$ on the bosonic side. 
Both lines separate the zero density vacuum from a finite density superfluid.
The unstable fixed point at $\nu=0$ is the joint endpoint of these two lines and is thus a multicritical point.
There is no phase transition at finite density, only a smooth crossover between the BCS and BEC limits. 
At finite temperature, 
the phase diagram at infinite scattering length displays a superfluid below a critical temperature $k_BT_c\simeq 0.4\,\mu$
which scales linearly with $\mu$. The associated numerical factor $0.4$ is taken from the experiment 
by Ku et.al.~\cite{ku12superfluid} which will be discussed below.
There is a quantum  critical regime $|\mu|\ll k_BT$ above the fixed point which crosses 
over to a non-degenerate gas and eventually to the vacuum state along 
the line $T=0,\,\, \mu<0$.   Approaching the fixed point at $\mu,\nu,T=0$ from either the superfluid 
or the normal phase, the correlation length diverges. 
In the superfluid, the fact that $\mu=\xi_s\,\hbar^2k_F^2/2m$ at zero temperature with a universal
value $\xi_s\simeq 0.37$ of the Bertsch parameter implies that $\xi_{\mu}=1/(\sqrt{\xi_s}k_F)\simeq 1.6/k_F\simeq 0.5\, n^{-1/3}$. 
Similarly, the property of scale invariance at vanishing chemical potential fixes the correlation length in the 
quantum critical regime to 
be $\xi_T\simeq 1.43\, n^{-1/3}$ (for a derivation of this result see Eq.~(\ref{eq:t-dependent}) below).
In both cases, the correlation length diverges 
like the average interparticle spacing itself, i.e. the fixed point is a strong coupling one.

In contrast to conventional quantum critical points, which separate two phases of finite density,
the quantum critical points discussed above are quite unusual  because one of the phases
is a trivial vacuum state.
Nevertheless, these fixed points describe the thermodynamics and 
correlation functions of dilute gases in the three possible universality classes 
which show up in the present context: a) the weak coupling BCS superfluid in the 
regime $a<0$ and $k_F|a|\ll 1$ i.e. $\mu/\bar{E}\ll\nu^2$, 
b) the dilute gas of bosonic dimers if $a>0$ and $n_Ba_{B}^3\ll1$  i.e. $\mu_B/\bar{E}\ll\nu^2$
and - finally -  c) the unitary superfluid characterized by $\bar{a}\ll n^{-1/3}\ll |a|$ or
 $\nu^2\ll \mu_{(B)}/\bar{E}\ll 1$. In particular, the unstable fixed point at unitarity, 
where $T_c(\mu)\sim \mu$ rises linearly with $\mu$, 
governs the complete thermodynamics at finite temperature $T$, chemical potential
$\mu$ and the detuning $\nu$ of dilute, ultracold gases
near a scattering resonance. The thermodynamic functions in the vicinity of this fixed point 
therefore exhibit scaling. Specifically, the dimensionless pressure $\beta\lambda_T^3\, p$
of the gas is determined by a universal function
\footnote{Here we do not discuss imbalanced Fermi gases, which may arise at finite values of
the chemical potential difference  $h=(\mu_{\uparrow}-\mu_{\downarrow})/2$. For an introduction to
this quite rich subject see the reviews by Chevy and Salomon and by Recati and Stringari in~\cite{zwer12book}.} 
 
\begin{equation}
\beta\lambda_T^3\, p(\mu,T,1/a)=f_p\,(\beta\mu, \frac{\lambda_T}{a})
\label{eq:pressure-scaling}
\end{equation}
of the dimensionless variables $\beta\mu$ and $\lambda_T/a$. 
As will be discussed below, precisely this function
has been measured by Ku et.al.~\cite{ku12superfluid} for the unitary gas at $a=\pm\infty$ in the relevant range 
of $\beta\mu$ between the non-degenerate limit at $\beta\mu\simeq - 1.6$ down to and below the 
superfluid transition at $(\beta\mu)_c\simeq 2.5$. From  Eq.~(\ref{eq:pressure-scaling}) all other thermodynamic 
properties can be deduced by differentiation. For example, using Eq.~(\ref{eq:pressure}),
the density equation of state and the Tan contact follow from 
\begin{equation}
\lambda_T^3\, n(\mu, T,1/a)=\frac{\partial f_p\,(x, y)}{\partial x} \qquad {\rm and} \qquad
\lambda_T^4\,\mathcal{C}(\mu,T,1/a)=8\pi^2\, \frac{\partial f_p\,(x, y)}{\partial y}
\label{eq:contact-scaling}
\end{equation}
The result~(\ref{eq:pressure-scaling}) appears to be obvious from dimensional arguments: it just states
that the dimensionless combination  $\beta\lambda_T^3\, p$ can only depend on the dimensionless 
ratios $\beta\mu$ and $\lambda_T/a$ provided $a$ is the single relevant length scale. In this
form, it has been written down first by Ho~\cite{ho04uni}. One may thus ask what is the
additional insight gained by realizing the connection to the unstable fixed point at zero density discussed in some
detail above. The answer is apparent from Fig.~\ref{fig:phase}: 
just as in critical phenomena, the universality of the unitary gas is related to a RG fixed point at which a
correlation length diverges. The dimensionless variables $\beta\mu$ and $\lambda_T/a$
describe the two relevant perturbations away from this fixed point. Eq.~(\ref{eq:pressure-scaling}) 
is therefore analogous to the existence of a universal scaling function $f_{\rm sing} (h,t)$ for the 
singular contribution to the free energy density near a continuous, finite temperature 
ferromagnetic transition, where a non-zero magnetic field $h$ or a deviation $t=(T-T_c)/T_c$ 
away from the critical temperature are relevant perturbations. That simple
dimensional analysis is sufficient for the scaling function is a consequence of the fact 
that in the present case thermodynamic variables like pressure or density do not exhbit 
an anomalous dimension, in contrast to observables like the three-body loss
rate $L_3\sim (k_Fl_{\rm vdw})^{2\gamma}$ which involves the additional microscopic length $l_{\rm vdw}$.  
A further simplification which appears in the present case as a result of the fact 
that the fixed point is at zero density is that it is the {\it full} thermodynamic potentials
which exhibit scaling rather than a singular contribution on top of a smooth background
as usual, where scale invariance at the fixed point only appears for a suitable order parameter
but not for the complete microscopic Hamiltonian.      

Apart from this conceptual insight, the analysis of the unitary gas fixed 
point in dimensions away from $d=3$ is also useful for quantitative calculations
of universal numbers like the Bertsch parameter or the critical 
temperature for the superfluid transition 
within a systematic expansion around the upper and lower 
critical dimensions $d=4$ and $d=2$.     
This possibility has first been realized by Nishida and Son~\cite{nish06epsilon}.
It is based on the observation~\cite{nuss06becbcs}
that the unitary gas in four dimensions is an ideal Bose gas
while in two dimensions, it is an ideal Fermi gas. 
This surprising statement can be understood in physical terms
by noting that in four dimensions  a two-particle bound state in a zero range potential
only appears at infinitely strong attraction. Thus, already at an arbitrary small value of the
binding energy, the associated dimer size vanishes, quite in contrast to the
situation in $d=3$, where the size of the two-particle bound state is infinite at
unitarity. The unitary Fermi gas in four dimensions  is thus a non-interacting BEC,
similar to the limit $a\!\to\! 0^+$ in three dimensions. The $d = 4 - \epsilon$ expansion
may be complemented by an expansion around the lower critical dimension, which is two for the
present problem~\cite{nuss06becbcs,nish07eps}. 
Indeed, for $d\leq 2$ a bound state at zero binding energy
appears for an arbitrary weak attractive interaction. An expansion around $d = 2 + \epsilon$ is
thus effectively one around the non-interacting Fermi gas. 
Since pairing is an effect that only
appears at order $\exp{-1/\epsilon}$, this expansion only covers quantities which are not sensitive to superfluidity, 
for instance the equation of state and the Bertsch 
parameter, which is zero in $d=4$ and equal to one in $d=2$. 
For quantitatively reliable results in the
relevant case $d=3$,  the $\epsilon=4-d$ expansion has been extended up to three loops \cite{arno07bertsch}.
Within a Pad\'e resummation that takes into account
the exactly known limits $\xi_s(d\!\to\! 2)=1$ and  $\xi_s(d\!\to\! 1)=4$, the resulting
value in  3D is  $\xi_s=0.365\pm 0.01$~\cite{nish09eps,nish12book}. This
is perfectly consistent both with experiment (see below) and with the result $\xi_s=0.36$ 
obtained from a diagrammatic calculation based on the Luttinger-Ward
approach \cite{haus07bcsbec}.  \\

Before discussing the experimental results for the
thermodynamics of the unitary gas, it is important to distinguish 
the origin of universality in this context from the more familiar 
but quite different one which underlies the standard BCS description of
fermionic pairing. The origin of universality in BCS theory
relies on two assumptions:  the attractive interaction is weak and, moreover, 
non-vanishing only in a thin shell of thickness $\hbar\omega_c\ll\varepsilon_F$ 
around the Fermi surface. As a result, there is a separation of energy scales
$k_BT_c,\Delta \ll \hbar\omega_c\ll \varepsilon_F$. Here, the first inequality 
arises from the fact that both the critical temperature and the gap 
$\Delta\simeq\hbar\omega_c\,\exp{(-1/gN(0))}$ 
are suppressed by an exponentially small factor which only involves   
the product  $gN(0)\ll 1$ of the strength $g$ of the pairing interaction at the Fermi energy and the 
associated density of states $N(0)$.  With these conditions, the BCS description
of the paired superfluid is universal in the sense that 
the thermodynamic functions are independent of the cutoff $\omega_c$ 
and identical for all weak coupling superconductors if energies are 
measured in units of $\Delta$ or $k_BT_c$. A particular consequence of BCS
universality is that the compressibility is
not affected at all by the superfluid transition~\cite{legg65}.   
In ultracold atoms, the situation is completely different. Indeed, 
as discussed in section 1.2, the effective range of the interactions in
cold gases is $r_e\simeq 3\, l_{\rm vdw}$ for open or $|r_e|\simeq 2\, r^{*}$ for 
closed channel dominated Feshbach resonances. The relevant  
energy cutoff $\hbar\omega_c$ is thus either the 
van der Waals energy $E_{\rm vdw}$ or the much smaller energy  
$\epsilon^{*}=\hbar^2/m(r^{*})^2$. Since $E_{\rm vdw}\gg\varepsilon_F$
in the standard regime of dilute gases defined in Eq.~(\ref{eq:hierarchy}),
the effective interaction between Fermions obeys $k_F|r_e|\ll 1$ if 
it is due to an open channel dominated resonance. 
In contrast to standard BCS theory, the gap therefore exhibits no energy 
dependence in the relevant range below $\varepsilon_F$ and the characteristic scale for the critical temperature is
set by $T_F$.
In weak coupling, where $k_F|a|\ll 1$, one obtains
\begin{equation}
\label{eq:GorkovT_c}
T_c=\frac{8\exp{(\gamma_E)}}{(4e)^{1/3}\pi e^2}\, T_F\,\exp\bigl(-\pi/2k_F|a|\bigr)=0.277\, T_F \,\exp\bigl(-\pi/2k_F|a|\bigr)\, ,
\end{equation}
a result which has been derived by Gorkov and Melik-Barkhudarov  
in 1961~\cite{gork61} ($\gamma_E= 0.577$ is Euler's constant). Since the cutoff scale $E_{\rm vdw}$
is much larger than the Fermi energy, the ratio $T_c/T_F$ is a universal function of 
$k_Fa$. In particular, the slope $(k_BT/\mu)_c\simeq 0.41$ of the superfluid 
transition line in Fig.~\ref{fig:phase} fixes the corresponding linear 
dependence $T_c/T_F\simeq 0.16$ of $T_c$ on the Fermi energy at unitarity.  More generally,  
as emphasized above, the universality relevant for dilute gases is related to fixed points at zero density
and relies on being able to take the zero range limit $k_F|r_e|\!\to\! 0$. 
A description equivalent to that of BCS is possible only in the {\it opposite} limit $k_F|r_e|\gg 1$,
which is incompatible with our basic definition~(\ref{eq:hierarchy}) of a dilute gas. 
In practice, the regime $k_F|r_e|\gg 1$ becomes
relevant for ultracold gases if one considers closed channel
dominated resonances, for which the condition $k_Fr^{*}\gg 1$ of a 'narrow' Feshbach resonance 
can be achieved. 
The theoretical description is then very much simplified because
the bosonic field $\hat{\Phi}$ in the basic two channel Hamiltonian~(\ref{eq:BFM})
can be replaced by a c-number gap function $\Delta(\mathbf{x})$ via
\begin{equation}
\tilde{g}\,\int_{x'}\,  \chi(|\mathbf{x}-\mathbf{x'}|)\, \hat{\Phi}^{\dagger}(\frac{\mathbf{x}+\mathbf{x'}}{2})\;\to\;\Delta(\mathbf{x})\, .
\label{eq:replace}
\end{equation}
The  two channel model is thus reduced to the exactly solvable BCS Hamiltonian.  
Physically, the mean field replacement~(\ref{eq:replace})
is legitimate because for $s_{\rm res}\ll 1$ the closed channel state
is responsible for the interaction between the Fermions in the open channel
but is unaffected by their condensation, similar to phonons in
a conventional superconductor.  Motivated by its relevance 
in the regime $k_Fr^{*}\gg 1$ of narrow Feshbach resonances, 
an extended  BCS  description of fermionic pairing at arbitrary coupling strength
has been used in many publications, in particular in connection with
imbalanced Fermi gases, see e.g. the detailed analysis in Ref.~\cite{shee06phase}.
It should be kept in mind, however, that the extended BCS description 
does not account for the collective excitations in neutral superfluids associated
with the Bogoliubov-Anderson mode~\cite{bloc08review}. It only captures
fermionic excitations and is thus inapplicable at finite temperature beyond
the weak coupling limit.  
In the following, we focus on fermionic pairing in dilute gases in the relevant regime 
$k_F|r_e|\ll 1$ associated generically with open channel 
dominated resonances. 
They realize a novel universality class which is
associated with the fixed point structure discussed in Fig.~\ref{fig:phase}.

\subsection{Thermodynamics of the unitary Fermi gas}

Following the first experimental realizations of strongly interacting Fermi 
gases near a Feshbach resonance~\cite{ohar02science,bour03,bart04,rega04,zwie04rescond},
a lot of effort has been spent to measure their thermodynamic properties 
and in particular to determine the associated universal numbers
which characterize the unitary gas like the Bertsch parameter $\xi_s$
(the subscript in $\xi_s$ is a reminder of the fact that the parameter refers to the superfluid state).  
As mentioned in the context of Eq.~(\ref{eq:zeta}) above, 
it may be defined by the ratio $\xi_s=\epsilon_0/\epsilon^{(0)}=p_0/p^{(0)}$
of the ground state energy density or pressure to its  values
$\epsilon^{(0)}=3p^{(0)}/2=3n\varepsilon_F/5$ in the non-interacting Fermi gas. 
The Bertsch parameter also determines the enhancement of the zero temperature compressiblity
$\kappa_0/\kappa^{(0)}=1/\xi_s$ due to the attractive interactions compared to its value $\kappa^{(0)}=3/(2n\varepsilon_F)$
in the absence of interactions. 
Superfluid properties of the unitary gas are associated with new and independent universal numbers. 
Of particular interest are the ratios  $T_c/T_F$ and $\Delta/\varepsilon_F$ which 
determine the critical temperature for the
superfluid transition and the zero temperature gap
 for fermionic quasiparticles. 
%

Experimentally, the first measurements of the Bertsch parameter relied on determining the 
reduction  of the release energy~\cite{bour04coll} or of the cloud size observed by in-situ 
imaging of the density distribution~\cite{bart04}.
The values $\xi_s=0.36\pm 0.15$ and $\xi_s=0.32\pm 0.13$ obtained were 
smaller than those claimed in subsequent experiments~\cite{kina05heat,part06phase,stew06pot, nasc10thermo,hori10energy}.
They are close, however, to the value $\xi_s=0.37\pm 0.01$ obtained in the most precise measurements
to date by M. Zwierlein and coworkers at MIT~\cite{ku12normal, ku12superfluid}. 
In these measurements, all thermodynamic functions are determined from the 
density profile $n(V)$ as a function of the trap potential $V(\mathbf{x})$.
Since the latter is cylindrically symmetric, 
with harmonic confinement along the axial direction, the 3D density $n(V)$ may be obtained from an
inverse Abel transform of the measured column density. 
To determine the equation of state from $n(V)$ one uses the local density approximation (LDA),
where thermodynamic quantities like the pressure $p(\mathbf{x})$ are 
given by the corresponding equilibrium values in the {\it uniform} system
evaluated at the local density $n(\mathbf{x})$
\footnote{LDA 
is essentially the leading order in a semiclassical approximation~\cite{brac08}.
For the unitary gas in a harmonic trap, 'bulk' properties like the total
energy become exact in LDA for large particle number $N$, with 
corrections which vanish like $N^{-2/3}$~\cite{son06symmetry,haus08}.}.
Within LDA, the change in the local chemical potential 
$d\mu = -dV$ is just the negative of the change $dV$ in the local potential. 
Using $dp=n\, d\mu$, the pressure therefore 
follows from an integration $p(\mu)=\int_{-\infty}^{\mu} d\mu' n(\mu')=\int_{V}^{\infty} dV' n(V')$.  
In turn, the compressibility $n^2\kappa(\mu)=-dn/dV$ requires to differentiate $n(V)$ once.
Remarkably, using the pressure $p$ and compressibility $\kappa$ as variables, 
the complete thermodynamics of the unitary gas may be inferred just from the density 
distribution $n(V)$, with no other input whatsoever~\cite{ku12superfluid}. This relies on
the fact that both the normalized pressure $\tilde{p}(\theta)=p/p^{(0)}$ and 
compressibility $\tilde{\kappa}(\theta)=\kappa/\kappa^{(0)}$
only depend on the dimensionless temperature $\theta=T/T_F$. Since the latter cannot be
directly measured in an ultracold gas, one eliminates $\theta$ from  $\tilde{\kappa}$ and
$\tilde{p}$, thus arriving at a compressibility equation of state  which -   for the unitary gas -
is a universal function $\tilde{\kappa}(\tilde{p})$ defined for $\tilde{p}\geq\xi_s$. 
The Bertsch parameter $\xi_s$ may be obtained from the limit $\tilde{\kappa}(\xi_s)=1/\xi_s$ 
of the dimensionless compressibility at the lowest possible value of the pressure or, alternatively, from
$\tilde{\kappa}=1/\tilde{p}$ at $T=0$.  Since every experimental profile $n(V)$ at unitarity
must give rise to the same universal curve $\tilde{\kappa}(\tilde{p})$, this function
may be determined with high precision by averaging over many profiles.  
Moreover, it also allows to determine
in a very precise manner the temperature of the gas and the dimensionless variable 
$\beta\mu$.  This is necessary to finally cast the results into the more conventional form 
of the universal scaling function defined in Eq.~(\ref{eq:pressure-scaling}). 

The possibility 
to infer the complete thermodynamics just from an analysis of density profiles crucially relies
on the fact that the unitary gas is a scale invariant system. As will be discussed
in detail in section 3.4, this implies that pressure and energy density are related
by $p=2\epsilon/3$. An immediate consequence 
of this relation is that both the pressure and the thermal expansion coefficient 
\begin{equation}
\beta_V=\left(\frac{\partial p}{\partial T}\right)_V=\gamma c_V\qquad {\rm and} \qquad 
\alpha_p=\frac{1}{V} \left(\frac{\partial V}{\partial T}\right)_p=\kappa_T\cdot\gamma\, c_V
\label{eq:Grueneisen}
\end{equation}
are directly proportional to the specific heat per volume $c_V$. The associated prefactor is a 
universal dimensionless number $\gamma=2/3$ which - in the context of thermal expansion 
in an anharmonic solid - is called the Gr\"uneisen parameter~\cite{ashc76}. 
A second consequence of scale invariance in the form $U=3pV/2$ follows from the 
quite general thermodynamic relation
\begin{equation}
T\left(\frac{\partial p}{\partial T}\right)_V=p+\left(\frac{\partial U}{\partial V}\right)_T \; \longrightarrow \; 
T\left(\frac{\partial p}{\partial T}\right)_V=\frac{5}{2}\,p -\frac{3}{2}\frac{1}{\kappa_T}\, .
\label{eq:dpbydT}
\end{equation}
For a scale invariant system, this connects the pressure coeffcient with $p$ itself and the inverse compressibility $1/\kappa_T=-V(\partial p/\partial V)_T$.  
In terms of the variables $\tilde{p}$ and $\tilde{\kappa}$ introduced above, which depend on temperature only via $\theta=T/T_F$, 
this can be rewritten in the form
\begin{equation}
\frac{d\tilde{p}}{d\theta}=\frac{5}{2\theta}\left(\tilde{p}-\frac{1}{\tilde{\kappa}}\right)= \frac{5}{3} \frac{C_V}{Nk_B}\,\;(>0) \, ,
\label{eq:specific-heat}
\end{equation}
connecting the pressure coefficient and the specific heat per particle with the dimensionless temperature,
pressure and compressibility. In order to convert back from the pressure thermometer that is 
used in the function  $\tilde{\kappa}(\tilde{p})$
to the actual dimensionlesss temperature $\theta$ which monotonically decreases from
the edge of the cloud towards its center, one integrates the first equation in~(\ref{eq:specific-heat}). The relation
\begin{equation}
\theta(\tilde{p})=\theta(\tilde{p}_i)\,\exp{\left[\frac{2}{5}\int_{\tilde{p}_i}^{\tilde{p}} \frac{d\tilde{p}}{\tilde{p}-1/\tilde{\kappa}(\tilde{p})}\right]}
\label{eq:temperature}
\end{equation} 
then allows to determine the reduced temperature $\theta=T/T_F$ from the known function $\tilde{\kappa}(\tilde{p})$
provided an initial value $\theta(\tilde{p}_i)$ is known~\cite{ku12superfluid}. 
This crucial step, which avoids the uncertainties involved in any direct thermometry of the gas, relies on the fact 
that $\tilde{p}(\theta)$ is a continuous and monotonically increasing function of $\theta$. It can thus be uniquely 
inverted to give $\theta(\tilde{p})$. Note that this is possible despite the fact that in order to obtain 
$\theta(\tilde{p})$ in the temperature range below the superfluid transition, the integration
in Eq.~(\ref{eq:temperature}) includes the dimensionless pressure $\tilde{p}_c\simeq 0.51$~\cite{haus07bcsbec} 
below which the gas becomes superfluid. As is evident from Eq.~(\ref{eq:specific-heat}), however,
the function $\tilde{p}-1/\tilde{\kappa}(\tilde{p})=2\theta\,C_V/3Nk_B$ is positive and continuous. Its inverse
is thus integrable even in the thermodynamic limit, where $C_V/Nk_B$ exhibits a singularity (see below). 

In practice, one chooses the initial value $\theta(\tilde{p}_i)$ in the regime $\tilde{p},\theta\gg 1$ 
where the virial expansion (note that $4/(3\sqrt{\pi}\,\theta^{3/2})=n\lambda_T^3/2$)
\begin{equation}
\label{eq:virial2}
\tilde{p}(\theta)=\frac{5}{2}\theta\cdot\sum_{l=1}^{\infty}\, a_l \left(\frac{4}{3\sqrt{\pi}\,\theta^{3/2}}\right)^{l-1}=
\frac{5}{2}\theta\cdot\left(1-\frac{1}{\sqrt{2\pi}\,\theta^{3/2}} +\frac{16 a_3}{9\pi\,\theta^3}+\ldots\right)
\end{equation}
 for the pressure of the two component Fermi gas applies. The  associated 
 dimensionlesss coefficients $a_l$ are 
 fixed by the standard virial coefficients $b_l$ via $a_1=b_1=1; \, a_2=-b_2; \, a_3=4b_2^2-2b_3 $ etc.. 
In particular, for the unitary gas, the Beth-Uhlenbeck formula for the second virial coefficient gives 
$a_2=-3/(4\sqrt{2})$~\cite{ho04virial} while $a_3\simeq 1.71$
\footnote{The large and positive value of $a_3$ is a consequence of the repulsive atom-dimer
interaction for Fermions mentioned in section 1.3.}. 
 The latter follows from $b_3\simeq -0.29$ which has been inferred 
from a numerical calculation of the energy spectrum of 
three Fermions in a harmonic trap~\cite{liu09virial} or a diagrammatic expansion in powers of the fugacity~\cite{kapl11virial}.     
Finally, the dimensionless parameter $\beta\mu$
may be determined by converting the measured function $\tilde{\kappa}(\tilde{p})$ to $\tilde{\kappa}(\theta)$ 
and using the relation
\begin{equation}
\frac{d\theta}{d(\beta\mu)}=-\theta^2\cdot\tilde{\kappa}\;\longrightarrow\;
(\beta\mu)(\theta)=(\beta\mu)(\theta_i) - \int_{\theta_i}^{\theta}\, \frac{d\theta}{\theta^2\tilde{\kappa}(\theta)}
\label{eq:chemical-potential}
\end{equation} 
which follows from $n^2\kappa_T=(\partial n/\partial\mu)_T$. As in Eq.~(\ref{eq:temperature})
above, the initial value $(\beta\mu)(\theta_i)$ for the integration can be assumed to be in
the range where the virial expansion $n\lambda_T^3=2\,\sum_{l=1}^{\infty}\, l b_l z^l $ 
with fugacity $z=\exp{(\beta\mu)}$ allows to analytically connect $\beta\mu$ with $\theta$.
Note also that knowledge of the chemical  potential of the unitary gas as a function of $T/T_F$ directly determines
the entropy per particle via $S/Nk_B=\tilde{p}/\theta - \beta\mu$ as a simple consequence of the 
Gibbs-Duhem relation with $U=3pV/2$ from scale invariance. \\

 \begin{figure}
\includegraphics[width=1.0\linewidth]{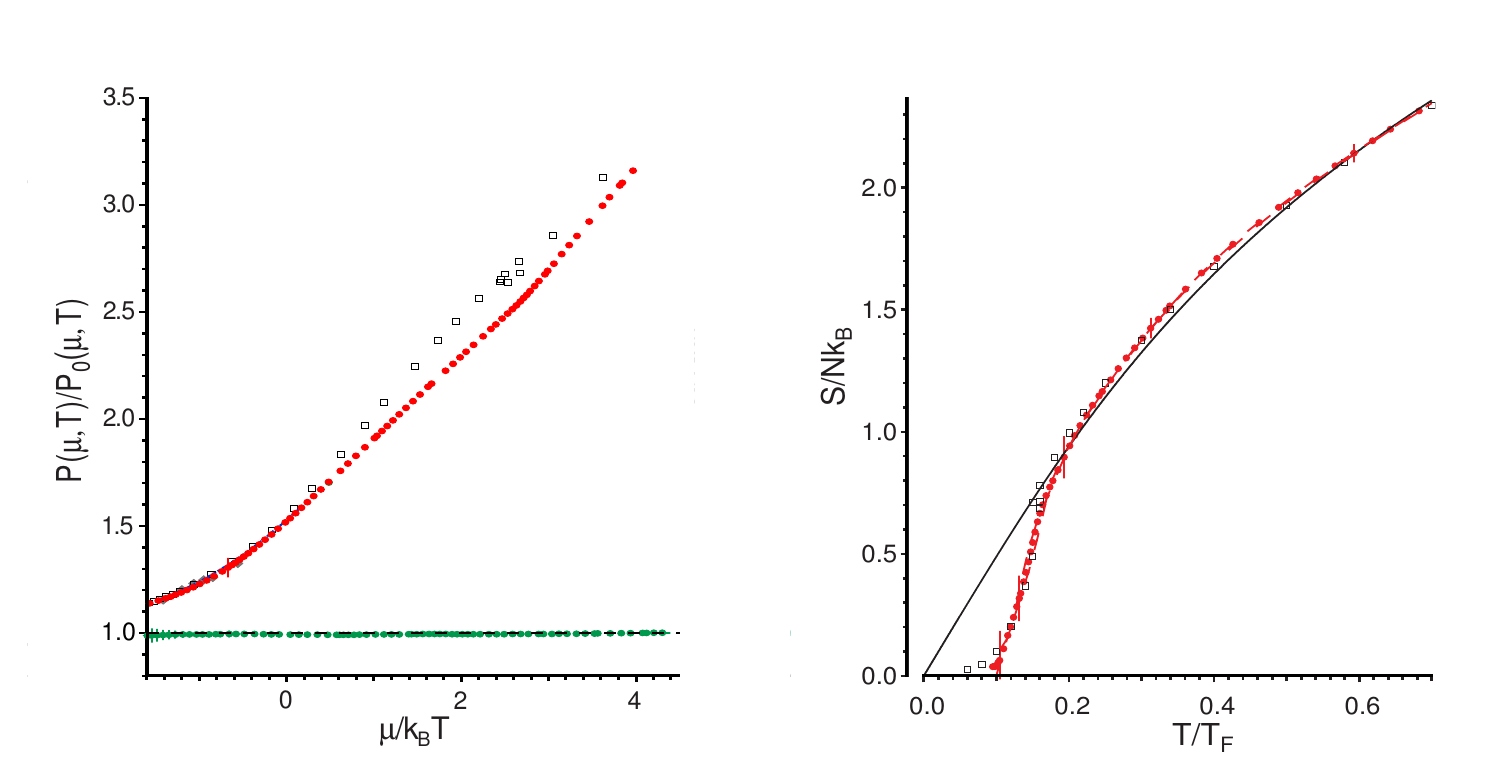}
\caption{Pressure and entropy of the unitary Fermi gas as a function of $\beta\mu$ and $T/T_F$ respectively.
The experimental data are compared with the theoretical predictions from a Luttinger-Ward approach, which 
are shown as open squares (from~\cite{ku12superfluid}). The solid line is the entropy of an ideal
Fermi gas which - surprisingly - is almost identical with that of the unitary gas above $T_c$.
\label{fig:MIT-Data}}
\end{figure}

The experimental results obtained in this manner are shown in Fig.~\ref{fig:MIT-Data}.
Here, on the left, the pressure $p(\beta\mu)$ of the unitary gas normalized to that of a non-interacting
two-component Fermi gas at the same value of $\beta\mu$ is plotted in the 
experimentally accessible range $-1.6\lesssim\beta\mu\lesssim 4.2$.  The horizontal line 
are the results for the non-interacting gas at zero scattering length. They agree
perfectly with the textbook prediction and thus indicate the level of accuracy achieved in
these measurements. For the unitary gas, the experimental data directly determine 
the universal scaling function $f_p(\beta\mu,0)$ defined in Eq.~(\ref{eq:pressure-scaling}).
The theoretically predicted results based on the Luttinger-Ward approach are quite close to
the measured data except near  $\beta\mu\simeq 2$. There is, moreover, a small range of multivaluedness 
in the theory near $\beta\mu\simeq 2.5$ which is associated with the weak first order nature of the  superfluid transition
in this approach.  This problem is hardly visible in the entropy per particle as a 
function of $T/T_F$, which agrees extremely well with the experimental results over the
complete range of temperatures. In particular, the entropy per particle at the transition 
$S_c=(0.73\pm 0.13)\, Nk_B$ is very close to the predicted value $0.71$~\cite{haus07bcsbec}.
Overall, therefore, the Luttinger-Ward approach provides a quantitatively reliable description of the data, 
including the range where the gas is superfluid. 
Note that there is no adjustable parameter at all in both the experimental and the theoretical results.
In the normal fluid regime $\beta\mu\lesssim  2.5$,
an essentially perfect agreement between experiment
and theory is achieved within the Bold Diagrammatic Monte Carlo method~\cite{ku12normal},
which has not, however, been extended into the superfluid regime so far.
The Bertsch parameter requires an extrapolation towards zero temperature which is obtained
most conveniently from either the compressibility $\kappa$ or the chemical potential~\cite{ku12superfluid}.
After correcting for the position of the Feshbach resonance, which is at $B_0=832.18\,$G rather 
than the value $834.15\,$G assumed in Ref.~\cite{ku12superfluid}, one obtains $\xi_s=0.37\pm 0.01$~\cite{zuer13Li-resonance}.    
Now, as mentioned in section 2.4, the Bertsch parameter is a universal number only in the
zero range limit. Its finite range corrections have been obtained numerically 
and are of the form $\xi_s(r_e)=\xi_s+0.12\, k_F|r_e|+\ldots$~\cite{carl11bertsch}.
Since $r_e=3\bar{a}$ for the open channel dominated resonance in $^6$Li and with $k_F\bar{a}\simeq 10^{-2}$, 
these corrections are of order $0.003$. They indicate that the true universal value $\xi_s$ of the 
Bertsch parameter is slightly smaller than $0.37$, in agreement
with the prediction of both the $\epsilon$-expansion and Luttinger-Ward.  \\ 

 \begin{figure}
\includegraphics[width=1.1\linewidth]{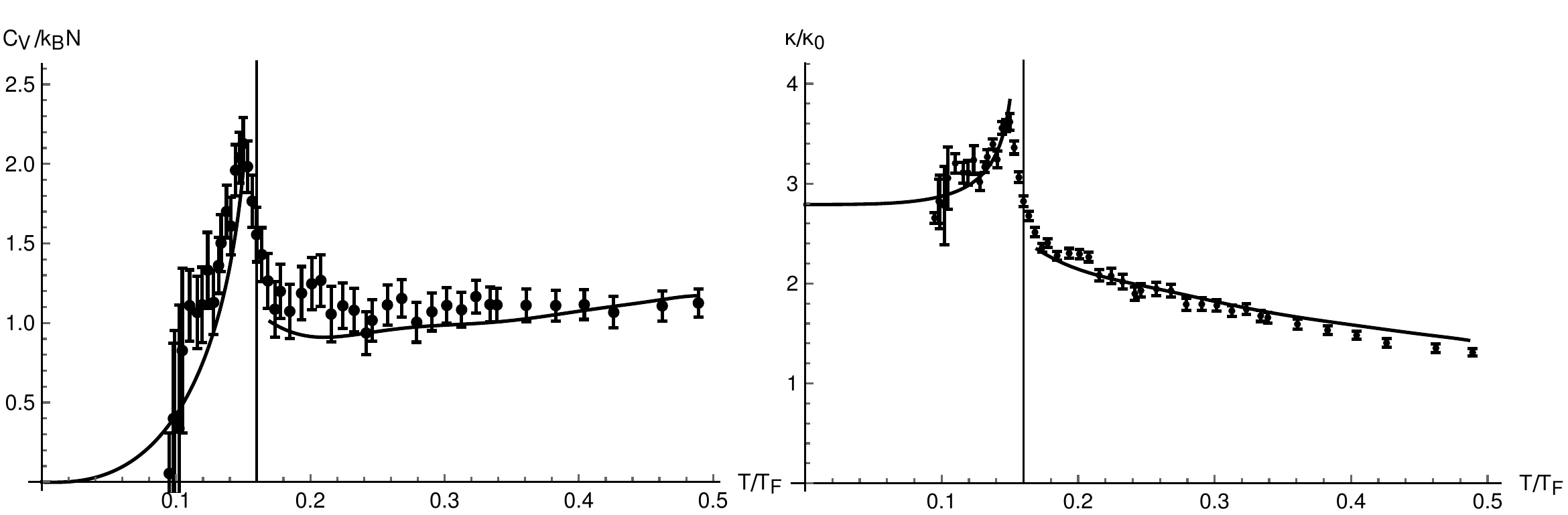}
\caption{Specific heat and compressibility of the unitary Fermi gas as a function of $T/T_F$.
The experimental data are compared with the theoretical predictions from a Luttinger-Ward approach, which 
are indicated by the solid line. 
\label{fig:specific-heat}}
\end{figure}

A quite sensitive measure for the critical temperature of the 
superfluid transition is obtained by considering the specific heat per particle
and the compressibility 
of the gas as a function of $T/T_F$. The results are shown in Fig.~\ref{fig:specific-heat}.
They clearly exhibit a sharp feature in both $C_V$ and $\kappa$ which coincides
with the sudden drop of the entropy in Fig.~\ref{fig:MIT-Data}.
The experimental estimate $T_c/T_F=0.16\pm 0.01$~\cite{ku12superfluid} for the critical temperature is consistent again with
the predicted value from the Luttinger-Ward approach and with results obtained via quantum Monte Carlo,
which give $T_c/T_F=0.152(7)$~\cite{buro06TC} or $T_c/T_F=0.171(5)$~\cite{goul10}.   
The fact that both the specific heat and the compressibility exhibit a 
pronounced maximum at the transition is expected from the 
critical behavior in the infinite system. For the specific heat the associated
singularity is of the form
\begin{equation}
\frac{C_V(T\simeq T_c)}{Nk_B} =\tilde{c}_V(T_c) - A_{\pm}\cdot |t|^{-\alpha} +\ldots\, \qquad t=(T-T_c)/T_c\, .
\label{eq:critical}
\end{equation}
The critical exponent $\alpha\simeq -0.01$ and the amplitude ratio  
$A_{+}/A_{-}\simeq 1.05$ are known from the theory of the 3D XY-model~\cite{zinn10}.
They have been measured very precisely~\cite{lipa03} in the case of the superfluid transition in 
$^4$He, which is in the same universality class. The fact that $-\alpha$ is 
rather small and positive leads to a specific heat which exhibits a very sharp cusp 
but remains finite at $T_c$. The associated peak value $\tilde{c}_V(T_c)$  at the tip
is non-universal in the case of $^4$He.  
For the unitary Fermi gas, however, it is 
again a universal constant.  An upper bound for this is obtained by using the second relation 
in Eq.~(\ref{eq:specific-heat}).  Specifically, the fairly well known values $\theta_c\simeq 0.16$  for the critical temperature 
and the critical pressure ratio $p_c/p^{(0)}\simeq 0.51$ together with the positivity of the compressibility
imply an upper bound $\tilde{c}_V(T_c)\lesssim 4.8$ which is about a factor two larger than the 
maximum value observed in the trapped system. This is expected, since the 
singularity~(\ref{eq:critical}) is in practice rounded both due to 
finite size effects and the finite resolution of the imaging system. The former provide a cutoff 
in a temperature range $|t|_R\lesssim (\xi_0/R)^{1/\nu}$, where
the correlation length $\xi=\xi_0\cdot|t|^{-\nu}$ of the infinite system is larger 
than the trap size $R$. 
In order to infer the singular behavior of the compressibility in the infinite system, one notes 
that the pressure $\tilde{p}=\tilde{p}_c+\mathcal{O}(|t|^{1-\alpha})$ has a weaker 
singularity than the specific heat. Using again the second relation 
in Eq.~(\ref{eq:specific-heat}), this implies that $\tilde{\kappa}$ is 
finite at $T_c$ and has the same type of singular behavior as the specific heat.
 Finally, we mention that at very low temperatures the specific heat per volume $c_V(T)= 2\pi^2 k_B\, (k_BT/\hbar c)^3/15$
 is again a universal function which only involves the velocity $c= v_F\sqrt{\xi_s/3}\simeq 0.36\, v_F$ of
 the Bogoliubov sound modes of the superfluid~\cite{bloc08review}. Unfortunately, the regime
 of temperatures where phonons dominate has not been accessible so far.

Universal numbers also characterize the quantum critical regime $|\mu|\ll k_BT$
of the unitary gas in the normal phase above the quantum critical point of Fig.~\ref{fig:phase}. 
In particular, Eq.~(\ref{eq:contact-scaling}) shows that  both particle and contact density 
\begin{equation}
n(\mu=0, T)=f_n(0)\cdot\lambda_T^{-3} \qquad {\rm and} \qquad
\mathcal{C}(\mu=0,T)=f_{\mathcal{C}}(0)\cdot \lambda_T^{-4}
\label{eq:density-QCP}
\end{equation}
vanish with simple power laws in the inverse thermal wavelength $1/\lambda_T$.
The associated prefactors $f_n(0)=\partial_x f_p\vert_{x,y=0}$ and 
$f_{\mathcal{C}}(0)=8\pi^2 \partial_y f_p\vert_{x,y=0}$ are universal constants and are 
determined by the basic scaling function $f_p (x,y)$ defined in Eq.~(\ref{eq:pressure-scaling}). 
 In Table~\ref{tab:vals}, the experimental results for $f_n(0) \simeq 3$ from the density equation of state~\cite{ku12normal}
and other thermodynamic properties in the quantum critical regime are compared 
with various theoretical approaches, 
including the self-consistent Luttinger-Ward theory \cite{haus94, haus07bcsbec, enss11viscosity} and Bold Diagrammatic Monte Carlo (BDMC)
\cite{ku12normal}. We also include the results of a large-$N$ approach~\cite{enss12QCP} extrapolated to $N=1$.
This approach is equivalent to an
approximation originally introduced by Nozi\`{e}res and Schmitt-Rink~\cite{nozi85} to describe the 
BCS-BEC crossover problem at finite temperature. Remarkably, the approximation 
gives reasonable results in the quantum critical regime $\beta|\mu|\ll 1$ 
even though it is not reliable for a quantitative description of the 
crossover and the superfluid regime in particular. 
\begin{table}[t]
  \centering
  \begin{tabular}{lllll}
    \hline
    & Experiment & Large-$N$ & LuttWard & BoldDiagMC \\
    \hline\hline
    $n(\mu=0)\, \lambda_T^3$ & 2.966(35) \cite{ku12superfluid} & 2.674 \cite{enss12QCP} & 3.108
    \cite{haus07bcsbec} & 2.90(5)\;\: \cite{ku12normal} \\
    $p\;[nk_BT]$ & 0.891(19) \cite{ku12superfluid} & 0.928 \cite{enss12QCP} &
    0.863 \cite{haus07bcsbec} & 0.90(2)\;\: \cite{ku12normal} \\
    $s\;[nk_B]$ & 2.227(38) \cite{ku12superfluid} & 2.320 \cite{enss12QCP} & 2.177
    \cite{haus07bcsbec} & 2.25(5)\;\: \cite{ku12normal} \\
    $\mathcal{C}(\mu=0)\,\lambda_T^4$ & & 26.84 \cite{enss12QCP} & 28.54 \cite{enss11viscosity} &
   27.2(1.7) \cite{vanh13contact} \\
       \hline
  \end{tabular}
  \caption{Thermodynamic properties of the
    unitary Fermi gas in the quantum critical regime $\beta|\mu|\ll 1$:
    density $n=n(\mu=0,T)$, pressure $p$, entropy density $s$ and contact
    density $\mathcal{C}$. 
    The Large-$N$ results are extrapolated to $N=1$ (from Ref.~\cite{enss12QCP}).} 
  \label{tab:vals}
\end{table}


\subsection{Luttinger-Ward theory}  The diagrammatic formalism
for many-body problems at finite temperature is based on Green functions
in imaginary time $\tau\in [0,\beta\hbar]$. Following standard notation~\cite{fett71,abri75},
they are defined by 
\begin{equation}
\mathcal{G}(\mathbf{x},\tau)=-\langle  \mathcal{T}\, \hat{\psi}_{\sigma}(\mathbf{x},\tau)\hat{\psi}^+_{\sigma}(\bm{0},0)\rangle \qquad {\rm and} \qquad
\mathcal{F}(\mathbf{x},\tau)=-\langle  \mathcal{T}\, \hat{\psi}_{\uparrow}(\mathbf{x},\tau)\hat{\psi}_{\downarrow}(\bm{0},0)\rangle
\label{eq:GF}
\end{equation}
where $ \mathcal{T}$ is the time ordering operator. Since we consider a balanced system,
the Green function $\mathcal{G}$ does not depend on the spin index $\sigma$.
The appearance of a superfluid phase below a critical temperature $T_c$ is 
accounted for by a non-vanishing anomalous Green function $\mathcal{F}$.
For Fermions with a zero range attractive interaction only singlet pairing
is possible. This allows to again restrict $\mathcal{F}$ to a simple scalar. 
Moreover, in the absence of any competing instabilities, the relevant 
two-particle Green or vertex function, which in general depends on three independent 
momenta and frequencies, can be reduced to a function $\Gamma(\mathbf{x},\tau)$ which only 
involves the center of mass dynamics of an up-down pair via
\begin{equation}
\Gamma(\mathbf{x},\tau)=\bar{g}(\Lambda)\,\delta(\tau)\delta(\mathbf{x})-\bar{g}^2(\Lambda)\,
\langle  \mathcal{T}\, \left(\hat{\psi}_{\downarrow}\hat{\psi}_{\uparrow}\right)(\mathbf{x},\tau)\,\left(\hat{\psi}_{\uparrow}^+\hat{\psi}_{\downarrow}^+\right)
(\bm{0},0)\rangle\, .
\label{eq:Gamma-def}
\end{equation}
The associated  pair propagator $\Gamma(\mathbf{Q},\Omega_n)$ in Fourier space
depends on the center of mass momentum $\mathbf{Q}$ of a pair and a bosonic Matsubara frequency 
$\Omega_n=2 \pi n/\beta \hbar$ with $n \in \mathbb{Z}$. 
The behavior of the one particle Green functions at short distances and times determines 
the density and gap in the standard manner~\cite{fett71,abri75} 
\begin{equation}
n_{\sigma}=\mathcal{G}(\mathbf{x}=\bm{0},\tau=0^-) \qquad {\rm and} \qquad
\Delta=\lim_{\Lambda\to\infty}\bar{g}(\Lambda)\mathcal{F}(\mathbf{x}=\bm{0},\tau=0^-)\, .
\label{eq:gap}
\end{equation}
Similarly, the associated limit of the pair propagator is connected with
the contact density via the relation~\cite{haus09rf} 
\begin{equation}
\Gamma(\mathbf{x}=\bm{0},\tau=0^-) =- \lim_{\Lambda\to\infty}\,\bar{g}^2(\Lambda)\,\langle \hat{\psi}_{\uparrow}^{\dagger}(\mathbf{x})
\hat{\psi}_{\downarrow}^{\dagger}(\mathbf{x})
 \hat{\psi}_{\downarrow}(\mathbf{x})\hat{\psi}_{\uparrow}(\mathbf{x})\rangle = |\Delta|^2  -\frac{\hbar^4}{m^2}\mathcal{C}\, .
 \label{eq:contact-Delta}
 \end{equation}
 This follows from the definition of the contact density in Eq.~(\ref{eq:contact-def}) by noting that in the
 presence of finite, anomalous averages it is only the connected part of the four Fermion expectation value
 which defines $\mathcal{C}$. 
Since the pair propagator $\Gamma$ at short distances and times is a smooth function of temperature,
the relation~(\ref{eq:contact-Delta}) implies that the contact density exhbits an additive, singular contribution
$\delta\mathcal{C}=m^2|\Delta|^2/\hbar^4$ at the superfluid transition, as mentioned in section 2.2 above.

An explicit calculation of the Green functions is conveniently formulated in terms of the Luttinger-Ward formalism~\cite{lutt60, baym61}. 
It provides a systematic method to calculate the single-particle Green 
functions $G$ of quantum many-body systems from an exact, self-consistent
Dyson equation
\begin{equation}
G^{-1}=G_0^{-1}-\Sigma\,[G]\, .
\label{eq:Dyson}
\end{equation}
The self-energy is a functional of the interacting Green function $G$
and may be obtained from the functional derivative
\begin{equation}
\Sigma[G]=\frac{\delta\Phi[G]}{\delta G}
\label{eq:sigma1} 
\end{equation}
of a Luttinger-Ward functional $\Phi[G]$ which appears in
the formally exact representation
\begin{equation}
\Omega[G] = \beta^{-1} \bigl(
\text{Tr} \{ \ln[G] + [1-G_0^{-1} G ] \} + \Phi[G] \bigr ) \, ,
\label{eq:LW-functional}
\end{equation}
of the grand canonical thermodynamic potential 
as a functional of the full Green function $G$. 
Using the definition~(\ref{eq:sigma1}) of the self energy, 
the Dyson equation~(\ref{eq:Dyson}) is equivalent to 
the condition $\delta \Omega[G] / \delta G = 0$ that 
the thermodynamic potential is stationary with respect to 
variations of $G$. While the exact functional $\Phi[G]$ 
is unknown, the formalism guarantees that
the self-consistent solution of the functional equation~(\ref{eq:Dyson}) 
based on an {\it approximate} Luttinger-Ward functional gives rise
to a conserving approximation~\cite{baym61}.

The Luttinger-Ward formalism can be generalized to superfluid systems
by including both the normal and anomalous functions $ \mathcal{G}$ 
and $\mathcal{F}$. Using a Nambu-index $\alpha$, they can 
be combined into a matrix Green function~\cite{fett71}
\begin{equation}
G_{\alpha \alpha^\prime}(\mathbf{k},\omega_n) = 
\begin{pmatrix}
\mathcal{G}(\mathbf{k},\omega_n) &\mathcal{F}(\mathbf{k},\omega_n)\cr
\mathcal{F}(\mathbf{k},\omega_n)^* &-\mathcal{G}(\mathbf{k},\omega_n)^* 
\end{pmatrix}
\label{B_100}
\end{equation}
with momentum variable $\mathbf{k}$ and fermionic Matsubara frequencies 
$\omega_n=2 \pi (n+1/2)/\beta \hbar$.
Similarly, there is a matrix vertex function $\Gamma_{\alpha \alpha^\prime}(\mathbf{Q},\Omega_n)$
whose diagonal element is just the pair propagator introduced in Eq.~(\ref{eq:Gamma-def}). 
In general, $\Gamma$ carries four Nambu indices $\alpha$.
It can be reduced to a two-by-two matrix  $\Gamma_{\alpha \alpha^\prime}$ if the interactions are restricted to 
particle-particle scattering~\cite{haus93}. This is the standard ladder 
approximation which is known to provide the leading contribution
to the pairing instability for Fermions at low density~\cite{fett71}. 
Within this approximation, one obtains a closed set of equations 
for the matrix of single particle Green functions which reads~\cite{haus94,haus07bcsbec}
\begin{equation}
\Sigma_{\alpha \alpha^\prime}(\mathbf{k},\omega_n)=
\begin{pmatrix}
0 &\Delta \\
\Delta^* &0 \\
\end{pmatrix} \,
+  \int_Q  \, \frac{1}{\beta} \sum_{\Omega_n}
G_{\alpha^\prime \alpha}(\mathbf{Q}-\mathbf{k},\Omega_n-,\omega_n) \, \Gamma_{\alpha \alpha^\prime}(\mathbf{Q},\Omega_n)\ , 
\label{eq:sigma2}
\end{equation}

\begin{equation}
\Gamma^{-1}_{\alpha \alpha^\prime}(\mathbf{Q},\Omega_n)= \frac{\delta_{\alpha \alpha^\prime}}{g} 
+ \int_k
\Bigl[ \frac{1}{\beta} \sum_{\omega_n}
G_{\alpha \alpha^\prime}(\mathbf{Q}-\mathbf{k},\Omega_n-\omega_n)
G_{\alpha \alpha^\prime}(\mathbf{k},\omega_n) - \frac{m}{\hbar^2 \mathbf{k}^2} 
\, \delta_{\alpha \alpha^\prime} \Bigr] \ .
\label{eq:Gamma}
\end{equation}
The gap parameter $\Delta$, which appears in an 
anomalous contribution to the self-energy, has to be determined 
self consistently from the gap equation
\begin{equation}
\Delta = \lim_{\Lambda\to\infty}\bar{g}(\Lambda)\, \mathcal{F}(\mathbf{x}=\bm{0},\tau=0^{-}) = g\,\int_k \Bigl[ \mathcal{F}(\mathbf{k},\tau=0^{-}) +
\Delta\, \frac{m}{\hbar^2 \mathbf{k}^2} \Bigr] \ .
\label{eq:gap}
\end{equation}
Note the specific 
structure of the $GG$ term in~(\ref{eq:Gamma})  with respect to the Nambu indices $\alpha,\alpha^\prime$,
which implies that the particle-particle ladder is summed up.
In the weak coupling limit, this approximation
reproduces the standard BCS results. Since no particle-hole fluctuations
are included, however, the resulting critical temperature is
larger by a factor   $(4e)^{1/3}\simeq 2.22$ compared to the exact result~(\ref{eq:GorkovT_c}). 
Fortunately, at unitarity, particle-hole fluctuations are suppressed because
the chemical potential $\mu=\xi_s\varepsilon_F$ is substantially reduced. In fact, as 
noted above, the critical temperature $T_c/T_F=0.16$ which follows from
the Luttinger-Ward approach is in very good agreement both with experiment
and precise numerical calculations.  In the BEC limit, where the Fermions
form a Bose gas of strongly bound pairs, the ladder approximation correctly accounts 
for the formation of pairs.  Their residual interaction, however,  is
described only in an approximate manner. Indeed, in the BEC limit the ladder 
approximation for the  functional $\Phi[G]$ gives rise to a theory for a dilute, repulsive
Bose gas with a scattering length $a_B=2\, a$ which differs substantially 
from the result $a_{dd}=0.6\,a$ obtained from an exact solution of 
the four-particle problem.  Despite these shortcomings in the BCS or BEC limit,  
the Luttinger-Ward approach outlined above provides an internally consistent and quantitatively reliable description
of the thermodynamics in the most interesting unitary regime, both in the normal and in the superfluid phase.
Compared to a variety of other 
analytical approaches that have been used in this context, it  
\begin{itemize}
\item is a conserving approximation, i.e. thermodynamic relations are obeyed~\cite{haus07bcsbec} 
\item respects all symmetries, in particular the relation $p=2\epsilon/3$ 
at infinite scattering  length due to scale invariance is fulfilled at the percent level~\cite{haus07bcsbec} 
\item  obeys the exact Tan relations, which follow from $\delta\, (\beta\Omega) = -{\rm Tr}\, (G\, \delta G_0^{-1})$~\cite{enss12QCP}.
\end{itemize}
There are, however, two major problems which have not been resolved so far: the first one is related 
to the issue of the gapless Bogoliubov-Anderson mode which always exists in a neutral superfluid. Formally, this is 
guaranteed by a Ward identity which can be derived from the 
Luttinger-Ward formalism for any gauge invariant functional $\Phi[G]$. 
This functional defines an exact inverse vertex function  
$\Gamma^{-1}_{\rm ex}(\mathbf{Q},\Omega_n)$ which has a vanishing eigenvalue 
associated with the existence of a Bogoliubov-Anderson mode~\cite{haus99}.
In the non-superfluid phase, the divergence of $\Gamma_{\rm ex}$ at 
$\mathbf{Q}=\Omega_n=0$ is equivalent to the well known Thouless 
criterion which signals the onset of superfluidity~\cite{abri75}.
Unfortunately, in the presence of a finite anomalous average $\Delta$,
the vertex function in Eq.~(\ref{eq:Gamma}) does not obey the Ward identity. 
 The requirement 
of a gapless Bogoliubov-Anderson mode must therefore be imposed as 
an additional constraint by choosing a modified coupling constant in 
the renormalized gap equation Eq.~(\ref{eq:gap}). As shown in Ref.~\cite{haus07bcsbec},
this modified approach can be formulated in a manner which is still 
compatible with the Luttinger-Ward formalism, thus retaining the 
conserving and gapless nature
\footnote{For a recent discussion of Ward identities in $\Phi$-derivable theories 
for superfluid pairing and a comparison to different forms of non-conserving T-matrix approximations,
see Ref.~\cite{he14ward-identity}.}

A second problem is related to the precise nature of the normal to superfluid transition.
For a homogeneous gas in the thermodynamic limit, it is a continuous 
transition of the 3D XY type along the full BCS to BEC crossover.
By contrast, the selfconsistent solution of the equations~(\ref{eq:sigma2}),
~(\ref{eq:Gamma}) and ~(\ref{eq:gap}) above gives rise to a weak 
first-order transition.  This is evident in Fig.~\ref{fig:MIT-Data}, where the the 
theoretical results for the pressure are not single valued near the transition at $(\beta\mu)_c\simeq 2.5$. 
Within the Luttinger-Ward theory, the superfluid phase of the unitary gas in fact disappears 
at a critical temperature  $\theta_c =T_c /T_F=0.1604$ which is above
the lowest temperature $\theta_c=0.1506$ down to which the normal-fluid phase
is stable~\cite{haus07bcsbec}.
Fortunately, the range where the thermodynamic functions are multivalued is confined to a narrow regime of temperatures
of the order of the present experimental uncertainty in determining $T_c$. The first order nature
of the transition is clearly an artefact of self-consistent Green function methods.
It is an unsolved challenge to develop conserving approximations that
properly account for both the gapless nature of excitations in the symmetry broken
phase and the continuous nature of the superfluid transition, a problem that
appears already in the theory of weakly interacting Bose gases~\cite{hohe65helium}.
In principle, the bold diagrammatic Monte Carlo method~\cite{ku12normal,vanh13BDMC},
which may be viewed as a Luttinger-Ward theory including diagrams with an 
arbitrary number of vertices, might solve both problems. At present, however, it has 
not been extended into the superfluid regime, so the question is open.  
The problem with the first order nature of the transition does not arise in a number of alternative approaches to the BCS-BEC crossover
problem, in particular in $1/N$-expansions~\cite{nozi85,niko07renorm} or the functional
renormalization group~\cite{dieh07bcsbec}. On a quantitative level, however, these 
approaches give results e.g. for $T_c/T_F$ or the Bertsch parameter $\xi_s$ which are rather far
from the experimentally observed values.   \\

\subsection{Scale invariance}

 A non-relativistic many-body problem 
is scale invariant if under a rescaling $\mathbf{x}\to\lambda\mathbf{x}$ of lengths, its Hamiltonian
$\hat{H}\to\hat{H}/\lambda^2$ is reproduced up to a scale factor $1/\lambda^2$,
as happens trivially for free particles
\footnote{The possibility of a more general form of scale invariance $\hat{H}\to\hat{H}/\lambda^z$
with $z\ne 2$ is apparently not possible for Hamiltonians with
a non-relativistic kinetic energy. It appears, however, for electrons in graphene in the limit
of strong Coulomb repulsion $e^2/(\epsilon_0\hbar v_F)\to\infty$~\cite{son07graphene}.}. 
Apart from the obvious non-interacting case, the only realization of a 
scale invariant many-body Hamiltonian seems to require interactions 
proportional to $1/|\mathbf{x}|^2$ at arbitrary separation or a delta-function interaction $\bar{g}_2\,\delta^{(2)}(\mathbf{x})$
 in two dimensions. The latter case actually turns out {\it not} to be scale invariant,
 as will be discussed in section 3.5 below. \\

In the following, we will show that the unitary Fermi gas in 3D provides 
an exact realization of a scale invariant, non-relativistic many-body system.
An elementary argument for why a system is scale invariant in the case of resonant 
two-body interactions can be given following a discussion by Holstein~\cite{hols93anomalies}
of examples where analogs of an anomaly in QFT  - i.e. a symmetry of the classical Lagrangian does 
not survive quantization - appear in single particle quantum mechanics.
The argument relies on the observation that 
scale invariance requires interparticle interactions which 
are not associated with any intrinsic length scale. Correspondingly, the phase shifts $\delta_l(k)$
of  two-body scattering need to be independent of $k$ as in the trivial 
case of free particles, where they vanish identically.
For interacting particles,  $k$ - independent phase shifts arise if the
scattering is resonant with $\delta_l(k)=\pi/2$ for one or possibly multiple values of $l$ 
and zero otherwise. This is precisely the situation reached for low energy scattering
of ultracold atoms in the zero range limit $r_e\to 0$. Indeed, 
the expansion $k\cot{\delta_0(k)}\to -1/a+r_ek^2/2+\ldots$ shows that at infinite scattering length $1/a=0$
the single non-vanishing phase shift  $\delta_0(k)=\pi/2 -r_ek/2+\ldots$ becomes independent 
of $k$ provided the effective range correction is negligible.

 For a formal derivation of scale invariance in the unitary Fermi gas
 consider the microscopic action $S=\int_{\tau}\int_{\mathbf{x}}\,\mathcal{L}$  
 with a Lagrange density $\mathcal{L}$. Scale invariance 
 is clearly present in the absence of interactions since the action
 \begin{equation}
 S_0=\int_{\tau}\, \int_{\mathbf{x}} \,\sum_{\sigma}\left[ \psi_{\sigma}^\star\partial_{\tau}\psi_{\sigma} +\frac{\hbar^2}{2m}\left| \nabla\psi_{\sigma}\right|^2\right]
 \label{eq:action1}
 \end{equation}
is invariant under $\mathbf{x}\to\mathbf{x} e^{-l}, \tau\to\tau e^{-zl}$ with a dynamical exponent $z=2$ 
provided the fields are rescaled in their canonical form $\psi\to\psi\, e^{dl/2}$.
Adding a zero range interaction 
\begin{equation}
\mathcal{L}_{\rm int}=\bar{g}(\Lambda)\,\bar{\psi}_{\uparrow}\bar{\psi}_{\downarrow}\psi_{\downarrow} \psi_{\uparrow} (\mathbf{x},\tau)\, ,
\label{eq:action2}
\end{equation}
the coupling constant scales like  $\bar{g}\to\bar{g} e^{(2-d)l}$. 
For a fixed value of $\bar{g}$,  zero range interactions in $d>2$ thus flow to the 
non-interacting fixed point shown in Fig.~\ref{fig:flow} for $d=3$, which is trivially scale invariant. 
Now, in $d\geq 2$ a finite value
of the low energy scattering amplitude requires to consider a cutoff dependent coupling constant $\bar{g}(\Lambda)$ which - in the 3D case - 
is determined by Eq.~(\ref{eq:abar}). At finite values of the scattering length, this depends non-trivially on the cutoff. 
At unitarity, however, $\bar{g}(\Lambda)=-2\pi^2\hbar^2/m\Lambda$ is simply proportional to  $1/\Lambda$. Thus,
$\mathcal{L}_{\rm int}\to \mathcal{L}_{\rm int} e^{5l}$ scales precisely like the terms $\mathcal{L}_0$ which appear in the 
Lagrange density of the non-interacting system described by Eq.~(\ref{eq:action1}).
At this special value of the scattering length, the
action is scale invariant even in the presence of interactions.
This is just the observation made in section 3.1
that a system at infinite scattering length is at a fixed point. As far as equilibrium properties
are concerned, the associated universal scaling functions are contained in
the dimensionless form~\eqref{eq:pressure-scaling} of the pressure equation of state. More
generally, scaling arguments also constrain time dependent correlations  near this fixed point
like the time dependent one particle density matrix. At infinite scattering length, this can be written in the form
\begin{equation}
\big\langle \hat{\psi}^{+}_{\sigma} (\mathbf{x},t) \hat{\psi}_{\sigma} (\bm{0},0) \big\rangle\to
\lambda_T^{-3}\, \Phi \left(\beta\mu , |\mathbf{x}|/\lambda_T , t/\beta\hbar\right)\, ,
\label{eq:t-dependent}
\end{equation}
where $\Phi(x,y,z)$ is again a universal function.  In particular, in the quantum critical regime $\beta|\mu|\to 0$,
the equal time correlation function only depends on $\mathbf{x}/\lambda_T$. 
Scale invariance at the fixed point $\mu=1/a=0$ thus implies that the
correlation length $\xi_T$ in the quantum critical regime is equal to the thermal length $\lambda_T$. 
Using the relation~(\ref{eq:density-QCP}) which connects $\lambda_T$ with the average interparticle
spacing in this regime gives the result $\xi_T\simeq 1.43\, n^{-1/3}$ for the divergence of the 
correlation length stated in section 3.1.\\

An important consequence of scale invariance, which has been used in 
the discussion of the thermodynamics at unitarity in section 3.2, is the fact that 
pressure $p$ and energy density $\epsilon$ are related by $p=2\,\epsilon/3$, 
just as for an ideal quantum gas. 
Moreover, the bulk
viscosity $\zeta(T)\equiv 0$ vanishes at arbitrary temperatures
\footnote{In the superfluid regime there are actually three different bulk viscosities $\zeta_{1,2,3}$~\cite{fors75}.
The combination of scale and conformal invariance requires two of them to vanish. In particular,
$\zeta_2\equiv 0$, which is the one which takes the role of the standard bulk viscosity in the
normal fluid.}
~\cite{son07bulk}.
To prove these statements, consider the operator $\hat{D}$ which generates dilatations.
Since $\hat{H}\to\hat{H}/\lambda^2$ under a length rescaling, the standard
argument which shows that generators of symmetries commute with
the Hamiltonian yields in the case of scale invariance the relation
\begin{equation}
\frac{i}{\hbar} \left[\hat{H},\hat{D}\right]=2\,\hat{H}=\frac{d}{dt}\hat{D}\, .
\label{eq:Dilaton1}
\end{equation}   
Noting that $\hat{D}=\int x_i\hat{j}_i$ is equal the spatial integral of the scalar 
product of $\mathbf{x}$ with the momentum density operator $\hat{j}$ (repeated
indices are summed over), the momentum balance equation $\partial_t\hat{j}_i=-\partial_j\hat{\Pi}_{ij}$
and a partial integration lead to 
\begin{equation}
\frac{d}{dt}\hat{D}=2\int\,\hat{\epsilon}=\int\, x_i\partial_t\hat{j}_i=-\int\, x_i\,\partial_j\hat{\Pi}_{ij}=\int\,\hat{\Pi}_{ii}
\label{eq:Dilaton2}
\end{equation}   
 where $\hat{\Pi}_{ij}$ is the stress tensor.
 Scale invariance thus implies that $2\,\hat{\epsilon}=\hat{\Pi}_{ii}$ holds as an operator identity. This is analogous 
 to the vanishing of the trace $\rm{Tr}\,\hat{T}=\hat{\epsilon}-\hat{\Pi}_{ii}=0$ of the stress-energy tensor
 for scale invariant relativistic systems~\cite{zee10book}.  
 The fact that the energy density in~(\ref{eq:Dilaton2}) appears with a factor $2$ is a consequence
 of the rescaling of $\hat{H}$ with $1/\lambda^2$ instead of $1/\lambda$ in the relativistic case. 
 In a thermal equilibrium state $\langle\hat{\Pi}_{ij}\rangle=p\,\delta_{ij}$ is just the pressure.
 Scale invariance thus implies $p=2\,\epsilon/3$ as claimed. 
More generally, in a non-equilibrium situation with small but non-vanishing gradients
in the velocity field $\mathbf{v}(\mathbf{x})$, the stress tensor has the standard hydrodynamic form 
\begin{equation}
\langle\hat{\Pi}_{ij}\rangle=p\delta_{ij}+\rho v_iv_j-\eta\left( \partial_iv_j+\partial_jv_i-\frac{2}{3}\delta_{ij}\cdot {\rm div}\,\mathbf{v}\right)
-\zeta\,\delta_{ij}\cdot {\rm div}\,\mathbf{v}
\label{eq:stress-tensor}
\end{equation} 
where $\eta$ and $\zeta$ are the shear and bulk viscosities, respectively.  
The latter appears in a uniform expansion, where the finite value of ${\rm div}\,\mathbf{v}$ 
gives rise to an entropy production rate $\dot{s}=\zeta/T\cdot ({\rm div}\,\mathbf{v})^2$.
Now, in a scale invariant system, an increase in entropy can only occur in flows 
with finite shear. The bulk viscosity $\zeta$ therefore has to vanish identically.
This can be derived following an argument due to  Castin and Werner~\cite{zwer12book}.
It relies on the observation that for a unitary gas in a time dependent, isotropic harmonic trap, the many body 
wave function at time $t$ is obtained from its initial value at $t=0$ by the simple scaling transformation
defined in Eq.~\eqref{eq:ansatz} below,  which is identical with the result obtained for non-interacting
particles.  Both an ideal and the unitary gas therefore expand isentropically
if the trap potential is removed completely at $t=0$, which is possible only if $\zeta\equiv 0$. 
In a rather direct form, the result follows from the Kubo formula Eq.~\eqref{eq:kubo} below.  Indeed, 
the operator $\int_{\mathbf{x}}\hat{\Pi}_{ii}(\mathbf{x},t)=2\,\hat{H}$ which enters 
the commutator determining  the frequency dependent bulk viscosity $\zeta(\omega)$ is basically the conserved Hamiltonian 
and thus commutes with its value at $t=0$. As 
a result, $\zeta(\omega)$ vanishes at arbitrary frequencies, not only at $\omega=0$.
This is consistent with an exact sum rule due to Taylor and Randeria
\begin{equation}
  \label{eq:sumzeta}
  \frac{2}{\pi} \int_0^\infty d\omega\, {\rm Re}\,\zeta(\omega) 
  = \frac{\hbar^2}{36\pi m\, a^2}\, \frac{\partial\mathcal{C}}{\partial (1/a)}
\end{equation}
which connects the bulk viscosity of the two-component Fermi gas at an arbitrary value of 
the scattering length with the derivative of the contact density with respect to $1/a$~\cite{tayl10visc}. 
Since the latter is finite at infinite scattering length,
the right hand side vanishes at unitarity.  Due to ${\rm Re}\,\zeta(\omega)\geq 0$, this 
is possible only if $\zeta(\omega)\equiv 0$. \\

A subtle point in the arguments above is related to the fact that 
scale invariance of the unitary gas, i.e. $S_0+S_{\rm int}$ is 
unchanged under the transformation $\mathbf{x}\to\mathbf{x} e^{-l}, \tau\to\tau e^{-2l}$,
is violated in the presence of a finite chemical potential $\mu\ne 0$.  It is only the point 
$\mu=1/a=0$ in Fig.~\ref{fig:phase} which is a fixed point. 
Indeed, a non-vanishing chemical potential or a finite difference $h\ne 0$ which favors 
one of the two spin species gives rise to additional contributions 
\begin{equation}
\mathcal{L}_{\mu}=-\mu \sum _{\sigma}\, \bar{\psi}_{\sigma}\psi_{\sigma} (\mathbf{x},\tau) \qquad {\rm or} \qquad
\mathcal{L}_{h}=- h \sum _{\sigma} \sigma\, \bar{\psi}_{\sigma}\psi_{\sigma} (\mathbf{x},\tau) 
\label{eq:Lagrange-mu}
\end{equation}
to the Lagrange density. Under a rescaling of $\mathbf{x}$ and $\tau$, 
both the chemical potential and a possible difference between 
the two species change  according to $\mu\to \mu e^{2l}$ and $h\to h e^{2l}$. 
They are therefore relevant perturbations at the zero density fixed point of the unitary gas 
with ${\rm dim} \left[\mu, h\right]=2$, as anticipated in section 3.1. Since a finite value of $\mu$ 
violates scale invariance,  it is not evident why the relation $p=2\,\epsilon/3$ or the
vanishing of the bulk viscosity holds for the unitary gas at
finite density. This is a consequence of the fact that  the basic symmetry in Eq.~(\ref{eq:Dilaton1}) 
implies that $2\,\hat{\epsilon}=\hat{\Pi}_{ii}$ holds as an operator identity.
It is therefore legitimate to take expectation values of this in states with a finite density of particles. 
An additional subtlety with the argument above for scale invariance is that it relies on just the classical 
Lagrangian.  On a superficial level, it should thus also hold for Bosons at unitarity. 
Now, as discussed in section 1.4, this is an unstable system 
with an infinite number of three-body - and probably also N-body for arbitrary N - 
bound states through the Efimov effect. 
Despite the invariance of the action, scale invariance for Bosons at unitarity is 
therefore broken because a discrete set of levels is incompatible with 
the result that for any state with energy $E$ there is another one with energy $E/\lambda^2$.
The breaking of scale invariance for a gas of Bosons is an example of an anomaly:
quantum fluctuations may break a symmetry of the classical Lagrangian~\cite{zee10book}. 
Specifically, in the context of the Efimov effect, the continuous scale invariance of the classical
action is replaced by a discrete scaling symmetry
$\mathbf{x}\to\lambda_0^n\,\mathbf{x}, \;\tau\to\lambda_0^{2n}\,\tau\; {\rm with}\; n=1,2,3\ldots$~\cite{braa06}.
Here, $\lambda_0=\exp{(\pi/s_0)}$ is a number which depends on both the statistics 
of the particles involved and their mass ratio. For identical Bosons it is $\lambda_0\approx 22.69\ldots$
as mentioned in section 1.4.
Now, for a Fermi gas with equal masses of the two-components, there is no Efimov effect. 
Scale invariance of the full action $S$ 
thus indeed entails the consequences derived above. 
In a mass imbalanced case, however, the Efimov effect appears even for Fermi gases provided
 the mass ratio $M/m$ exceeds a critical value $\simeq 13.6$~\cite{petr03three_body}. 
On a qualitative level this can be understood by considering two heavy atoms and a light one
within the Born-Oppenheimer approximation~\cite{petr13}. At infinite scattering length, the
light atom with mass $m$ experiences an attractive potential $-0.16\, \hbar^2/mR^2$ 
which depends like an inverse square on the distance $R$ between the heavy atoms.
Such a potential is known to be unstable towards a 'fall to the center'~\cite{land77qm}.
For Bosons, where this can occur in a $l=0$ configuration, this happens in fact for 
arbitrary values of the mass ratio, even for $M/m\to 0$. For Fermions, which can get 
close only in a relative p-wave state, there is a critical value of the mass ratio. In the limit $M/m\gg 1$,
where the Born-Oppenheimer approximation becomes exact, this is obtained from the condition that
the parameter~\cite{petr13}
\begin{equation}
s_0=\sqrt{0.16\,\frac{M}{m}-(l+1/2)^2} \qquad {\rm for} \qquad M\gg m
\label{eq:s_0}
\end{equation}
is real for $l=1$. The associated critical value is $14.06$, not far from the exact result. 
A quite different approach to determine the critical mass ratio for the appearance of an 
Efimov effect for Fermi gases relies on calculating 
 the anomalous dimension $\Delta_{\mathcal{O}}=9/2+\gamma$ from Eq.~(\ref{eq:a-dimension}) 
 of the unequal mass generalization of 
 the operator in Eq.~(\ref{eq:three-body-operator}). As discussed by Nishida and Son~\cite{nish12book},
 this is a decreasing function of the mass ratio $M/m$
 and it becomes complex of the form $\Delta_{\mathcal{O}}=5/2+is_0$ for $M/m>13.607\ldots$.  \\

A consequence of scale invariance which is closely related to the vanishing bulk viscosity of the uniform case
appears in the dynamics of the gas in an isotropic harmonic trap. Scale invariance implies 
the existence of an infinite number of exact excited states at multiples of twice 
the trap frequency.  Physically, the excitations correspond to a breathing mode which
may,  for instance, be excited by changing the trap frequency $\omega(t)=\omega+\delta\omega(t)$ 
by a small amount $\delta\omega(t)\sim\epsilon$ during a finite time interval.  
In the context of the unitary Fermi gas, this was first pointed out by Werner and Castin~\cite{wern06unitary}, 
extending earlier work by Pitaevskii and Rosch for Bose gases in two dimensions~\cite{pita97symmetry}. 
As mentioned above, in a scale invariant system, the 
many-body wave function $\psi(\mathbf{X},t)$ for an arbitrary time dependence of $\omega(t)$ 
is related to its initial value by a simple scale transformation~\cite{wern06unitary}
\begin{equation}
\label{eq:ansatz}
\psi(\mathbf{X},t) = \frac{e^{i\theta(t)}}{\lambda^{3N/2}(t)}
\exp\left[\frac{im\dot\lambda(t)}{2\hbar\lambda(t)} X^2\right]
\psi(\mathbf{X}/\lambda(t),0)\, .
\end{equation}
Here, $\mathbf{X}$ is a shorthand notation for all $N$ particle coordinates and the 
scaling factor obeys the simple differential equation 
\begin{equation}
\label{eq:russe}
\ddot\lambda(t) = \frac{\omega^2}{\lambda^3(t)} -\omega^2(t) \lambda(t) \qquad {\rm with} \qquad \omega=\omega(t=0)
\end{equation}
and initial conditions $\lambda(0) = 1$ and $\dot\lambda(0) = 0$ 
\footnote{The phase factor $\theta(t)$ in~\eqref{eq:ansatz} is irrelevant 
for observables which only involve the density.}.  
For the case of a small change of the trap frequency in a finite interval of time,
linearization of this equation around $\lambda=1$ gives rise to a time dependence
$\lambda(t)=1+\epsilon\cos{(2\omega t)}+\mathcal{O}(\epsilon^2)$. The  
gas therefore oscillates at twice the trap frequency without any damping.   
As realized by Pitaevskii and Rosch, the existence of such oscillations and the simple scaling solution~\eqref{eq:ansatz} 
for the many-body wavefunction is due to a hidden SO(2,1) symmetry which appears
generically for scale invariant systems in the presence of an isotropic trap.   
For a derivation of this symmetry, one notes that the associated Hamiltonian
can be written in the form
\begin{equation}
  \hat{H}_{\omega} = \hat{H} + \omega^2\,\hat{C} \qquad {\rm with} \qquad \hat{C}=\frac{m}{2}\int_{\mathbf{x}}\, \mathbf{x}^2\,\hat{n}(\mathbf{x})\, .
\end{equation}
The operator $\hat{C}$ turns out to be the generator of 
the special conformal transformation $\mathbf{x}\to\mathbf{x}/(1+\lambda t),\ t\to t/(1+\lambda t)$~\cite{son06symmetry, nish07CFT},
analogous to $\hat{D}$ which generates scale transformations.
Using the continuity equation, its commutator with the Hamiltonian of the 
translation invariant system is $i[\hat{H},\hat{C}]=\hbar\,\hat{D}$, while  
$i[\hat{D},\hat{C}]=2\hbar\, \hat{C}$.  
Using these commutators, it is straightforward to show that the operators defined by 
\begin{equation}
 \hat{L}_{\pm} = \frac{\hat{H}}{2\omega}-\frac{\omega}{2}\,\hat{C} \pm \frac{i}{2}\, \hat{D} \; = \hat{L}_1\pm i \hat{L}_{2} \qquad {\rm and} 
 \qquad \hat{L}_3=\hat{H}_{\omega}/2\omega
  \label{eq:ladder}
\end{equation}
generate Lorentz boosts in two directions and rotations in a plane.  Indeed, they obey 
\begin{equation}
[\hat{L}_1,\hat{L}_2]= - i\hbar\, \hat{L}_3\, , \;\;  [\hat{L}_2,\hat{L}_3]= i\hbar\, \hat{L}_1\, , \;\;   [\hat{L}_3,\hat{L}_1]= i\hbar\, \hat{L}_2\, ,
 \label{eq:L-algebra}
\end{equation}
which is  the algebra of the Lorentz group in $2+1$ dimensions. 
Moreover, the commutator  $ [\hat{H}_{\omega},\hat{L}_{\pm}]= \pm\, 2\hbar\omega\, \hat{L}_{\pm}$ shows that 
a tower of excited states with energy $2n\hbar\omega$ can be obtained from an arbitrary eigenstate $|\Psi\rangle$ of the trapped gas  
by repeating
\begin{equation}
\hat{H}_{\omega}(\hat{L}_{+}\, |\Psi\rangle) = (\hat{L}_{+}\hat{H}_{\omega} +2\hbar\omega \hat{L}_{+})\, |\Psi\rangle =
(E_{\Psi}+2\hbar\omega)\, (\hat{L}_{+}\, |\Psi\rangle)
 \label{eq:L-algebra}
\end{equation}
$n$ times. It may also be shown that 
 the exact ground state is annihilated by  $\hat{L}_{-}$. 
Similar to the case of a single particle in a harmonic oscillator potential, the 
 operators $\hat{L}_{\pm}$ thus act as raising and lowering operators, now for excitations at {\it twice} the trap frequency. 

\subsection{Broken scale invariance and conformal anomaly in 2D} 
 
 In principle, the observation of undamped breathing modes at multiples of twice the trap frequency 
 seems to be straightforward. In practice, this is not so because 
a perfectly isotropic trap in 3D is very difficult to realize. 
In two dimensions, however, an isotropic trap potential can readily be created. Moreover,  
a zero range interaction in 2D of the form $V(\mathbf{x})=\bar{g}_2\,\delta(\mathbf{x})$ 
seems like a perfect realization of scale invariance without the necessity for any 
fine tuning of the interaction strength $\bar{g}_2$. In particular, this invariance is expected 
also for Bose gases
because there is no Efimov effect in s-wave states in two dimensions. Following the idea of Pitaevskii 
and Rosch, therefore, a number of experiments have been performed 
to verify the existence of a breathing mode at twice the trap frequency
for 2D Bose gases. In practice,
this can be tested by observing the time dependent mean-square radius $\langle r^2\rangle (t)$
of the gas after preparing it in an arbitrary out-of-equilibrium state. As mentioned above, the 
resulting periodicity of the time dependent scale factor $\lambda(t)$ - which actually
holds even for a strong initial perturbation -   
implies that  $\langle r^2\rangle (t)$ oscillates at the frequency $2\omega$ 
without any damping, irrespective of the strength of the interaction.
This long lived breathing mode has been observed both in a quasi cylindrical geometry~\cite{chev02} 
and in a fast rotating gas~\cite{stoc04}.  \\ 
 
A careful analysis of scale invariance in two dimensions shows, 
however, that it is strictly valid only in the trivial limit of vanishing interactions.
To see this, it is convenient to consider the dependence of the scattering phase shift 
$\delta_0(k)$ on momentum. In a 2D situation, where both the chemical 
potential and the thermal energy are much less than the energy $\hbar\omega_z$
for excitations in the transverse direction, the scattering of an incoming plane wave  
$\exp{(i\mathbf{k}\cdot\mathbf{x})}$ gives rise to an outgoing {\it cylindrical} wave
which asymptotically can be written in the form~\cite{adhi86}
 \begin{equation}
\psi_{\mathbf{k}}(\mathbf{x}) \to e^{i\mathbf{k}\cdot\mathbf{x}}
-\sqrt{\frac{i}{8\pi}}\;f(k,\theta)\;\frac{e^{ikr}}{\sqrt{kr}}\, .
 \label{eq:scatt-2D}
 \end{equation}
At low energy,  the associated dimensionless scattering amplitude $f(k,\theta)$
becomes independent of the scattering angle $\theta$ and exhibits a logarithmic dependence
\footnote{Note that while $f(k)$ vanishes logarithmically as $k\to 0$, the total scattering cross section
$\lambda= -\, {\rm Im}\, f(k,\theta=0)/k\to |f(k)|^2/4k$ diverges in the low energy limit.}  
\begin{equation}
f(k)=\frac{4}{-\cot{\delta_0(k)}+i}\,\to\,\,
\frac{4\pi}{2\ln(1/k a_2)+i\pi}
\label{eq:s-amplitude-2D}
\end{equation}
on momentum, which defines the 2D scattering length $a_2$~\cite{petr01scattering-2D,bloc08review}. 
The argument in section 3.4 above that scale invariance requires phase shifts which do not
depend on $k$ thus immediately shows that for any non-vanishing low energy scattering amplitude in 2D,
there is no scale invariance. On a formal level, the violation of scale invariance arises from the fact that a delta function 
in 2D does not give rise to a finite low energy scattering amplitude, unless it is made cutoff dependent.  
Indeed, the solution of the associated Lippmann-Schwinger equation in the low energy limit takes the form
\begin{equation}
\frac{2\pi\hbar^2}{m\bar{g}_2(\Lambda)}=\ln{\left(\frac{1}{ka_2}\right)} - \ln{\left(\frac{\Lambda}{k}\right)}\qquad 
\to\qquad \bar{g}_2(\Lambda)=-\frac{2\pi\hbar^2}{m\ln{(\Lambda a_2)}}
\label{eq:LS-equation-2D}
\end{equation}
which is just the 2D analog of Eq.~(\ref{eq:gbar}). A finite value of the 2D scattering length $a_2$ thus requires
a coupling constant $\bar{g}_2(\Lambda)\sim -1/\ln(\Lambda a_2)$ which vanishes inversely with the logarithm 
of the cutoff.\\

In practice, the violation of scale invariance in 2D is difficult to observe because in the
standard situation where the 3D scattering length $a$ is much less than the
length $\ell_z$ associated with the transverse confinement, the 2D scattering length
\footnote{Note that $a_2$ is always positive which implies that there is a two-body bound state
at arbitrary values of the 3D scattering length associated with the pole of $f(k)$ at $k=i/a_2$.}
\begin{equation}
a_2(a)=\ell_z\sqrt{\frac{\pi}{B}}\exp{\left(
-\sqrt{\frac{\pi}{2}}\frac{\ell_z}{a}\right)} \qquad {\rm with} \qquad B=0.905\ldots
\label{eq:a_2(a)}
\end{equation}
is exponentially small~\cite{petr01scattering-2D,bloc08review}. 
For realistic parameters, with $\ell_z\simeq 0.1\,\mu$m and $a\simeq\bar{a}$ a few
nanometers, the scattering amplitude~(\ref{eq:s-amplitude-2D}) can thus be 
replaced by a constant $f(k)\approx \tilde{g}_2=\sqrt{8\pi}\, a/\ell_z$, which is 
equivalent to assuming a $k$-independent phase shift.  As a result, there 
is an undamped breathing mode at $2\omega$ and also a scale invariant equation of state
of the form $n\lambda_T^2=f_n(\beta\mu, \tilde{g}_2)$. 
The function $f_n(x,\tilde{g}_2)$  
depends only parametrically on the dimensionless interaction strength $\tilde{g}_2$.
Experimentally, this has been tested for a gas of Cesium atoms,
where a Feshbach resonance can be used to increase $\tilde{g}_2$ up to $0.26$~\cite{hung11} and also 
- in a precision measurement based on the method discussed in section 3.2 - 
for Rubidium~\cite{desb14} where $\tilde{g}_2\simeq 0.1$. For an understanding of what
are the necessary conditions to see deviations from this apparent scale invariance, it is convenient 
to consider the dimensionless coupling constant $u(\kappa)=m\bar{g}_2(\kappa)/(\pi\hbar^2)$
introduced in analogy to Eq.~(\ref{eq:ubar}) in the 3D case. Integration of the associated flow 
equation $du/dl=-u^2/2$ in two dimensions between a UV cutoff $\Lambda$ down to 
a momentum scale $\kappa$ gives rise to a logarithmic dependence  
\begin{equation}
u(\kappa)=\frac{u(\Lambda)}{1+\frac{u(\Lambda)}{2}\ln{(\Lambda/\kappa)}} \, \to \, \frac{2}{\ln{(1/(\kappa a_2))}}
 \label{eq:scaling-2D}
 \end{equation}
of the coupling constant. It becomes independent of the scale $\Lambda$ if the strength 
$\bar{g}_2(\Lambda)$ of the delta function potential is choosen to depend logarithmically
on $\Lambda$ as given in Eq.~(\ref{eq:LS-equation-2D}).
Now, as discussed by Rancon and Dupuis~\cite{ranc12scaling}, the density equation of state
in two dimensions is, quite generally, of the form $n\lambda_T^2=f_n(\beta\mu, \tilde{g}(T))$.  
The associated universal function $f_n(x,y)$ depends on temperature not only via $x=\beta\mu$ 
but also via $y= \tilde{g}(T)=\pi\, u(\kappa=1/\lambda_T)$. The relevant momentum scale $\kappa$
is thus just the inverse thermal length.  In practice, deviations from scale invariance 
appear if the temperature dependence of  $\tilde{g}(T)$ is appreciable. Using the result~(\ref{eq:a_2(a)}) 
for the 2D scattering length, Eq.~(\ref{eq:scaling-2D}) shows that $\tilde{g}(T)\simeq\tilde{g}_2$ is a constant unless the 
temperature is so small that $\ln(\lambda_T/\ell_z)$ becomes of order $\ell_z/a$. 
In weak coupling, where $\ell_z/a\gg 1$, this would require exponentially small 
temperatures which are far beyond reach. \\

Values of the 3D scattering length which are much larger than the 
confinement length $\ell_z$ can be reached easily with two component Fermi gases,
using standard Feshbach resonances in $^{40}$K~\cite{froh11} or in $^6$Li~\cite{somm12}.
The breaking of scale invariance in this context has been 
discussed by Hofmann~\cite{hofm12anomaly}.  It turns out that the 
basic operator relation~(\ref{eq:Dilaton1}) which expresses scale invariance
in quite general terms is replaced in 2D by 
\begin{equation}
\frac{d}{dt}\hat{D}=\frac{i}{\hbar} \left[\hat{H},\hat{D}\right]=2\,\hat{H} +\frac{\hbar^2}{2\pi m}\,\hat{I}\, .
\label{eq:anomaly1}
\end{equation}   
The appearance of the additional operator $\hat{I}$, which explicitely breaks scale invariance,    
is an example of an anomaly, similar to what has been discussed in the context of 
the Efimov effect above. The expectation value of the operator $\hat{I}$ is in fact the 
integrated contact. This can be seen by relating the time derivative
of $\hat{D}$ to the integral $\int \hat{\Pi}_{ii}$ of the trace of the stress tensor as in 
section 3.4. Introducing the contact density via $\langle\hat{I}\rangle =\int \mathcal{C}$, Eq.~(\ref{eq:anomaly1}) implies 
 \begin{equation}
p=\epsilon+\frac{\hbar^2}{4\pi m}\,\mathcal{C}\, ,
\label{eq:Tan-pressure-2D}
\end{equation}
which is just the 2D version of the Tan pressure relation~(\ref{eq:Tan-pressure}).
 Scale invariance in 2D requires $\mathcal{C}$ to vanish, which is 
 true only for non-interacting particles.\\
 
 Quite surprisingly, to probe the violation of scale invariance in 2D Fermi gases
turns out to be rather diffcult,
 even in the regime where $a\gg\ell_z$. In fact, experiments show that the breathing mode frequency
 $\omega_B$ stays close to the scale invariant value $2\omega$ over a rather wide range 
 of the dimensionless coupling $\ln{(k_Fa_2)}$~\cite{vogt12viscosity}. 
 This observation has been explained in detail by Taylor and Randeria~\cite{tayl12anomaly}. 
 Using sum rules, they have shown that $\omega_B\simeq 2\omega$ provided the deviation 
 \begin{equation}
\rho\left(\frac{\partial p}{\partial\rho}\right)_s\, -2\, p =
\frac{\hbar^2}{4\pi m}\,\left[\mathcal{C}+\frac{a_2}{2}\left(\frac{\partial\mathcal{C}}{\partial a_2}\right)_s\right]
 \label{eq:gamma_2}
\end{equation}
of the adiabatic compressibility from its value $2p$ in a scale-invariant system is small. This is
indeed the case because the two contributions on the right hand side of Eq.~(\ref{eq:gamma_2})
largely cancel. In the 
following section, it will be shown that the violation of scale invariance in 2D
can be observed in a more direct form by RF spectroscopy in the presence of non-vanishing final state interactions.

\newpage

\section{RF-SPECTROSCOPY AND TRANSPORT}
\label{sec:tan}

The final chapter is devoted to dynamical properties of the unitary Fermi gas.
In particular, we will discuss
the fermionic excitations accessible in RF-spectroscopy. They are described by a set of spectral
functions $A(\mathbf{k},\varepsilon)$ which only depend on the dimensionless ratios $k/k_F$, 
$\varepsilon/\varepsilon_F$ and $T/T_F$.  Surprisingly, far below $T_c$,  
the excitations have a structure similar to that of a BCS superfluid. The unitary gas is therefore
still basically a Fermionic system. Universal features also appear in
transport properties of the unitary gas, especially in its quantum critical regime, where
the viscosity or spin diffusion constant are determined by fundamental constants.

\subsection{RF-spectroscopy}

\begin{figure}
\includegraphics[width=0.85\linewidth]{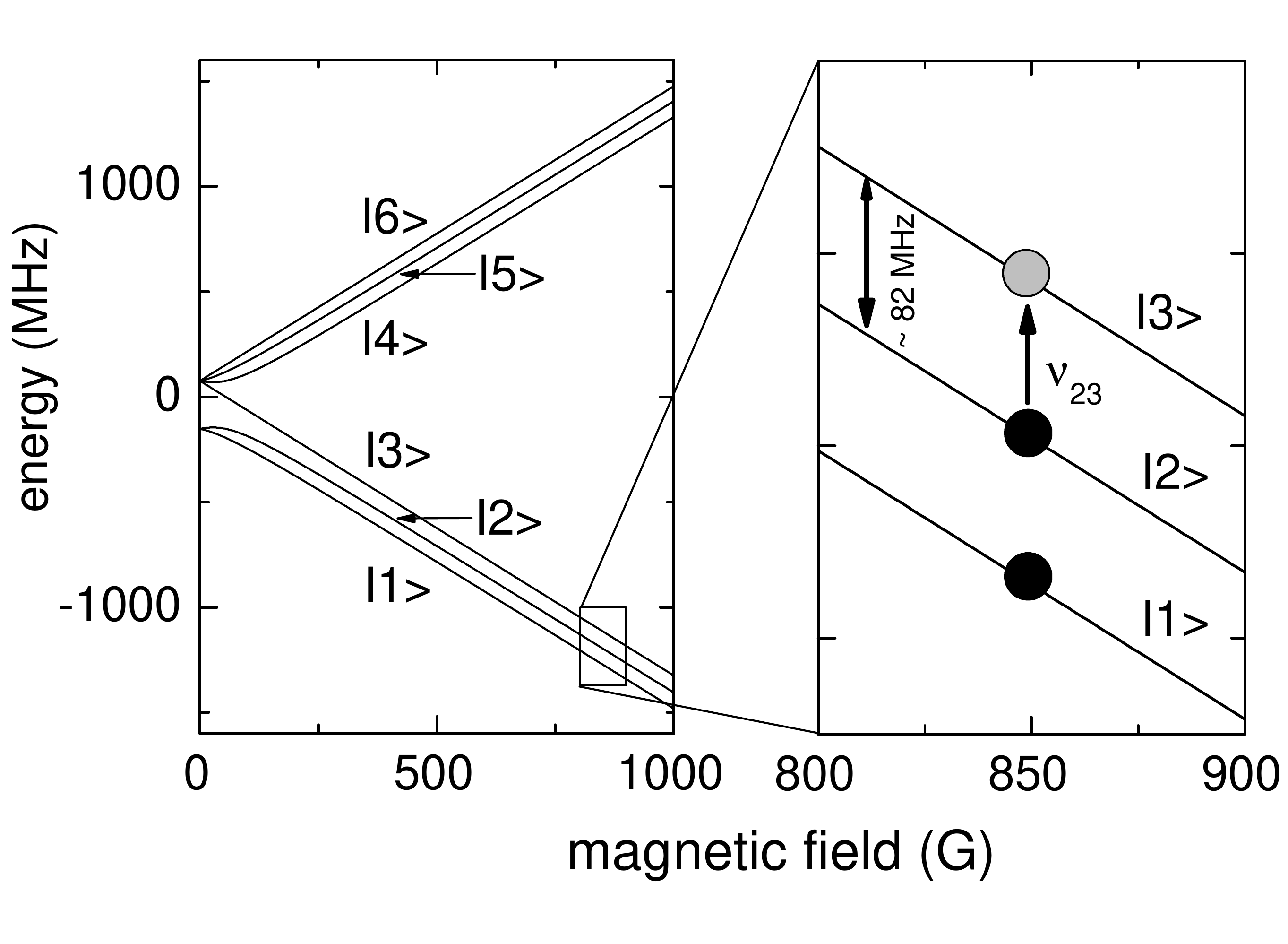}
\caption{Hyperfine level scheme of $^6$Li and a schematic sketch of the RF-transition 
from one of the two lowest hyperfine states $\left|2\right>$ into an empty state $\left|3\right>$.} 
\label{fig:rf-excitation}
\end{figure}

As first suggested by T\"orm\"a and Zoller~\cite{torm00}, interactions in ultracold Fermi gases can
be probed by RF-spectroscopy. It relies on transferring atoms from one of the internal hyperfine states, usually labelled as spin-down or $\left|2\right>$ into an empty state $\left|3\right>$ by a  radiofrequency pulse, which is tuned near the energy difference $\hbar\omega_{23}$ between these states in a free atom.  In a situation where the atom in the initial state $\left|2\right>$ interacts with other atoms in state $\left|1\right>$, for example by forming a molecular bound state, the RF photon has to supply - beyond the energy offset $\hbar\omega_{23}$ between states $\left|1\right>$ and 
$\left|2\right>$ in the free atom -  at least the binding energy of the molecule to break the bond and transfer the atom from state $\left|2\right>$ to state $\left|3\right>$, see Fig.~\ref{fig:rf-excitation}. 
Initial RF-experiments in the superfluid regime of Fermi gases near a Feshbach resonance were done by
C. Chin, R. Grimm and coworkers in Innsbruck~\cite{chin04gap}. They have traced the evolution of the molecular dissociation spectrum
which appears in the BEC limit $0<k_Fa\ll 1$, where the binding energy between the hyperfine levels $\left|1\right>$ and $\left|2\right>$  
is larger than the Fermi energy, all the way across the Feshbach resonance. Although there is no two-body bound state for $a<0$, 
the spectra were still shifted and broad, providing evidence for pair formation at weak attractive interactions due to the presence of  a Fermi sea. 
A quantitative interpretation of these spectra turned out to be quite difficult, however, since the final state $\left|3\right>$ employed in these 
experiments was still strongly interacting with atoms in state $\left|1\right>$. 
Indeed, as will be discussed below, the average 'clock' shift of the resonance away from its value in free space 
is a measure of the contact, not of superfluid pairing.  
The influence of final state interactions can be minimized by a proper choice of the initial and final states of the RF transfer, 
for instance working with a  $\left|1\right>-\left| 3\right>$ mixture and transfer to state $\left| 2\right>$, as shown by
W. Ketterle and coworkers at MIT~\cite{schu08pairsize}. 
Moreover, it is possible to measure the loss of atoms outcoupled through the RF pulse
{\it locally}~\cite{shin07rf}. This avoids the difficulties associated with analyzing trap averaged spectra,
in which the response from a superfluid center and the normal-fluid edge appear 
simultaneously.  In this manner, local spectra essentially free from final state effects can be obtained.
A further major step has been taken by D. Jin and coworkers at JILA, who have succeeded to 
perform RF spectroscopy in a momentum resolved manner by measuring the momentum 
distribution of the outcoupled atoms from state selective time-of-flight images~\cite{stew08arpes}. 
As will be shown in Eq.~(\ref{eq:C_220}) below,
this gives direct access to the spectral function $A_{-}(\mathbf{k},\varepsilon)$ for the removal of
a particle as a function of both momentum and energy. The method may be viewed as a cold
atom analog of angle-resolved photoemission spectroscopy, which allows
to determine the elementary excitations of strongly
correlated electronic systems~\cite{dama03arpes}. Unfortunately, so far momentum resolved
RF spectroscopy has not been combined with local resolution. Moreover, the short RF pulse limits the energy resolution to about 
$20\%$ of the Fermi energy~\cite{gaeb10pseudo}. At the present stage, therefore, a comparison of the trap averaged 
spectra with theory is not at the level of precision necessary to distinguish different 
predictions for the detailed spectral functions, in particular to clarify the issue of a possible pseudogap of the unitary gas. \\

A theoretical description of momentum resolved RF-spectroscopy is based on the
concept of spectral functions, which essentially measure the distribution
of energies associated with a given momentum $\mathbf{k}$. 
Their microscopic definition and physical interpretation is most easily understood at zero temperature.
Denoting the exact many-body ground state with $\left|\psi_0 \right>$ and suppressing the spin index $\sigma$, 
the functions $A_{\pm}(\mathbf{k},\varepsilon)$ are just the Fourier transform of the matrix elements
\begin{equation}
A_{+}(\mathbf{k}, t)= \langle \hat{c}_{\mathbf{k}}^{+}\psi_0\vert e^{-i\hat{H}t/\hbar}\, \hat{c}_{\mathbf{k}}^{+}\psi_0\rangle \qquad
{\rm and} \qquad A_{-}(\mathbf{k}, t)= \langle \hat{c}_{\mathbf{k}} \psi_0\vert e^{-i\hat{H}t/\hbar}\, \hat{c}_{\mathbf{k}}\psi_0\rangle^{*}
\label{eq:C_200}
\end{equation}
which determine the overlap between the many-body states $ \hat{c}_{\mathbf{k}}^{+}\, |\psi_0\rangle$ or $\hat{c}_{\mathbf{k}}\, |\psi_0\rangle$
associated with the addition or removal of a bare particle with momentum $\mathbf{k}$ and the 
corresponding state after time evolution. In Fourier space, the functions are positive. Their sum 
\begin{equation}
A(\mathbf{k},\varepsilon) = A_+(\mathbf{k},\varepsilon) + A_-(\mathbf{k},\varepsilon)
\label{eq:C_030}
\end{equation}
is normalized according to $\int_{\varepsilon}\, A(\mathbf{k},\varepsilon) = 1$, which is a 
simple consequence of the anti-commutation relation $\{ \hat{c}_{\mathbf{k}}, \hat{c}_{\mathbf{k}}^{+} \}=1$.  
In thermal equilibrium, the partial spectral functions are related by
the detailed balance condition 
\begin{equation}
A_-(\mathbf{k},\varepsilon) =e^{-\beta(\varepsilon -\mu)}\, A_+(\mathbf{k},\varepsilon) \qquad {\rm or} \qquad
 A_-(\mathbf{k},\varepsilon)=f({\varepsilon})\, A(\mathbf{k},\varepsilon)
\label{C_035}
\end{equation}
with $f({\varepsilon})$ the Fermi function. At zero temperature, therefore, the hole part $A_-(\mathbf{k},\varepsilon)$ 
of the spectral function vanishes for $\varepsilon >\mu\;$: it is not possible to remove a particle from the ground state
at energies above the chemical potential. Vice versa, the particle part $A_+(\mathbf{k},\varepsilon)$
vanishes for $\varepsilon <\mu$. Using the definition in Eq.~(\ref{eq:C_200}),
the total spectral weight in the hole part
\begin{equation}
\int d\varepsilon\, A_-(\mathbf{k},\varepsilon) = A_-(\mathbf{k},t=0)= n(\mathbf{k})
\label{eq:C_062}
\end{equation}
is equal to the momentum distribution.
For non-interacting Fermions with a free particle dispersion $\varepsilon_{\mathbf{k}}$,
 there is a sharp energy for a given momentum $\mathbf{k}$. The full spectral function 
 $A^{(0)}(\mathbf{k},\varepsilon)=\delta(\varepsilon-\varepsilon_{\mathbf{k}})$ thus exhibits a single peak.
In the interacting case, one finds in general a continuous distribution of energies
associated with a given momentum.\\

 For a description of the RF spectrum, we assume that the final state, which is 
denoted by an index $f$, has a negligible interaction with the initial one. It can thus 
be described by the free particle spectral function
\begin{equation}
A_f(\mathbf{k},\varepsilon) = \delta( \varepsilon - [E_f + \varepsilon_{\mathbf{k}}] ) \ ,
\label{eq:C_210}
\end{equation}
where $E_f=\hbar\omega_{23}$ is the bare excitation energy of the final state.
The rate of transitions out of the initial state induced by the RF field with frequency $\omega$ 
into free atoms with wave vector $\mathbf{k}$, which is measured in momentum resolved RF~\cite{stew08arpes} is given by~\cite{haus09rf}
\begin{equation}
I(\mathbf{k},\omega) =\ \hbar \int d\varepsilon\,
 A_{f}(\mathbf{k}+\mathbf{q},\varepsilon+\hbar\omega) A_-(\mathbf{k},\varepsilon)\, =\, \hbar\, A_-(\mathbf{k},\varepsilon_{\mathbf{k}}-\hbar\omega)\, .
\label{eq:C_220}
\end{equation}
Here, an unknown prefactor that depends on the interaction parameters for the coupling 
to the RF field has been set equal to $\hbar$, which fixes the
normalization for the total weight integrated over all frequencies (see~(\ref{eq:D_010}) below).  
Note that the wave vector $\mathbf{q}$ of the RF field is 
much smaller than those of the atoms, which allows to set $\mathbf{q}=\mathbf{0}$
in~(\ref{eq:C_220}). 
Moreover, for convenience we have taken $E_f=0$, which just redefines the 
position of zero frequency $\omega=0$ in the RF spectrum.
Measuring the number of transferred atoms without momentum resolution
gives rise to a spectrum  
\begin{equation}
I(\omega) =\hbar\int_{\mathbf{k}} \,
A_-(\mathbf{k},\varepsilon_{\mathbf{k}}-\hbar\omega)
\label{eq:C_230}
\end{equation}
which is a function only of the RF frequency $\omega$. 
In view of the fact that 
the integral of $A_{-}$ over all energies is equal to the momentum distribution
by Eq.~(\ref{eq:C_062}), our choice of normalization implies that the total weight integrated over all frequencies 
\begin{equation}
\int d\omega\, I(\omega)=n_{2}
\label{eq:D_010}
\end{equation}
is equal to the density $n_{2}$ of atoms from which
the transfer to the empty final state occurs. \\

\begin{figure}
 \includegraphics[width=1.1\linewidth]{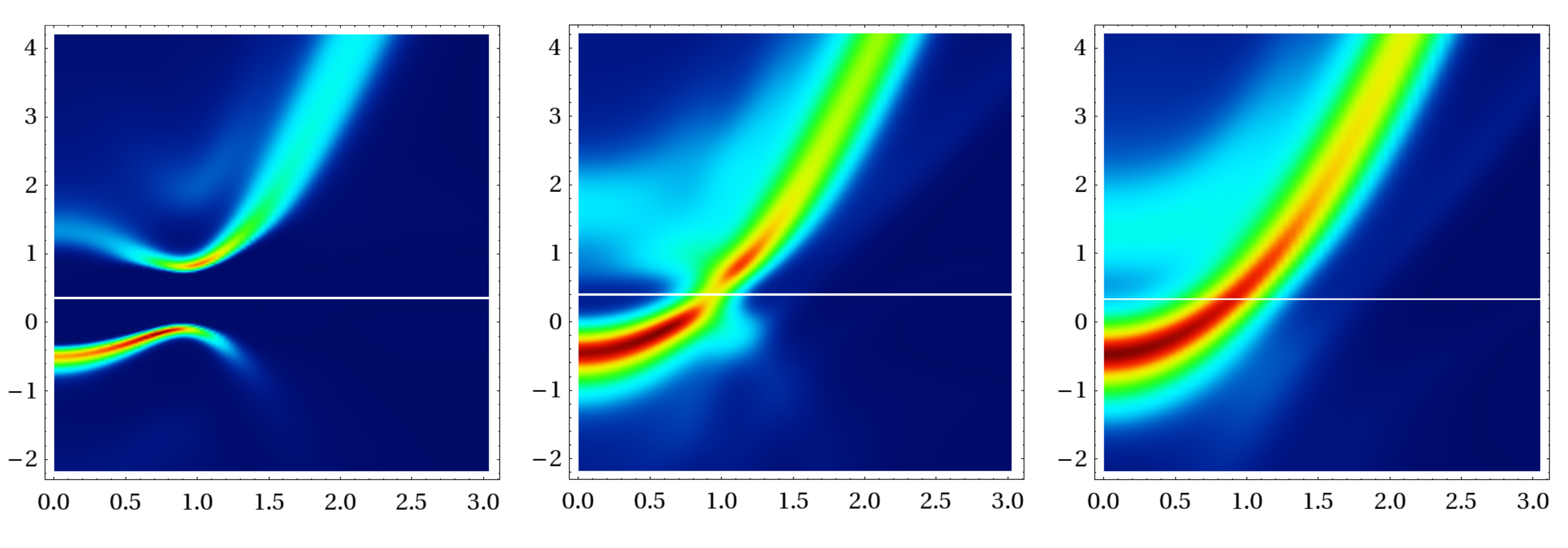}
\caption{Spectral functions of the unitary Fermi gas for different temperatures:  at $T/T_F=0.01$ (left)
at $T_c=0.16 T_F$ (middle) and at $T=0.3 T_F$ (right). Energy on the vertical axis is given 
in units of $\varepsilon_F$, momentum $|\mathbf{k}|$ on the horizontal axis in units of $k_F$. The 
horizontal line denotes the position $\varepsilon=\mu$ of the chemical potential  (from~\cite{haus09rf}).
\label{fig:RF-spectra}}
\end{figure}

The full spectral function $A(\mathbf{k},\varepsilon)$ at finite temperature 
is determined by the  
single particle temperature Green function $\mathcal{G}(\mathbf{k},\omega_n)$ 
via the spectral representation~\cite{fett71}
\begin{equation}
\mathcal{G}(\mathbf{k},\omega_n) = \int d\varepsilon
\,\frac{A(\mathbf{k},\varepsilon)}{i\hbar\omega_n-(\varepsilon-\mu)} \ .
\label{eq:C_010}
\end{equation}
Specifically, the results shown in Fig.~\ref{fig:RF-spectra}
are obtained from the Green function $\mathcal{G}(\mathbf{k},\omega_n)$
within the Luttinger-Ward approach  
by inverting this representation~\cite{haus09rf}. While unique in principle, the 
inversion is not a stable procedure in a mathematical sense. The 
consistency of the results for the spectral functions 
have, therefore, been checked carefully
by making sure that both sides of Eq.~(\ref{eq:C_010}) agree at the level of $10^{-5}$
over the whole relevant range of momenta and frequencies~\cite{haus09rf}.  
Since the thermodynamic properties obtained from the Luttinger-Ward approach are in
very good agreement with precision measurements for the unitary gas (see section 3.2),
the associated spectral functions are also expected to be quantitatively reliable.  
The numerical results for 
the spectral functions $A(\mathbf{k},\varepsilon)$ of 
the unitary gas are shown in Fig.~\ref{fig:RF-spectra}.  
Deep in the superfluid regime, at $T/T_F=0.01$, the excitation spectrum 
has a structure similar to that obtained within a BCS description, 
where the spectral function~\cite{fett71}
\begin{equation}
A_{\rm BCS}(\mathbf{k},\varepsilon) = u_{\mathbf{k}}^2\,
\delta(\varepsilon - E^{(+)}_{\mathbf{k},\rm BCS})
+ v_{\mathbf{k}}^2\,\delta(\varepsilon-E^{(-)}_{\mathbf{k},\rm BCS})
\label{eq:C_080}
\end{equation}
consists of two infinitely sharp peaks. The associated energies 
\begin{equation}
E^{(\pm)}_{\mathbf{k},\rm BCS}=\mu\pm\sqrt{\left(\varepsilon_{\mathbf{k}}-\mu\right)^2+\Delta^2}
\label{eq:C_085}
\end{equation}
describe the standard dispersion of Bogoliubov quasiparticles. 
The appearance of
two separate peaks in  $A(\mathbf{k},\varepsilon)$ is a consequence of the
fact that these quasiparticles are a coherent superposition of particle creation and
annihilation with amplitudes $u_{\mathbf{k}}$ and $v_{\mathbf{k}}$ according to
$\hat{\gamma}_{\mathbf{k}\uparrow}^{+}=u_{\mathbf{k}}\hat{c}_{\mathbf{k}\uparrow}^{+} -v_{\mathbf{k}}\hat{c}_{-\mathbf{k}\downarrow}$.
At zero temperature, the spectral function $A_-(\mathbf{k},\varepsilon)$ associated with the removal of
an atom is just the part below the line $\varepsilon=\mu$. Within BCS,   
this coincides with the second term in Eq.~(\ref{eq:C_080}). 
Its prefactor $v_{\mathbf{k}}^2=n_{\sigma}(\mathbf{k})$ is non-vanishing at arbitrary 
values of the momentum.  Particles can therefore be removed {\it above} $k_F$
even at zero temperature. The minimum distance of $E^{-}_{\mathbf{k},\rm BCS}$ to 
the chemical potential defines the excitation gap $\Delta$. Within conventional BCS theory,
this minimum is reached at $k_{0,\rm BCS}=k_F$, i.e. right at the Fermi surface
because the deviation between the chemical potential 
$\mu$ and the bare Fermi energy $\varepsilon_F$ is of order $\Delta^2$ and
thus exponentially small in the BCS limit~\cite{fett71}. \\

For the unitary gas, 
a number of important differences 
compared to~(\ref{eq:C_080}) show up even near $T=0$, where the structure 
of the spectrum appears to be similar to BCS.  First of all, 
the minimum value of the excitation gap is at a wave vector $k_{0}\simeq 0.92\, k_F$.
It is below but still close to the Fermi surface of the non-interacting gas. This is quite remarkable
since the $T=0$ momentum distribution of the unitary gas has no sharp features at any $k$. 
Moreover, the chemical potential $\mu=\xi_s\varepsilon_F$ is much less than $\varepsilon_F$.
A BCS spectrum of the form~(\ref{eq:C_085}) would therefore predict a minimum at $k_0=\sqrt{\xi_s}\, k_F\simeq 0.6\, k_F$.  
As a second point, the fermionic excitations of the unitary gas 
exhibit a finite lifetime away from the extrema of the dispersion, even at vanishing temperature.
This is in stark contrast to Bogoliubov excitations, which 
have infinite lifetime at {\it arbitrary} momenta $k$ due to 
the presence of delta functions in~(\ref{eq:C_080}). For the unitary gas, considering e.g. the lower (hole) branch in
Fig.~\ref{fig:RF-spectra} a, there is an appreciable broadening which increases upon 
moving away from the dispersion maximum. A hole deep in Fermi sea therefore has a short lifetime. 
It is only near the minimum of
the dispersion curve for the particles or close to the maximum of the dispersion
for holes where the lifetime broadening has 
to vanish because there are no available final states for decay.
The numerically calculated spectral functions are not precise enough to 
account for that but they show at least the expected tendency of an increasing
broadening away from the extrema of the dispersion 
\footnote{Exact results for the spectrum of fermionic excitations, including their damping,  
have been obtained by Nishida in the regime $k\gg k_F$ from an operator product expansion~\cite{nish12OPE}.}
As a third point, in the vicinity of the extrema, the dispersion of the fermionic excitations can be parametrized by
\begin{equation}
E^{(\pm)}_{\mathbf{k}}=\mu\pm \left[ |\Delta| +\frac{\hbar^2}{2m^{*}}(|\mathbf{k}|-k_{0})^2 +\ldots\right] \,.
\label{eq:C_086}
\end{equation}
As discussed in~\cite{haus09rf}, the associated values of the effective mass $m^{*}$ and the value $k_0$ 
turn out to be slightly different for particle and hole excitations. 
The particle-hole symmetry of the standard BCS description
of the quasiparticle dispersion is therefore broken at these
large coupling strengths. In quantitative terms,  the gap $\Delta\simeq 0.46\, \varepsilon_F$ of the unitary gas 
is almost half the bare Fermi energy~\cite{haus07bcsbec}. 
This value, which is consistent with Monte Carlo results~\cite{carl08gap}, is in very good agreement
with experiments at MIT, where the pairing gap $\Delta\simeq 0.44\,\varepsilon_F$
at unitarity was directly measured by injecting unpaired atoms into the superfluid in a slightly imbalanced Fermi mixture~\cite{schi08gap}. 
Paired and unpaired atoms respond at different frequencies, allowing to read off the pairing gap.\\

\begin{figure}
\begin{center}
\includegraphics[width=2.8in]{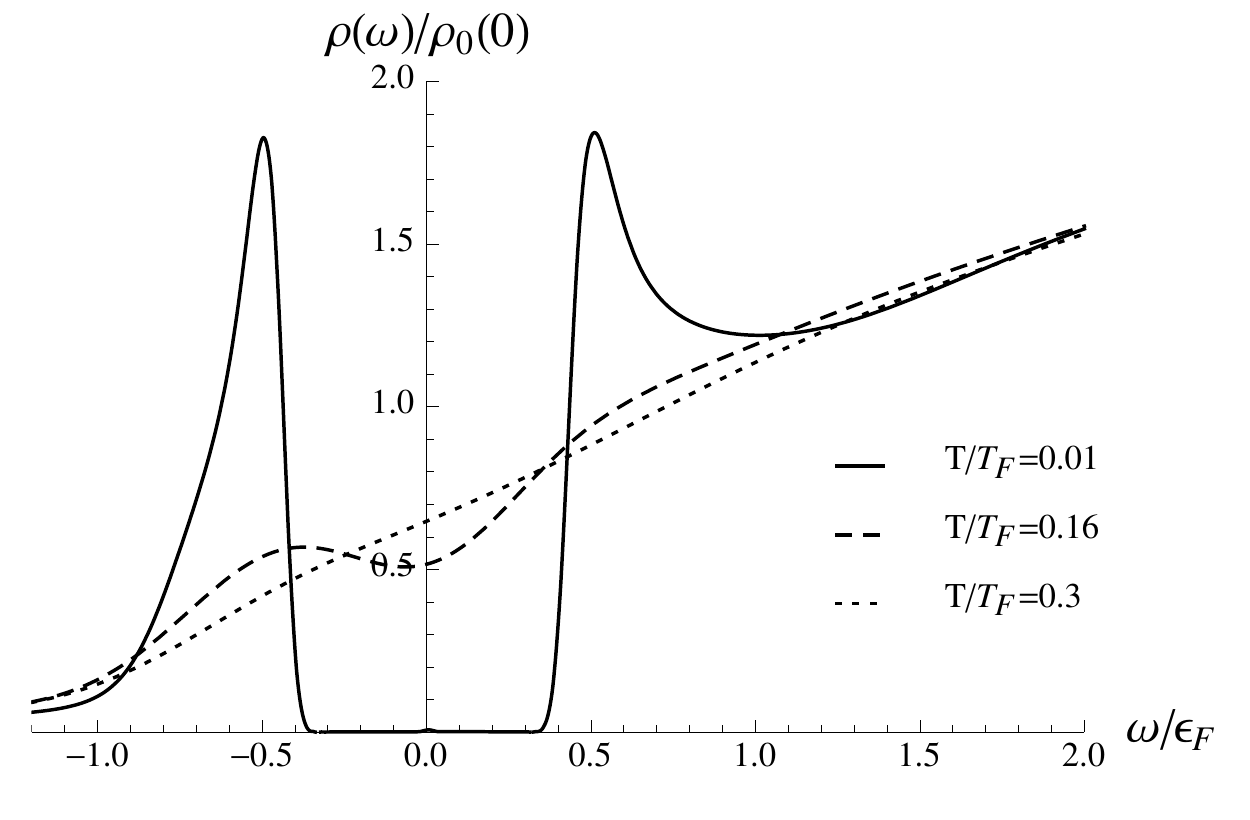} \vspace*{0.3cm}\includegraphics[width=2.2in]{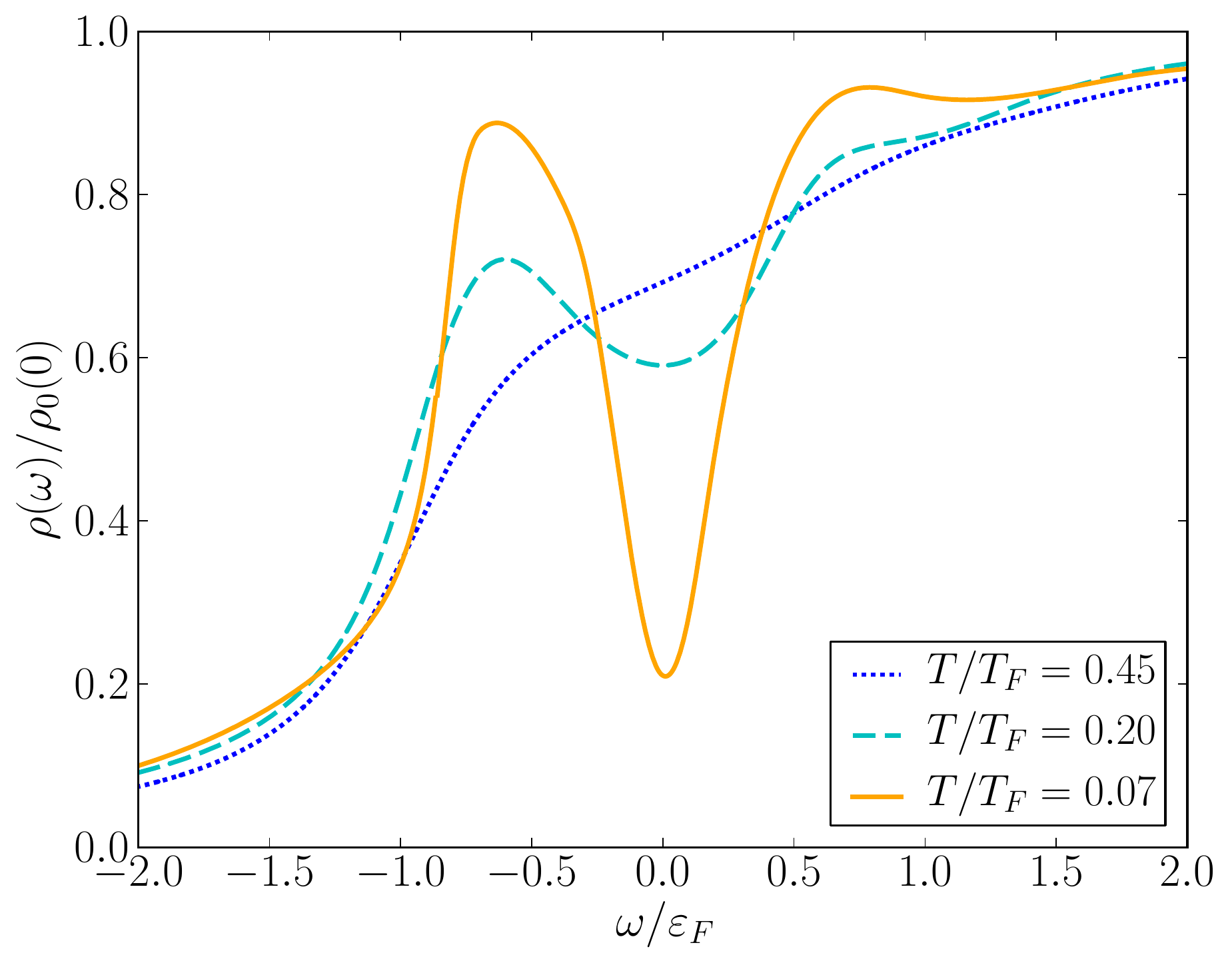}
\caption{Left: Normalized density of states for the unitary Fermi gas in 3D at different temperatures.
No pseudogap is present above the critical temperature $T_c=0.16\, T_F$. 
Right:  Normalized density of states in the normal state of a 2D Fermi gas at $\ln{(k_Fa_2)}=0.8$.
Here, a pronounced pseudogap appears above the critical temperature $T_c\simeq 0.07\, T_F$ (from~\cite{baue14}). }  
\label{fig:pseudogap}
\end{center}
\end{figure}

With increasing temperature, the two separate quasiparticle branches,
which are a signature of a coherent mixing of particle and hole excitations, 
merge into a single excitation structure. This is seen already at $T_c$ where
there is still some suppression of spectral weight near $\varepsilon=\mu$ 
and in fully developed form at $T=0.3\, T_F$. The dispersion of 
the associated fermionic excitation is close to quadratic. It is, however, 
shifted downwards quite substantially compared to a 
free particle dispersion. This is due
to the strong attractive interaction of a Fermion with its 
environment at unitarity and has been observed as a 'Hartree'
shift of the RF spectra above $T_c$~\cite{schi08gap}. It is 
similar to the shift $-0.6\, \varepsilon_F$ which appears
in the Fermi polaron problem where a single down spin
is added to a sea of up-spin Fermions~\cite{prok08polaron}.   
Apart from the dominant peak, the spectral functions at and above $T_c$ 
show some additional structure at positive energies.
Specifically, a rather broad structure is visible for $k \lesssim k_F$ 
and energies in the range between $\varepsilon_F$ and $2\,\varepsilon_F$ in Fig.~\ref{fig:RF-spectra}, 
which contains $\simeq17$\% of the spectral weight at $T=0.3\, T_F$. 
The physical origin of this structure 
is quite likely due to the {\it repulsive} branch of the Feshbach resonance which 
has been studied in the context of a possible ferromagnetic state of
the unitary Fermi gas~\cite{jo09ferro}. Unfortunately, the lifetime of 
the repulsive branch is very short, consistent with the rather broad 
feature seen in Fig.~\ref{fig:RF-spectra}. 
Finally, we briefly adress the issue of a possible pseudogap phase of the unitary gas 
due to preformed pairs above $T_c$. This concept is supported by 
a number of theories which 
indicate a gapped excitation spectrum with two branches of the form in
Eq.~(\ref{eq:C_085}) even above the critical temperature, see e.g.~\cite{bulg09,chen09rf,pier09rf}. 
Momentum  resolved RF spectra involving a trap average appear consistent with the pseudogap scenario~\cite{gaeb10pseudo}.
It is not found in the Luttinger-Ward approach, however, which is consistent with measurements of  the thermodynamic 
properties~\cite{ku12superfluid,nasc11fermiliquid}
including, in particular, the absence of a strong suppression of the spin susceptibility $\chi_s$ above $T_c$~\cite{somm11}.  
A related, quantitative criterion for a possible pseudogap is provided by the spectral density of states
\begin{equation}
\rho(\omega)=\int_{\mathbf{k}} A(\mathbf{k}, \varepsilon=\mu+\hbar\omega)\, .
\label{eq:C_280}
\end{equation}
It reflects the density of excitations summed over all possible wave vectors,
measured with respect to the chemical potential. A pseudogap is present if this density
of states has a strong suppression near $\omega=0$ even above $T_c$.
Now, as shown in Fig.~\ref{fig:pseudogap}, within the Luttinger-Ward approach,
the unitary gas right at $T_c$ shows some suppression of spectral weight near $\omega=0$ 
but not a pronounced gap.  Moreover, the suppression disappears rather quickly  
above $T_c$. As a result, there is no clear separation of temperatures in the unitary
gas between the superfluid transition and a pair formation temperature $T^{*}$.  
This behavior should be contrasted
with the situation in 2D, where a clear pseudogap is found to be present at $T_c$ 
and appreciably above~\cite{baue14} even for a coupling constant $\ln{(k_Fa_2)}=0.8$ 
on the BCS side of the crossover, where the Fermi energy is still larger
than the two-body binding energy $\hbar^2/(ma_2^2)$. The
presence of a pseudogap in the normal fluid phase of attractive Fermi gases in
2D is consistent with momentum resolved RF spectroscopy performed by M. K\"ohl
and coworkers in Cambridge~\cite{feld11}.  \\

A simple limit in which the momentum integrated RF spectrum can be given in analytical form 
describes dissociation of single bound pairs of atoms, which applies in the BEC limit. 
The resulting spectrum
\begin{equation}
I_{BEC}(\omega) = \frac{n}{\pi a\sqrt{m}}\frac{(\hbar\omega-\varepsilon_b)^{1/2}}{\omega^2}\cdot\theta(\hbar\omega-\varepsilon_b)
\label{eq:C_290}
\end{equation}
at zero temperature has a sharp onset at the molecular binding energy $\varepsilon_b=\hbar^2/ma^2$.
At large frequencies the RF spectrum falls off like 
$\omega^{-3/2}$. As realized by Schneider and Randeria~\cite{schn10shortrange} and by Braaten et.al.~\cite{braa10rf},
this is a special case of an exact result which connects the behavior of the RF spectrum
at large frequencies with the Tan contact density $\mathcal{C}$ of the strongly correlated initial state. 
For a derivation of this connection, it is convenient to rewrite Eq.~(\ref{eq:C_230})
in the form
\begin{equation}
I(\omega) =\hbar\int d\varepsilon\int_{\mathbf{k}} \,
A_-(\mathbf{k},\varepsilon)\,\delta(\varepsilon-\varepsilon_{\mathbf{k}}+\hbar\omega)
\label{eq:C_240}
\end{equation}
and to realize that large frequencies require that $\varepsilon_{\mathbf{k}}\simeq\hbar\omega$ becomes 
large too, such that $\varepsilon$ is negligible in the argument of the delta-function.
Using Eq.~(\ref{eq:C_062}), which connects the integral of  $A_-$ with respect to $\varepsilon$ with the momentum distribution,
one obtains
\begin{equation}
I(\omega\to\infty) =\hbar\int_{\mathbf{k}} \,
n_{\sigma}(\mathbf{k})\,\delta(\varepsilon_{\mathbf{k}}-\hbar\omega)\to 
\frac{\mathcal{C}}{4\pi^2}\left(\frac{\hbar}{m}\right)^{1/2}\cdot\omega^{-3/2}
\label{eq:rf-tail1}
\end{equation}
in three dimensions by simple power counting, since the delta-function fixes $k=\sqrt{2m\omega/\hbar}$ 
and $n_{\sigma}(\mathbf{k})$ can be replaced by its asymptotic behavior $\mathcal{C}/k^4$
derived in section 2.5.  The power law decay of the RF spectrum predicted by Eq.~\eqref{eq:rf-tail1} 
has been observed by Stewart et.al.~\cite{stew10contact} over an appreciable range of 
$k_Fa$ - values. In particular, the experiments have verified that the contact extracted from
the tail of the RF spectrum agrees with the one obtained from the asymptotics~\eqref{eq:Tan-momentum} 
of the momentum distribution. In practice, the power law
holds for frequencies larger than the intrinsic characteristic scales $\hbar k_F^2/m$,  $\hbar /ma^2$
and $k_B T/\hbar$ of the strongly interacting many-body system.
It extends up to frequencies of order $\hbar/mr_e^2\simeq E_{\text{vdw}}/\hbar$ beyond which it 
is cutoff by finite range effects. Remarkably, the scale at which finite range effects become 
appreciable depends on the  strength of final state interactions, which
have been neglected so far.  This becomes evident from an
explicit calculation of the RF spectrum in the molecular limit by Chin and Julienne~\cite{chin05rf}.
They have shown that in the presence of a nonzero
scattering length $a_f\ne 0$ between the hyperfine state 
that is not affected by the RF pulse and the final state 
of the RF transition, the spectrum decays like $\omega^{-5/2}$ at 
large frequencies. The short range part of the interaction,  
which is responsible for the slow decay of the spectrum, is 
thus cancelled by the interaction between the final state and the state
that remains after the RF transition~\cite{baym07,zhan08rf}. 
A quite general result for the RF spectrum in the presence of 
final state interactions has been
derived by Braaten, Kang and Platter~\cite{braa10rf}. Using a short time and short distance 
expansion of an operator product which involves the RF transition
operator $\hat{\psi}_3^{+}\hat{\psi}_2$ at points separated 
by small times and distances $t, |\mathbf{x}|\to 0$, they have shown that
the asymptotic behavior of the RF spectrum is given by
\begin{equation}
I(\omega\to\infty)  =
\frac{(1/a_f - 1/a)^2}
    {4 \pi^2 \omega \sqrt{m \omega/\hbar} (1/a_f^2 + m \omega/\hbar)}\,\mathcal{C}\, .
\label{eq:rf-tail2}
\end{equation}
The RF spectrum therefore decreases like $\omega^{-5/2}$ at frequencies  
$\omega \gg \hbar/m a_{f}^2$. The behavior found in  
Eq.~(\ref{eq:rf-tail1}), in turn,  is recovered in the regime $\omega \ll \hbar/m a_{f}^2$
which always applies in the limit of negligible final state interactions $a_{f} \to 0$.
For finite $a_f$, the asymptotic $\omega^{-5/2}$ decay 
guarantees that the RF spectrum has a first moment which is not sensitive to the
range of the interactions. The resulting average clock shift~\cite{punk07rf, baym07}  
\begin{equation}
\bar{\omega}=\frac{\int d\omega\, \omega I(\omega)}{\int d\omega\, I(\omega)}=
\frac{\hbar\,\mathcal{C}}{4\pi mn_{2}} \left(\frac{1}{a}-\frac{1}{a_{f}}\right)
\label{eq:clock-shift}
\end{equation}
is again determined by the contact coefficient $\mathcal{C}$.
In the special case of equal scattering lengths $a=a_f$ 
of the initial and final states with the unperturbed state $\left|1\right>$,
both the high-frequency tail in Eq.~\eqref{eq:rf-tail2} and the 
clock shift~\eqref{eq:clock-shift} vanish identically. 
In this case, the RF spectrum just consists of an unshifted peak
$I(\omega)=n_{2} \delta(\omega)$, because the RF pulse merely 
rotates $\left|2\right>$ and $\left|3\right>$ in spin space~\cite{baym07}.
The result~\eqref{eq:clock-shift} has played an important role in understanding
an initially surprising observation in the RF shift of Fermi gases at a finite imbalance~\cite{schu07pair}.
In fact, this shift did hardly change between the balanced superfluid and 
a normal state beyond a critical value of the imbalance,
where superfluidity is destroyed by the large 
mismatch of the Fermi energies.  Now, combining Eq.~\eqref{eq:clock-shift} with 
the result  $s=\mathcal{C}/k_F^4\simeq 0.1$ for the dimensionless contact density of the balanced gas at unitarity $1/a=0$
from section 2.2, gives a positive clock shift $\bar{\omega}\simeq -0.46\, v_F/a_f>0$ 
due to the negative final state scattering length $a_f=a_{13}<0$. In the strongly imbalanced 
limit of a vanishing density $n_2=k_{F,\downarrow}^3/(6\pi^2)\to 0$ of minority atoms
which are transferred by the RF pulse, the contact density $\mathcal{C}=\tilde{s}\, k_{F,\downarrow}^3k_{F,\uparrow}$
vanishes linearly with the minority density $n_2$. The associated coefficient $\tilde{s}$
has been calculated from a variational solution of the Fermi polaron problem~\cite{punk09molaron}
and turns out to be $\tilde{s}\simeq 0.08$ at unitarity. The clock shift 
therefore approaches $\bar{\omega}\simeq -0.43\, v_{F,\uparrow}/a_f$ in the strongly imbalanced  
limit which is close to the value in the balanced gas. Clock shifts are therefore a measure of 
the contact, i.e. the probability of finding Fermions of opposite spin close together, but not of 
superfluidity. In fact, as discussed below Eq.~(\ref{eq:contact-Delta}), the contact 
is only weakly affected by the superfluid transition.  \\

\begin{figure}
\begin{center}
\includegraphics[width=3.5in]{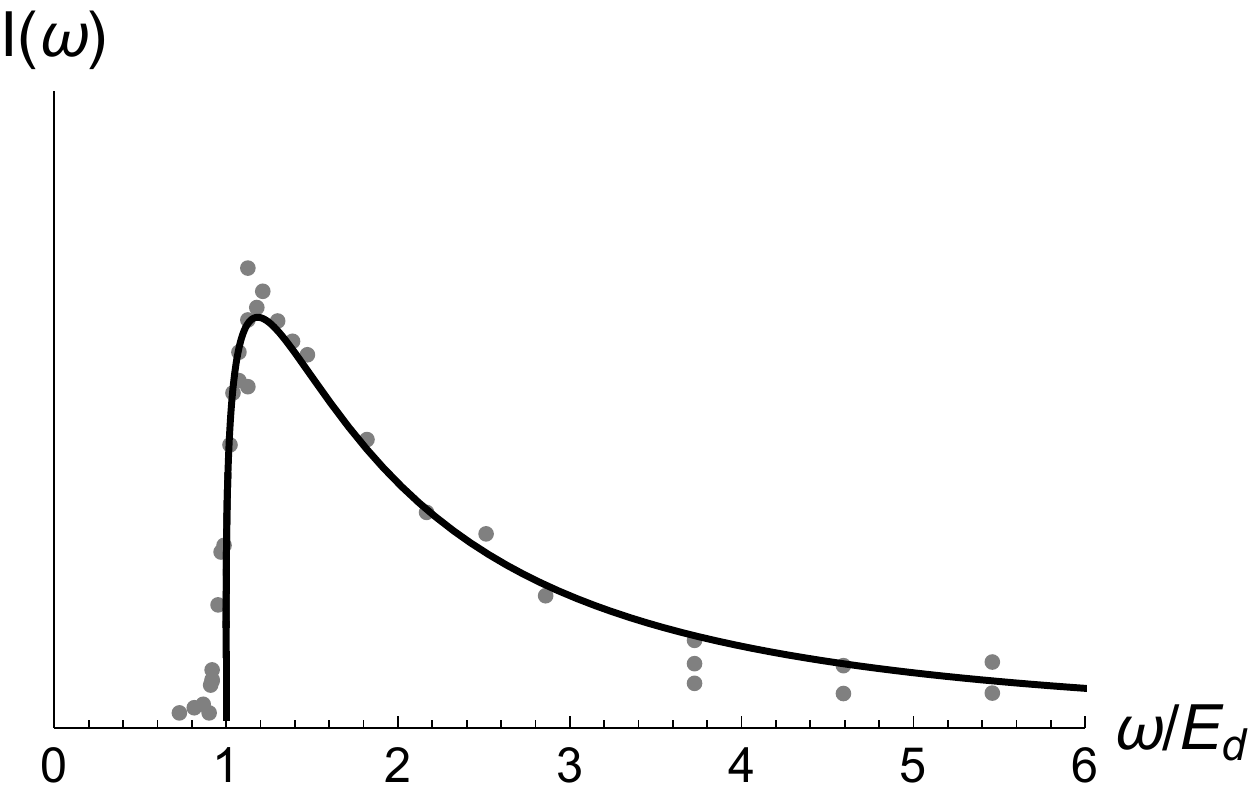} 
\caption{RF spectrum for the continuum of bound-free transitions in a two dimensional
Fermi gas at  $\ln{(k_Fa_2)}\simeq -0.55$. The threshold for breaking a molecule is at $\omega=E_d$.
Experimental data from~\cite{somm12} are the gray dots, the full line is the theory prediction~\cite{lang12}. }  
\label{fig:RF-2D}
\end{center}
\end{figure}
An extension of the result~(\ref{eq:rf-tail2}) for the asymptotics of the RF spectrum to strongly interacting Fermi 
gases  in two dimensions has been derived by Langmack et.al.~\cite{lang12}. In this case, a non-vanishing
final state interaction gives rise to logarithmic corrections
\begin{equation}
I(\omega\to\infty)= \frac{\hbar\, \ln^2(E_d'/E_d)} {4\pi m \omega^2 \big[ \ln^2(\omega/E_d') + \pi^2 \big]} \,\, \mathcal{C} 
\label{eq:rf-tail3}
\end{equation}
 to the naive $1/\omega^2$ scaling expected from dimensional analysis as in~(\ref{eq:rf-tail1}). 
 They involve the two-body binding energies $E_d=\hbar^2/(ma_2^2)$ and $E_d'=\hbar^2/(ma_2'^2)$ associated 
 with the 2D scattering lengths in the initial and final state
 \footnote{ The expression $\hbar^2/ma_2^2$ for the binding energy is consistent with  
the value~(\ref{eq:a_2(a)}) of the 2D scattering length only for very weak binding at negative values of $a$,
where $a_2$ is exponentially large. In turn, 
for $0<a\ll\ell_z$, the binding energy approaches the 
3D result $\hbar^2/ma^2$ while the 2D scattering length which describes the 
interaction of particles in the continuum, is exponentially small. For a discussion, see section V.A in~\cite{bloc08review}.}. 
The logarithmic corrections are a consequence of the violation of scale invariance due to the 
cutoff dependence of the bare coupling constant $\bar{g}_2(\Lambda)$ discussed in Eq.~(\ref{eq:LS-equation-2D}) above. 
Experimentally, these corrections show up in the RF spectrum of a collection of essentially decoupled 2D Fermi gases which
are generated by adding a deep 1D optical lattice to a cloud of $^6$Li atoms~\cite{somm12}. At 
the 3D Feshbach resonance, the binding energy of a two-body bound state is $E_d=0.244\,\hbar\omega_z$~\cite{bloc08review}.
For the parameters used in the experiment, this is about six times the Fermi energy. 
The confined, resonant Fermi gas is therefore essentially in the BEC limit, with a 
dimensionless coupling constant $\ln{(k_Fa_2)}\simeq -0.55$. Due the strong final state interactions
with $a_2'\simeq 0.32\, a_2$, the detailed form of the dissociation spectrum
exhibits a smooth onset $\sim 1/\ln^2[(\omega-E_d)/E_d']$ 
near the threshold $\omega=E_d$ instead of a jump associated with the one in the 2D density of states~\cite{lang12}.
Moreover, this factor also leads to a decay of the RF spectrum at large frequencies which is
faster than the $1/\omega^2$ tail obtained in a Fermi golden rule approach. As shown in Fig.~\ref{fig:RF-2D},
both features are consistent with experiment. The detailed form of the RF spectrum thus
provides a clear signature of the violation of scale invariance in 2D.

\newpage

\subsection{Quantum limited viscosity and spin diffusion}

A quite unexpected connection between the physics of ultracold atoms
and recent developments in field theory has opened up with the realization
that not only equilibrium but also transport properties of the
unitary Fermi gas exhibit universal features~\cite{gelm05viscosity, son07bulk}. 
An example, discussed already in section 3.4, is the vanishing of the bulk viscosity.
More generally, it turns out that the existence of universal scaling functions associated
with the zero density quantum critical point of the unitary gas implies that transport 
coefficients like shear viscosity or the spin diffusion 
constant exhibit minimal values
which are determined only by the mass of the particles and
constants of nature like $\hbar$ and $k_B$.  \\

Experimentally, transport in strongly interacting Fermi gases has been investigated by J. Thomas 
and coworkers and by M. Zwierlein and his group who have measured 
the shear viscosity and the spin diffusion constant, respectively. 
The latter can be determined from the late stage equilibration
dynamics of two initally separated spin components of the unitary gas~\cite{somm11}.
The diffusion constant is a decreasing function of temperature, reaching a minimum
close to, but still above, the superfluid transition. By extracting a spin diffusivity 
of the homogeneous system from the measured trap-averaged data,
the minimum value is $D_s\simeq 1.3\,\hbar/m$. 
Similar results have been obtained in recent measurements 
of transverse spin diffusion by J. Thywissen and coworkers in 3D and by M. K\"ohl et.al. in 2D Fermi gases.
The associated diffusion constant $D_s^{\perp}$ is obtained from the decay of a magnetization $\mathbf{m}=m\,\mathbf{e}$
due to an inhomogenity of its direction $\mathbf{e}(\mathbf{x},t)$ at constant  
magnitude $m\sim n_{\uparrow}-n_{\downarrow}$. In the 3D case, the transverse spin diffusion constant 
of an initially fully polarized Fermi gas
approaches a minimum value $D_s^{\perp}\simeq 1.1\, \hbar/m$ at the lowest temperature $T\simeq 0.25\, T_F$~\cite{bard14},
consistent with theoretical results based on a solution of the quantum kinetic equation~\cite{enss13spindiffusion}. 
In 2D, a much lower value  $D_s^{\perp}\simeq 0.006\, \hbar/m$ is found~\cite{kosc13}, for which no
quantitative explanation has been given so far. The shear viscosity of the unitary Fermi gas has been determined 
from the damping of the radial breathing mode~\cite{turl08perfect}
or the precise time dependence of the inversion of the aspect ratio in the expansion of 
the gas from a strongly anisotropic trap~\cite{cao11viscosity}.  
As a consequence of scale invariance, the aspect ratio exhibits an elliptic flow
while the mean square radius expands ballistically. This provides a direct signature 
of the vanishing bulk viscosity of the unitary gas~\cite{elli14conformal}. In addition,  the power law dependence
$\eta(T)\sim T^{3/2}$ in the non-degenerate limit, 
where the Boltzmann equation applies~\cite{bruu07viscosity},
has been verified in the temperature range above $\simeq T_F$. 
Using the precision data for the equation of state discussed in section 3.2, recent measurements
have succeeded to obtain the local shear viscosity as a function of density and temperature from 
trap-averaged results. The minimum value 
in the normal state just above the superfluid transition is $\eta\simeq 0.5\,\hbar n$~\cite{jose15viscosity}
\footnote{In the superfluid, the measured shear viscosity continues to decrease, showing no sign
of the expected eventual rise as $T\to 0$. For a discussion see below.}.
In terms of the kinematic viscosity $\nu=\eta/\rho=D_{\eta}$,  this implies a shear diffusion 
constant $D_{\eta}\simeq 0.5\,\hbar/m$.
A qualitative argument for the fact that various diffusion constants of the unitary Fermi gas at low 
temperatures exhibit a minimum value of order $\hbar/m$ can be given within kinetic theory, 
where $D_{\rm kin}=\langle v\rangle \ell/3$
is determined by an average velocity times the relevant mean free path $\ell$. 
At low temperatures, $\langle v\rangle$ must clearly be of order $v_F=\hbar k_F/m$. Moreover, 
for a unitarity limited interaction, the mean free path is expected to be bounded below 
by the mean interparticle spacing. This gives $\ell\simeq 1/k_F$ and thus a diffusion
constant of order $\hbar/m$.  While correct on purely dimensional grounds, the argument is based on kinetic
theory which only applies if the quasiparticles whose transport leads to equilibration are well defined.
In the relevant regime, where the diffusion constants take their minimal values, this is not 
a valid assumption, however. Indeed, the minima appear within the 
quantum critical regime above the zero density quantum critical point discussed in section 3.1.
As will be shown below, the appearance of universal values for the 
transport coefficients of the unitary gas is a consequence of the universality of scaling functions 
and amplitude ratios associated with this quantum critical point rather than a simple saturation of 
some ill defined mean free path.
Specifically, the spin diffusion constant is related to a spin conductivity $\sigma_s$ and 
the equilibrium spin-susceptibility $\chi_s$ by an Einstein relation $D_s=\sigma_s/\chi_s$~\cite{somm11}.
Near the zero density quantum critical point $\mu=T=0$
both $\sigma_s=f_{\sigma}/(\hbar\,\lambda_T)$ and $\chi_s=f_{\chi}/(k_BT\,\lambda_T^3)$ 
can be expressed in terms of universal scaling functions $f_{\sigma,\chi}(x=\beta\mu)$.
Their ratio exhibits a minimum at $x_{\rm min}\simeq 0.3$ within the quantum critical regime
which gives rise to a universal prefactor of the spin diffusion constant $D_s=(2\pi\, f_{\sigma}/ f_{\chi})_{\rm min} \, \hbar/m$.
On a quantitative level, the observed temperature dependence and minimum value $D_s\simeq 1.3\,\hbar/m$ 
can be explained by evaluating the exact Kubo formula within a Luttinger-Ward approach~\cite{enss12spindiffusion}. \\

Part of the motivation to study the shear viscosity or spin diffusion constant of the unitary Fermi gas
did arise from the observation that  universal numbers in transport appear in 
scale invariant {\it relativistic} quantum field theories like the $\mathcal N=4$ supersymmetric Yang-Mills theory, whose shear viscosity 
to entropy density ratio turns out to be $\eta/s =\hbar/(4\pi k_B)$~\cite{poli01viscosity}.
Perturbations away from this exactly soluble model give rise to larger values of $\eta/s$.
Kovtun, Son and Starinets (KSS) have thus conjectured that the constant
$\hbar/(4\pi k_B)$ is a lower bound on the shear viscosity to entropy density ratio
for a large class of strongly interacting quantum field theories~\cite{kovt05visc}.
Remarkably, despite some formal counterexamples~\cite{brig08viscosity},
the KSS conjecture turns out to be valid for all fluids that are known in nature
\footnote{It holds even for water, where both $\eta$ and $s$ in the regime where
$\eta/s$ exhibits its minimum are determined by {\it classical} physics. For a simple argument
why the mininum value of $\eta/s$ for water is only a factor $\simeq 25$ above the KSS bound - which 
contains $\hbar$ - see Appendix A in~\cite{enss11viscosity}.}. This observation has motivated
a search for 'perfect fluids' which realize or a least come close to the KSS bound. 
As will be shown below,  the ratio $\eta/s$ for the unitary Fermi gas 
is a factor of about seven above this bound, a value close to that of the quark-gluon plasma~\cite{scha09perfect}. 
The unitary Fermi gas therefore appears to be the most perfect 
among all {\it non}-relativistic fluids that have been studied so far.
For a further discussion of relativistic theories and - in particular - 
the possibility of obtaining exact results for transport coefficients of strongly coupled field theories via the
AdS-CFT correspondence, see the reviews by Sachdev~\cite{sach12review} and by Adams et.al.~\cite{adam13review}.  \\

For the unitary gas at fixed density $n=k_F^3/(3\pi^2)$, purely dimensional
arguments require the static shear viscosity to be of the form~\cite{son07bulk}
\begin{equation}
  \label{eq:scaling}
  \eta(T)=\hbar n\, \alpha(\theta)
\end{equation} 
where $\theta=T/T_F$ is the dimensionless temperature and
$\alpha(\theta)$ a universal scaling function.  In the high temperature limit,
where transport coefficients can be calculated using a Boltzmann
equation, this function is actually fixed up to a numerical constant of order one.  
This is based on the counter-intuitive but well known fact
that the viscosity of a classical gas is independent of its density
\cite{bali92}.  Due to $T_F\sim n^{2/3}$, the prefactor $\sim n$ in~(\ref{eq:scaling}) is 
cancelled if and only if $\alpha(\theta\gg 1)\sim\theta^{3/2}$. At sufficiently high temperatures, 
therefore, the shear viscosity of the unitary gas increases like $T^{3/2}$. From the explicit solution of the Boltzmann equation
or the classical limit of the relevant Kubo formula, the precise result turns out to be~\cite{bruu07viscosity, enss11viscosity} 
\begin{equation}
  \eta^{\rm cl}(T)=\frac{45\,\pi^{3/2}}{64\sqrt{2}}\,\hbar n\left(\frac{T}{T_F}\right)^{3/2}=\frac{15\,\pi}{8\sqrt{2}}\,\frac{\hbar}{\lambda_T^3}\, .
  \label{eq:eta-Boltzmann}
\end{equation} 
Note that $\eta^{\rm cl}$ scales like $1/\hbar^2$, provided the assumption of zero range s-wave
scattering remains valid at temperatures above $\simeq 2\, T_F$, where~(\ref{eq:eta-Boltzmann}) applies.
At low temperatures $\theta\lesssim 0.16$, the unitary Fermi gas is a
superfluid.  Contrary to naive expectations, a superfluid is not a
`perfect fluid' despite the fact that there is a vanishing viscosity
here.  In particular, superfluids do not provide trivial
counter-examples to the KSS conjecture.  Indeed, according to the
Landau two-fluid picture, any superfluid can be thought of 
as a mixture of a normal and a superfluid component
\footnote{For the unitary Fermi gas, the two-fluid picture has been 
verified nicely  through the observation of second sound, which is
a relative oscillation of the normal and superfluid component at constant pressure~\cite{sido13}.}.
The latter has both zero entropy and zero viscosity, so $\eta/s$ is undefined.  
In turn, the normal component which is present at any finite temperature, has 
a non-vanishing entropy and viscosity. In the context of superfluid $^4$He, this
viscosity was calculated by Landau and Khalatnikov in
1949~\cite{land49viscosity}.  In the low temperature, phonon-dominated
regime, the shear viscosity of the normal component
grows like $T^{-5}$ because the mean free path for phonon-phonon
collisions, which are necessary for the relaxation of shear, diverges.
Specifically, with the assumption that the phonon dispersion has
negative curvature and thus no Beliaev decay is possible, 
Landau and Khalatnikov have shown that the low temperature shear viscosity is given by
\begin{equation}
  \label{eq:landau}
  \eta(T\to 0)=\rho_n(T)\,\frac{\rho^2
    c_s^3}{\hbar^2}\,\frac{2^{13}(2\pi)^7}{9(13)!(u+1)^4}
  \left(\frac{\hbar\, c_s}{k_BT}\right)^9\, .
\end{equation}
Here $\rho_n(T) =(2\pi^2\hbar/45 c_s)(k_BT/\hbar c_s)^4$ is the  mass density of the 
normal fluid while $u=d\ln{c_s}/d\ln{n}$ is the dimensionless
strength of the non-linear corrections to the leading-order quantum
hydrodynamic Hamiltonian which lead to phonon-phonon scattering~\cite{son06symmetry}.
The viscosity of the normal fluid component thus asymptotically diverges
like $T^{-5}$.  As realized by Rupak and Sch\"afer~\cite{rupa07viscosity}
a similar result is expected to hold for the unitary Fermi gas.  Indeed, at
temperatures far below $T_c$, the microscopic nature of the superfluid
is irrelevant and the linearly dispersing Bogoliubov-Anderson phonons
are the only excitations that remain. The result~\eqref{eq:landau} 
therefore applies also to the unitary Fermi
gas, provided the exact values of the sound velocity $c_s$ and
coupling constant $u$ are inserted.  At unitarity, the sound velocity
$c_s=v_F\sqrt{\xi_s/3}\simeq 0.36\, v_F\sim n^{1/3}$ is directly proportional to the
Fermi velocity $v_F$ with a prefactor, which is determined by the 
Bertsch parameter $\xi_s\simeq 0.37$. The dimensionless
coupling constant  $u=d\ln{c_s}/d\ln{n}$ which fixes the strength of the phonon-phonon
scattering amplitude thus has the universal value $u=1/3$.  Together with
the low-temperature expression $s=2\pi^2k_B\, (k_BT/\hbar
c_s)^3/45$ for the entropy density of a scalar phonon field, Eq.~\eqref{eq:landau} implies that the
viscosity to entropy density ratio
\begin{align}
  \label{eq:landau_ufg}
  \frac{\eta(T\to 0)}{s(T)} = \frac{\hbar}{k_B}\, 2.15 \times
  10^{-5}\, \xi_s^5\, \theta^{-8} 
  \end{align}
of the unitary Fermi gas at temperatures far below the superfluid
transition will diverge as $(T_F/T)^8$.  
Unfortunately, experiments so far 
have not been able to observe the predicted upturn in $\eta(T)$ 
below the superfluid transition, which should eventually display a $1/T^5$ power law
in the limit $k_BT\ll mc_s^2\simeq 0.2\, T_F$. This is not too
surprising since,  as mentioned in section 3.2, the phonon dominated regime is out of reach 
even as far as equilibrium properties are concerned, see e.g. Fig.~\ref{fig:specific-heat} for the specific heat.  
Moreover, in the finite geometry of an anisotropic trap, the effective mean free path 
is bounded above by the trap size.  As a result, one 
expects a saturation of the shear viscosity at a finite value as $T\to 0$ due to finite size effects, 
similar to what has been observed in the B-phase of superfluid $^3$He~\cite{einz87eta}.
The recent experimental data, however, show a monotonic decrease
of the shear viscosity down to the lowest measurable temperatures~\cite{jose15viscosity}. 
A quantitative understanding of this result has not been given so far and requires a proper description 
of the crossover from hydrodynamic behavior in the bulk and a kinetic theory at the edge of the cloud.
The situation is much more clear
for superfluid $^4$He, where the measured shear viscosity agrees with the theoretical expectation, at least at a qualitative level. 
Indeed, in this case, the shear viscosity exhibits a minimum at around 
$1.8\,$K just below the superfluid transition before it starts to rise again~\cite{heik55}. 
Based on this, it is quite certain that both the dimensionless shear viscosity $\eta/\hbar n$ and the ratio $\eta/s$ of
the {\it homogeneous} unitary gas will exhibit a minimum as a function of temperature, a behavior, which is in
fact typical for any fluid \cite{scha09perfect}. \\

In order to understand the microscopic origin of the minima in both 
$\eta/s$ or the spin diffusion constant for a unitary Fermi gas and the
appropriate limit in which these minima take universal values, it is 
convenient to switch from a constant density $n$ to a description 
with a given chemical potential $\mu$. As shown in Fig.~\ref{fig:phase}, the evolution from 
a non-degenerate gas to a superfluid can 
then be studied at a fixed low temperature $k_BT\ll\bar{E}$ by varying the chemical potential 
from negative values which obey $|\mu|\gg k_BT$ through the  quantum critical regime 
characterized by $|\mu|\ll k_BT$ into the superfluid where $\mu> 2.46\, k_BT$. 
It turns out that along such a trajectory both $\eta/s$ and the spin diffusion constant $D_s$ 
exhibit a  non-monotonic dependence on $x=\beta\mu$, with a minimum in the quantum critical regime.
On a microscopic level, transport coefficients are the zero frequency
limits of linear response functions~\cite{fors75}. They may be calculated from first
principles using a Kubo formula.  Specifically, both the shear and bulk
viscosities follow from the retarded correlation functions of the
stress tensor $\Pi_{ij}$~\cite{enss11viscosity}
\begin{align}
  \label{eq:kubo}
  \chi_{ij,kl}(\mathbf{q}=0,\omega)
  = \frac{i}{\hbar}\, \int dt\, \int d^3x\, e^{i\omega t}\, \theta(t)\,
  \bigl\langle [ \hat{\Pi}_{ij}(\mathbf{x},t), \hat{\Pi}_{kl}(\bm{0},0) ] \bigr\rangle \, .
\end{align}
The corresponding expression for the spin-conductivity involves
the retarded spin-current correlation function~\cite{enss12spindiffusion}
\begin{equation}
  \label{eq:spincorr}
  \chi_\text{js}(\mathbf q=0,\omega) = \frac{i}{\hbar} \int dt\, 
  \int d^3x \, e^{i\omega t}\,\theta(t)\, \left\langle\Bigl[(\hat{j}_\uparrow^z - \hat{j}_\downarrow^z)(\mathbf x,t),
    (\hat{j}_\uparrow^z - \hat{j}_\downarrow^z)(\mathbf 0,0)\Bigr]\right\rangle .
\end{equation}
The imaginary parts of these functions are odd in $\omega$ and determine the frequency
dependent shear viscosity and spin conductivity via 
\begin{equation}
  \label{eq:etaomega}
  {\rm Re}\;\eta(\omega) = \frac{{\rm Im}\, \chi_{xy,xy}(\mathbf q=0,\omega)}{\omega}\qquad {\rm and} \qquad
  {\rm Re}\; \sigma_\text{s}(\omega) =
  \frac{{\rm Im}\, \chi_\text{js}(\mathbf q=0,\omega)} {\omega}\, .
\end{equation}
Their static limits $\eta = \lim_{\omega\to 0}{\rm Re}\,\eta(\omega)$ and $\sigma_s= \lim_{\omega\to 0}{\rm Re}\,\sigma_s(\omega)$ 
are finite because the interactions between particles lead to a relaxation
of both transverse momentum and spin currents. In particular, scattering transfers momentum 
between $\uparrow$ and $\downarrow$ particles so that spin currents - in contrast to spin itself -
are not conserved. A short time expansion of the commutators involved in the Kubo formula 
gives rise to exact sum rules.
For the spin conductivity, this takes the simple form~\cite{enss12spindiffusion} 
\begin{align}
  \label{eq:fsum}
  \int_{-\infty}^\infty \frac{d\omega}{\pi}\,
 {\rm Re}\;  \sigma_\text{s}(\omega) = \frac{n}{m}
\end{align}
of a standard f-sum rule~\cite{fors75}. 
The related result for the shear viscosity is more 
subtle. Indeed, as discussed in the context of RF spectra in section 4.1, zero range interactions 
lead to response functions which exhibit a power law decay with a strength proportional to the contact density 
at large frequencies $\hbar\omega\gg\varepsilon_F, k_BT$~\cite{hofm11response, gold12sumrule}. Specifically,
the spin conductivity obeys ${\rm Re}\, \sigma_s(\omega)\to\hbar^{1/2}\mathcal{C}/(3\pi(m\omega)^{3/2})$~\cite{enss12spindiffusion},
a decay which is fast enough to guarantee a finite f-sum rule. The 
shear viscosity, in turn,  exhibits a power law decay $\sim \mathcal{C}/\sqrt{\omega}$~\cite{tayl10visc}.   
Thus, already the zeroth order moment of $\eta(\omega)$ diverges.
There is, however, a subtracted sum rule ~\cite{tayl10visc, enss11viscosity,gold12sumrule}  
\begin{align}
  \label{eq:etasum}
  \frac{2}{\pi} \int_0^\infty d\omega\, \Bigl[  {\rm Re}\;\eta(\omega) -
  \frac{\hbar^{3/2}\mathcal{C}}{15\pi\sqrt{m\omega}} \Bigr]
  = p - \frac{\hbar^2\mathcal{C}}{4\pi ma}
\end{align}
which relates the integrated frequency dependent shear viscosity at arbitrary values of the 
scattering length $a$ to the equilibrium pressure $p$ and the contact density $\mathcal{C}$. 
The relation~\eqref{eq:etasum} fixes the strength of the power law 
at large frequencies and may be proven by using a Ward identity due
to momentum conservation~\cite{enss11viscosity}. 
Using the expression~\eqref{eq:etaomega} and the fact that the commutators in the Kubo formula 
are purely imaginary and odd in time, the dc values of the shear viscosity and spin conductivity of the unitary gas 
can be obtained from 
\begin{equation}
\eta(\beta\mu) =   \frac{1}{\hbar}\int_0^{\infty} dt\, t\, g_{\eta}(\beta\mu, t) \qquad {\rm and} \qquad
\sigma_s (\beta\mu) =   \frac{1}{\hbar}\int_0^{\infty} dt\, t\, g_{\sigma}(\beta\mu, t)\, .
\label{eq:dc-values}
\end{equation}
The underlying time dependent correlation functions 
\begin{equation}
g_{\eta} (\beta\mu, t)=  i \int_{\mathbf{x}}\, \left\langle\Bigl[ \hat{\Pi}_{xy,xy}(\mathbf{x},t),
     \hat{\Pi}_{xy,xy}(\mathbf 0,0)\Bigr]\right\rangle
\label{eq:geta-definition}
\end{equation}
and the analogous function $g_{\sigma}(\beta\mu, t)$ for the spin conductivity, which 
is obtained by replacing the stress tensor $\hat{\Pi}_{xy,xy}$ in~\eqref{eq:geta-definition}
with the spin current density operator $\hat{j}_\uparrow^z - \hat{j}_\downarrow^z$, are real and odd in $t$. 
Now, as discussed in section 3.1, the existence of the zero density fixed point $\mu=T=1/a=0$ implies 
that both static and time dependent correlation functions of an ultracold, dilute gas near unitarity   
can be expressed in terms of universal scaling functions.  Suppressing the relevant variable $\nu=-\bar{a}/a$
by confining the discussion to the situation precisely at infinite scattering length, the analog of Eq.~\eqref{eq:t-dependent}
for the one-particle density matrix reads
\begin{equation}
g_{\eta} (\beta\mu, t) =\frac{1}{\beta^2\lambda_T^3}\, \Phi_{\eta} \left(\beta\mu ,  t/\beta\hbar\right) \;\; {\rm and} \;\;
g_{\sigma} (\beta\mu, t) =\frac{1}{(\beta\hbar)^2\lambda_T}\, \Phi_{\sigma} \left(\beta\mu ,  t/\beta\hbar\right)
\label{eq:gs-scaling}
\end{equation}
where $\Phi_{\eta}(x,y)$ and $\Phi_{\sigma}(x,y)$ are dimensionless, universal scaling functions.  This
result, which is the underlying microscopic basis for understanding the appearance of universal numbers for the 
transport coefficients of the unitary Fermi gas is based on one crucial assumption beyond the existence of the 
fixed point shown in Fig.~7: it is the assumption that the correlation functions defined microscopically in Eq.~\eqref{eq:geta-definition}
are finite in the zero range interaction limit. As a result, they do not exhibit an anomalous dimension and the 
prefactors in~\eqref{eq:gs-scaling} are fixed by simple dimensional analysis. Combining
Eqs.~\eqref{eq:dc-values} and~\eqref{eq:gs-scaling} then allows to express the transport coefficients 
\begin{equation}
\eta(\beta\mu) =\frac{\hbar}{\lambda_T^3}\, f_{\eta} (\beta\mu) \qquad {\rm and} \qquad
\sigma_s(\beta\mu) =\frac{1}{\hbar\lambda_T}\, f_{\sigma} (\beta\mu)
\label{eq:eta+sigma}
\end{equation}
of the unitary gas in terms of dimensionless scaling functions $f_\eta(x)$ and $f_{\sigma}(x)$, which are defined 
in an obvious manner via the 
associated integrals of $\Phi_{\eta}$ and $\Phi_{\sigma}$ over the dimensionless time variable $y=t/(\beta\hbar)$.
The requirement of a finite static shear viscosity and spin conductivity apparently requires the functions
$\Phi_{\eta}$ and $\Phi_{\sigma}$ to decay faster than $1/t^2$. More precisely, the expected singularity 
of ${\rm Re}\, \eta(\omega)=\eta(\omega=0)+b_{\eta}\,\sqrt{\omega}+\ldots$ at low frequencies~\cite{enss11viscosity} implies
that  $g_{\eta}(t\to\infty)=-3\hbar\, b_{\eta}/(\sqrt{8\pi}\, t^{5/2})$ exhibits an asymptotic power law decay.
Based on Eq.~\eqref{eq:eta+sigma}, it is now straightforward to discuss the qualitative behavior of the 
shear viscosity and spin diffusion constant between the limits of a non-degenerate gas and the superfluid.
In the former limit, where $x \ll -1$, the shear viscosity is given by Eq.~\eqref{eq:eta-Boltzmann}
and thus $f_{\eta}(x\ll -1)=15\,\pi/8\sqrt{2}\simeq 4.17$ is a constant. Similarly, the 
Boltzmann equation result~\cite{somm11, enss12spindiffusion}
\begin{equation}
  \sigma_s^{\rm cl}(T)=\frac{9\,\pi^{3/2}}{32\sqrt{2}}\,\frac{\hbar n}{mk_BT_F} \left(\frac{T}{T_F}\right)^{1/2}=\frac{3}{8\sqrt{2}}\,\frac{1}{\hbar\,\lambda_T} 
  \label{eq:sigma-Boltzmann}
\end{equation} 
for the spin conductivity implies $f_{\sigma}(x\ll -1)=3/8\sqrt{2}\simeq 0.27$.  Deep in
the superfluid, the divergence of the shear viscosity described by Eq.~\eqref{eq:landau}
gives rise to a power law increase $f_{\eta}(x\gg 1)=0.005\, x^{13/2}$. 
The spin conductivity, in turn, is expected to stay finite~\cite{einz91}.
As a result,  $f_{\sigma}(x\gg 1)\sim \sqrt{x}$ is again an increasing function of $x$ with a prefactor which has apparently not
been calculated so far. For intermediate values of $x$, quantitative results for the 
scaling functions $f_{\eta,\sigma}(x)$ are available only in the normal fluid regime $x\lesssim 2.5$.  
They are based on a diagrammatic evaluation of the exact Kubo-formula within a
Luttinger-Ward approach~\cite{enss11viscosity, enss12spindiffusion}. The formalism is consistent with the exact 
sum rules in Eqs.~\eqref{eq:fsum} and~\eqref{eq:etasum}.   
Moreover, it respects all symmetries of the problem. 
In particular, the requirement of a vanishing bulk viscosity due to scale invariance
is fulfilled exactly~\cite{enss11viscosity}.
The results show that both $f_{\eta}(x)$ and $f_{\sigma}(x)$ are monotonically increasing
functions of $x$, with $f_{\eta}(0)\simeq 4.76$ at $x=0$ 
and $f_{\eta}(x_c)\simeq 11$ at the superfluid transition $x_c\simeq 2.46$
\footnote{The value $x_c\simeq 2.46$ and the associated critical temperature 
$\theta_c\simeq 0.16$ obtained from the universal function $\theta(x)$ are taken from the Luttinger-Ward approach discussed in section 3.3.}. 
To determine the ratios $\eta/s$ and $D_s=\sigma_s/\chi_s$, one needs the 
corresponding behavior of the entropy density $s$ and the spin susceptibility $\chi_s$. 
Similar to~\eqref{eq:eta+sigma}, they can again be expressed in terms of universal scaling functions
\begin{equation}
s(\beta\mu) =\frac{k_B}{\lambda_T^3}\, f_{s} (\beta\mu) \qquad {\rm and} \qquad
\chi_s(\beta\mu) =\frac{1}{k_BT\,\lambda_T^3}\, f_{\chi} (\beta\mu)\, ,
\label{eq:s+chi}
\end{equation}
which is a simple consequence of dimensional analysis because neither $s$ nor $\chi_s$ 
exhibit an anomalous dimension. Using the Gibbs-Duhem relation, the scaling function for the 
entropy density is related to the function $f_p(x)=f_p(x,0)$ introduced
in~(\ref{eq:pressure-scaling}) by $f_s(x)=5f_p(x)/2\, - x\,\partial_x f_p(x)$. 
Now, at least in the limits of a non-degenerate gas and deep in the superfluid, these functions are again known exactly.
Regarding $f_{\chi}(x)$, the results $\chi_s^{\rm cl}=n/k_BT$
for a non-degenerate gas and $\chi_s^{\rm SF}\sim\exp{(-2\Delta/k_BT)}$ far below $T_c$
imply that $f_{\chi}(x\ll -1)=2\exp{(-|x|)}$ due to $n\lambda_T^3=2\exp{(\beta\mu)}+\ldots$ to leading order 
in the fugacity, while $f_{\chi}(x\gg1)\sim\exp{(-2.5\, x)}$ since the ratio of the zero temperature gap 
$\Delta\simeq 0.46\,\varepsilon_F$ and the chemical potential $\mu(T\to 0)\simeq 0.37\,\varepsilon_F$ is $2\Delta/\mu\simeq 2.5$. 
In both limits therefore, the scaling function $f_{\chi}(x)$ vanishes exponentially.  As a result,
it necessarily exhibits a maximum $\rm {Max}\, f_{\chi}\simeq 1.65$ which appears in the normal fluid regime 
at $\theta\simeq 0.3$~\cite{enss12spindiffusion} or $x\simeq 1.1$.  A similar behavior is obtained 
for the scaling function associated with the entropy density, whose maximum $\rm {Max}\, f_{s}\simeq 18.2$   
is reached close to the superfluid transition at $x_c\simeq 2.46$~\cite{fran15}.
The limiting results in the non-degenerate regime are
$f_{s}(x\ll -1)=2|x|\exp{(-|x|)}$, while $f_{s}(x\gg 1)\sim 1/x^{3/2}$ deep in the superfluid. This follows from
the asymptotic dependence $s(T)\sim k_B\left(k_BT/\hbar c_s\right)^3$ of the entropy density in the 
phonon dominated regime with sound velocity $c_s=\sqrt{\xi_s/3}\, v_F$. Apart from universal constants, the ratios 
\begin{equation}
\frac{\eta}{s}=\frac{\hbar}{k_B}\, \frac{f_{\eta}(\beta\mu)}{f_{s}(\beta\mu)} \qquad {\rm and} \qquad
D_s=2\pi\, \frac{f_{\sigma}(\beta\mu)}{f_{\chi}(\beta\mu)}\cdot\frac{\hbar}{m}
\label{eq:etas+Ds}
\end{equation}
therefore depend only on the dimensionless parameter $x=\beta\mu$ or - equivalently - on
the dimensionless temperature scale $\theta=T/T_F=\left(8/(3\sqrt{\pi}\, n\lambda_T^3)\right)^{2/3}$.
It is a monotonically decreasing function of $x$ with $\theta(x=0)\simeq 0.62$ from Eq.~\eqref{eq:density-QCP} and 
$\theta(x\gg 1)=\xi_s/x$. 
The fact that both $f_s(x)$ and $f_{\chi}(x)$ exhibit a pronounced maximum as $x$ is varied 
between the non-degenerate limit $x\ll -1$ and the superfluid at $x\gg 1$, while 
the functions $f_{\eta}(x)$ and $f_{\sigma}(x)$ are monotonically increasing,
necessarily implies minima of order $\hbar/k_B$ or $\hbar/m$ for the shear viscosity 
to entropy density ratio or the spin diffusion constant. The associated prefactors are ratios of scaling functions.
They are therefore universal numbers which characterize the quantum critical point 
of the unitary Fermi gas. 
\begin{figure}
\includegraphics[width=5.4in]{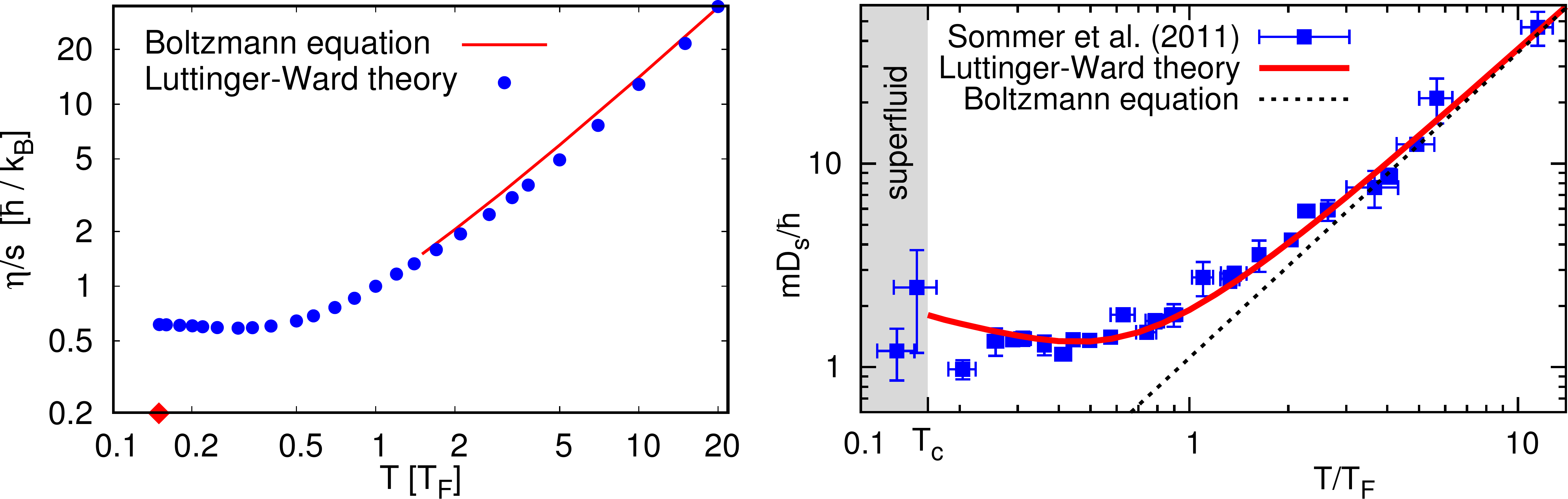} 
\caption{Left: Shear viscosity to entropy density ratio of the unitary Fermi gas 
within the Luttinger-Ward approach (from Ref.~\cite{enss11viscosity}).
Right:  The corresponding result for the spin diffusion constant, including a quantitative 
comparison with experimental data ( from Ref.~\cite{enss12spindiffusion}). }  
\label{fig:transport}
\end{figure}
In quantitative terms, the results for both  
$\eta/s$ and $D_s$ in the normal fluid phase of the unitary gas are shown in Fig.~\ref{fig:transport},
where the standard scaling variable $\beta\mu$ has been converted to the dimensionless
temperature $\theta$. Evidently, both ratios exhibit a shallow minimum
above the critical temperature of the superfluid transition
\footnote{The unknown behavior of the scaling functions close to the  
transition is therefore - fortunately - not relevant in determining the numerical values
at the minimum.} 
The minimum value $\eta/s\simeq 0.6\,\hbar/k_B$ is a factor of about seven above the
KSS bound. It is reached near $\theta\simeq 0.35$ or $x\simeq 0.9$. For the spin
diffusion constant the corresponding values are $D_s\simeq 1.3\,\hbar/m$ at 
$\theta\simeq 0.5$ or $x\simeq 0.3$. As emphasized above, these minima appear
within the quantum critical regime, where the unitary gas is a strongly coupled 
system with no proper quasiparticles. The concept of a mean free path is therefore not applicable.
This can be inferred in a direct form from the detailed behavior of the 
frequency dependent shear viscosity ${\rm Re}\, \eta(\omega)$. It 
has a Drude like peak at high temperatures $T\gtrsim 2\, T_F$ where the 
Boltzmann equation applies, while it is a rather structureless, 
monontonically decaying function of frequency in the quantum critical regime~\cite{enss11viscosity}.  
Note that Eq.~(\ref{eq:etas+Ds}) implies universal ratios
$\eta/s$ and $D_s=\sigma_s/\chi_s$ also if the temperature is lowered along the line $\mu\equiv 0$,
where $\theta\simeq 0.62$. 
The associated amplitudes, however, are larger than those at  $x_{\rm min}$. 
The minimum values of $\eta/s$ or $D_s$ are therefore realized 
by approaching the zero density fixed point along lines of finite slope, where 
both $T$ and $T_F$ vanish at a unique ratio $T/T_F=\theta(x_{\rm min})$. 
In quantitative terms, the difference is small, however, as is evident from the 
rather shallow minima in Fig.~\ref{fig:transport}. This is consistent with a  
calculation of the shear viscosity at $\mu =0$ within a $1/N$- expansion~\cite{enss12QCP}.
Extrapolating to $N=1$, the viscosity to entropy ratio $\eta/s = \hbar/k_B \cdot f_\eta(0)/f_s(0)\approx 0.74\,\hbar/k_B$ at $\mu=0$
turns out to be close to the minimum value $\simeq 0.6\, \hbar/k_B$ attained at $x_{\rm min}\simeq 0.9$.  \\

The discussion above shows that the universality for transport coefficients of 
the unitary Fermi gas can be extended by considering further dimensionless    
ratios like $\eta/\hbar \, n$. 
Combining the results in Eqs.~(\ref{eq:contact-scaling})
 and~(\ref{eq:eta+sigma}), the dimensionless shear viscosity  
\begin{equation}
\frac{\eta}{\hbar\, n}=\frac{f_{\eta}(\beta\mu)}{f_{n}(\beta\mu)} =\alpha(\beta\mu)
\label{eq:etan}
\end{equation}
is again determined by the ratio of two universal scaling functions. The 
well known degenerate gas limit, where $n\lambda_T^3=2z$ shows 
that $f_n(x\ll -1)=2\exp{(-|x|)}$ vanishes exponentially. Deep in the 
superfluid, the zero temperature equation of state $n(\mu)\sim\, (\mu/\xi_s)^{3/2}$ 
fixes the scaling function to increase like $f_n(x\gg 1)\sim x^{3/2}$.
Since $f_{\eta}$ approaches a constant at large negative $x$ and increases
faster than $f_n$ for $x\gg 1$, the ratio $f_{\eta}/f_n$ has again a minimum. 
Within the Luttinger-Ward approach, 
it turns out that the 
minimum value $\eta/\hbar\, n \simeq 0.37$ is
reached {\it below} the superfluid transition at $x\simeq 2.85$ or $\theta\simeq 0.13$~\cite{fran15}.
Apart from the predicted eventual increase of $\alpha(\theta)$ at very low temperatures 
which is {\it not} seen experimentally, this is consistent at least with the  observation, 
that the shear viscosity continues to decrease below the superfluid 
transition, where $\eta_c\simeq 0.5\, \hbar n$~\cite{jose15viscosity}. \\

Clearly,
the understanding of dynamical properties 
is still incomplete and many  questions remain open. Yet, strongly interacting, 
ultracold Fermi gases have opened a
new area in many-body physics whose implications reach far beyond the field
of dilute gases. In particular, they provide a unique opportunity to study well defined model systems in which 
a quantitative comparison between theory and experiment is possible. \\

{\bf Acknowledgement} I am deeply indepted to my collaborators in the
area of strongly interacting Fermi gases over the last ten years: M. Barth, E. Braaten,
S. Cerrito, T. Enss, B. Frank, R. Haussmann, J. Hofmann, C. Langmack, M. Punk, W. Rantner, S. Rath
and R. Schmidt. Moroever, it is a pleasure to acknowledge many discussions
with experimental colleagues on the subject of these Lectures, in particular with M. Greiner, R. Grimm, W. Ketterle,
G. Modugno, C. Salomon, C. Vale, M. Zaccanti and - last not least - with M. Zwierlein, whose 
deep insight and questions have contributed a lot to my own understanding of the subject.

\bibliographystyle{varenna}
\bibliography{References}

\begin{thebibliography}{100}
\expandafter\ifx\csname url\endcsname\relax\def\url#1{\texttt{#1}}\fi
\expandafter\ifx\csname urlprefix\endcsname\relax\def\urlprefix{URL }\fi

\bibitem{grib93scatt}
\NAME{Gribakin G. \atque Flambaum V.}, \IN{Phys. Rev. A}{48}{1993}{546}.

\bibitem{flam99scatt}
\NAME{Flambaum V.~V., Gribakin G.~F. \atque Harabati C.}, \IN{Phys. Rev.
  A}{59}{1999}{1998}.

\bibitem{chin10feshbach}
\NAME{Chin C., Grimm R., Julienne P. \atque Tiesinga E.}, \IN{Rev. Mod.
  Phys.}{82}{2010}{1225}.

\bibitem{bloc08review}
\NAME{Bloch I., Dalibard J. \atque Zwerger W.}, \IN{Reviews of Modern
  Physics}{80}{2008}{885}.

\bibitem{cast07}
\NAME{Castin Y.}, \TITLE{Basic theory tools for degenerate fermi gases}, in
  \TITLE{Proceedings of the International School of Physics {\it Enrico Fermi}
  on Ultra-Cold Fermi gases}, edited by \NAME{Inguscio M., Ketterle W. \atque
  Salomon C.} (SIF Bologna, Bologna) 2007, pp. 1--60.

\bibitem{schm12efimov}
\NAME{Schmidt R., Rath S.~P. \atque Zwerger W.}, \IN{The European Physical
  Journal B}{85}{2012}{386}.

\bibitem{wern09closed}
\NAME{Werner F., Tarruell L. \atque Castin Y.}, \IN{The European Physical
  Journal B}{68}{2009}{401}.

\bibitem{gora04shift}
\NAME{G\'oral K., K\"ohler T., Gardiner S.~A., Tiesinga E. \atque Julienne
  P.~S.}, \IN{J. Phys. B}{37}{2004}{3457}.

\bibitem{zuer13Li-resonance}
\NAME{Z\"urn G., Lompe T., Wenz A.~N., Jochim S., Julienne P.~S. \atque Hutson
  J.~M.}, \IN{Phys. Rev. Lett.}{110}{2013}{135301}.

\bibitem{kett08varenna}
\NAME{Ketterle W. \atque Zwierlein M.}, \TITLE{Making, probing and
  understanding ultracold {Fermi} gases}, in \TITLE{Ultracold Fermi Gases,
  Proceedings of the International School of Physics "Enrico Fermi", Course
  CLXIV, Varenna, 20 - 30 June 2006}, edited by \NAME{Inguscio M., Ketterle W.
  \atque Salomon C.} (IOS Press, Amsterdam) 2008.

\bibitem{petr04dimers}
\NAME{Petrov D., Salomon C. \atque Shlyapnikov G.}, \IN{Phys. Rev.
  Lett.}{93}{2004}{090404}.

\bibitem{tan04}
\NAME{Tan S.}, \IN{arXiv:cond-mat}{0412764}{2004}{}.

\bibitem{rega04lifetime}
\NAME{Regal C.~A., Greiner M. \atque Jin D.~S.}, \IN{Phys. Rev.
  Lett.}{92}{2004}{083201}.

\bibitem{bour04coll}
\NAME{Bourdel T., Khaykovich L., Cubizolles J., Zhang J., Chevy F., Teichmann
  M., Tarruell L., Kokkelmans S. \atque Salomon C.}, \IN{Phys. Rev.
  Lett.}{93}{2004}{050401}.

\bibitem{wern06traplevels}
\NAME{Werner F. \atque Castin Y.}, \IN{Phys. Rev. Lett.}{97}{2006}{150401}.

\bibitem{nish12book}
\NAME{Nishida Y. \atque Son D.~T.}, \TITLE{Unitary fermi gas, $\epsilon$
  expansion and nonrelativistic conformal field theories}, in \TITLE{The
  BCS-BEC Crossover and the Unitary Fermi Gas}, edited by \NAME{Zwerger W.}
  (Lecture Notes in Physics, Vol. 836, Springer, Berlin, Heidelberg) 2012, pp.
  233--275.

\bibitem{petr05dimer}
\NAME{Petrov D.~S., Salomon C. \atque Shlyapnikov G.~V.}, \IN{Phys. Rev.
  A}{71}{2005}{012708}.

\bibitem{cast12book}
\NAME{Castin Y. \atque Werner W.}, \TITLE{The unitary gas and its symmetry
  properties}, in \TITLE{The BCS-BEC Crossover and the Unitary Fermi Gas},
  edited by \NAME{Zwerger W.} (Lecture Notes in Physics, Vol. 836, Springer,
  Berlin, Heidelberg) 2012, pp. 127--191.

\bibitem{efim70}
\NAME{Efimov V.}, \IN{Phys. Lett. B}{33}{1970}{563}.

\bibitem{fedi96reco}
\NAME{Fedichev P., Reynolds M. \atque Shlyapnikov G.}, \IN{Phys. Rev.
  Lett.}{77}{1996}{2921}.

\bibitem{niel99boserecomb}
\NAME{Nielsen E. \atque Macek J.~H.}, \IN{Phys. Rev. Lett.}{83}{1999}{1566}.

\bibitem{flet13}
\NAME{Fletcher R.~J., Gaunt A.~L., Navon N., Smith R.~P. \atque Hadzibabic Z.},
  \IN{Phys. Rev. Lett.}{111}{2013}{125303}.

\bibitem{rem13}
\NAME{Rem B.~S., Grier A.~T., Ferrier-Barbut I., Eismann U., Langen T., Navon
  N., Khaykovich L., Werner F., Petrov D.~S., Chevy F. \atque Salomon C.},
  \IN{Phys. Rev. Lett.}{110}{2013}{163202}.

\bibitem{esry99recomb}
\NAME{Esry B.~D., Greene C.~H. \atque Burke J.~P.}, \IN{Phys. Rev.
  Lett.}{83}{1999}{1751}.

\bibitem{beda00boserecomb}
\NAME{Bedaque P.~F., Braaten E. \atque Hammer H.-W.}, \IN{Phys. Rev.
  Lett.}{85}{2000}{908}.

\bibitem{krae06efimov}
\NAME{Kraemer T., Mark M., Waldburger P., Danzl J., Chin C., Engeser B., Lange
  A., Pilch K., Jaakkola A., N\"agerl H.-C. \atque Grimm R.},
  \IN{Nature}{440}{2006}{315}.

\bibitem{dieh08}
\NAME{Diehl S., Krahl H.~C. \atque Scherer M.}, \IN{Phys. Rev.
  C}{78}{2008}{034001}.

\bibitem{braa06}
\NAME{Braaten E. \atque Hammer H.-W.}, \IN{Physics Reports}{428}{2006}{259}.

\bibitem{wang12}
\NAME{Wang J., D'Incao J.~P., Esry B.~D. \atque Greene C.~H.}, \IN{Phys. Rev.
  Lett.}{108}{2012}{263001}.

\bibitem{naid14}
\NAME{Naidon P., Endo S. \atque Ueda M.}, \IN{Phys. Rev.
  Lett.}{112}{2014}{105301}.

\bibitem{bern11}
\NAME{Berninger M., Zenesini A., Huang B., Harm W., N\"agerl H.-C., Ferlaino
  F., Grimm R., Julienne P.~S. \atque Hutson J.~M.}, \IN{Phys. Rev.
  Lett.}{107}{2011}{120401}.

\bibitem{wild12}
\NAME{Wild R.~J., Makotyn P., Pino J.~M., Cornell E.~A. \atque Jin D.~S.},
  \IN{Phys. Rev. Lett.}{108}{2012}{145305}.

\bibitem{petr13}
\NAME{Petrov D.}, \TITLE{The few-atom problem}, in \TITLE{Proceedungs of the
  Les Houches Summer Schools, Session 94}, edited by \NAME{Salomon C.,
  Shlyapnikov G. \atque Cugliandolo L.} (Oxford University Press, Oxford, UK)
  2013.

\bibitem{huan14efimov}
\NAME{Huang B., Sidorenkov L.~A., Grimm R. \atque Hutson J.~M.}, \IN{Phys. Rev.
  Lett.}{112}{2014}{190401}.

\bibitem{lang15}
\NAME{Langmack C., Schmidt R. \atque Zwerger W.}, \IN{in
  preparation}{}{2015}{}.

\bibitem{hamm07}
\NAME{Hammer H. \atque Platter L.}, \IN{Eur. Phys. Journal A}{32}{2007}{113}.

\bibitem{stec09tetramers}
\NAME{Stecher.~J v., D'Incao J.~P. \atque Greene C.~H.}, \IN{Nat.
  Phys.}{5}{2009}{417}.

\bibitem{schm10tetramers}
\NAME{Schmidt R. \atque Moroz S.}, \IN{Phys. Rev. A}{81}{2010}{052709}.

\bibitem{delt12}
\NAME{Deltuva A.}, \IN{Phys. Rev. A}{85}{2012}{012708}.

\bibitem{ferl09}
\NAME{Ferlaino F., Knoop S., Berninger M., Harm W., D'Incao J.~P., N\"agerl
  H.-C. \atque Grimm R.}, \IN{Phys. Rev. Lett.}{102}{2009}{140401}.

\bibitem{zene13}
\NAME{Zenesini A., Huang B., Berninger M., Besler S., N\"agerl H.-C., Ferlaino
  F., Grimm R., Greene C.~H. \atque von Stecher J.}, \IN{New Journal oof
  Physics}{15}{2013}{043040}.

\bibitem{stec10N-body}
\NAME{Stecher.~J v.}, \IN{J. Phys. B}{43}{2010}{101002}.

\bibitem{seir12}
\NAME{Seiringer R.}, \IN{J. Spec. Theo.}{2}{2012}{321}.

\bibitem{braa11bosons}
\NAME{Braaten E., Kang D. \atque Platter L.}, \IN{Phys. Rev.
  Lett.}{106}{2011}{153005}.

\bibitem{wern12bosons}
\NAME{Werner F. \atque Castin Y.}, \IN{Phys. Rev. A}{86}{2012}{053633}.

\bibitem{tan08energy}
\NAME{Tan S.}, \IN{Annals of Physics}{323}{2008}{2952}.

\bibitem{tan08momentum}
\NAME{Tan S.}, \IN{Annals of Physics}{323}{2008}{2971}.

\bibitem{tan08virial}
\NAME{Tan S.}, \IN{Annals of Physics}{323}{2008}{2987}.

\bibitem{punk07rf}
\NAME{Punk M. \atque Zwerger W.}, \IN{Phys. Rev. Lett.}{99}{2007}{170404}.

\bibitem{schn10shortrange}
\NAME{Schneider W. \atque Randeria M.}, \IN{Phys. Rev. A}{81}{2010}{021601}.

\bibitem{braa10rf}
\NAME{Braaten E., Kang D. \atque Platter L.}, \IN{Phys. Rev.
  Lett.}{104}{2010}{223004}.

\bibitem{tayl10visc}
\NAME{Taylor E. \atque Randeria M.}, \IN{Phys. Rev. A}{81}{2010}{053610}.

\bibitem{hofm11response}
\NAME{Hofmann J.}, \IN{Phys. Rev. A}{84}{2011}{043603}.

\bibitem{enss11viscosity}
\NAME{Enss T., Haussmann R. \atque Zwerger W.}, \IN{Ann. Phys.
  (NY)}{326}{2011}{770}.

\bibitem{braa12book}
\NAME{Braaten E.}, \TITLE{Universal relations for fermions with large
  scattering length}, in \TITLE{The BCS-BEC Crossover and the Unitary Fermi
  Gas}, edited by \NAME{Zwerger W.} (Lecture Notes in Physics, Vol. 836,
  Springer, Berlin, Heidelberg) 2012, pp. 193--231.

\bibitem{zhan09universal}
\NAME{Zhang S. \atque Leggett A.~J.}, \IN{Phys. Rev. A}{79}{2009}{023601}.

\bibitem{braa08contact}
\NAME{Braaten E. \atque Platter L.}, \IN{Phys. Rev. Lett.}{100}{2008}{205301}.

\bibitem{bali91}
\NAME{Balian R.}, \TITLE{From Microphysics to Macrophysics, Vol. I} (Spinger
  Berlin) 1991.

\bibitem{thom05virial}
\NAME{Thomas J.~E., Kinast J. \atque Turlapov A.}, \IN{Phys. Rev.
  Lett.}{95}{2005}{120402}.

\bibitem{thom08virial}
\NAME{Thomas J.~E.}, \IN{Phys. Rev. A}{78}{2008}{013630}.

\bibitem{dien08}
\NAME{Diener R.~B., Sensarma R. \atque Randeria M.}, \IN{Phys. Rev.
  A}{77}{2008}{023626}.

\bibitem{haus09rf}
\NAME{Haussmann R., Punk M. \atque Zwerger W.}, \IN{Physical Review
  A}{80}{2009}{063612}.

\bibitem{altm07precision}
\NAME{Altmeyer A., Riedl S., Kohstall C., Wright M.~J., Geursen R., Bartenstein
  M., Chin C., Hecker-Denschlag J. \atque Grimm R.}, \IN{Phys. Rev.
  Lett.}{98}{2007}{040401}.

\bibitem{navo10thermo}
\NAME{Navon N., Nascimb\`{e}ne S., Chevy F. \atque Salomon C.},
  \IN{Science}{328}{2010}{729}.

\bibitem{hoin13contact}
\NAME{Hoinka S., Lingham M., Fenech K., Hu H., Vale C.~J., Drut J.~E. \atque
  Gandolfi S.}, \IN{Phys. Rev. Lett.}{110}{2013}{055305}.

\bibitem{vanh13contact}
\NAME{van Houcke K., Werner F., E. K., Prokof'ev N. \atque Svistunov B.},
  \IN{arXiv:1303.6245}{}{2013}{}.

\bibitem{yu09correlations}
\NAME{Yu Z., Bruun G.~M. \atque Baym G.}, \IN{Phys. Rev. A}{80}{2009}{023615}.

\bibitem{kuhn11contact}
\NAME{Kuhnle E.~D., Hoinka S., Dyke P., Hu H., Hannaford P. \atque Vale C.~J.},
  \IN{Phys. Rev. Lett.}{106}{2011}{170402}.

\bibitem{part05}
\NAME{Partridge G.~B., Strecker K.~E., Kamar R.~I., Jack M.~W. \atque Hulet
  R.~G.}, \IN{Phys. Rev. Lett.}{95}{2005}{020404}.

\bibitem{bruu04eff}
\NAME{Bruun G. \atque Pethick C.}, \IN{Phys. Rev. Lett.}{92}{2004}{140404}.

\bibitem{land77qm}
\NAME{Landau L. \atque Lifshitz E.}, \TITLE{Quantum Mechanics: Non-Relativistic
  Theory} (Pergamon Press, New York) 1987.

\bibitem{haus94}
\NAME{Haussmann R.}, \IN{Phys. Rev. B}{49}{1994}{12975}.

\bibitem{fett71}
\NAME{Fetter A. \atque Walecka J.}, \TITLE{Quantum Theory of Many-Particle
  Systems} (McGraw-Hill, New York) 1971.

\bibitem{comb06}
\NAME{Combescot R., Giorgini S. \atque Stringari S.}, \IN{Europhys.
  Lett.}{75}{2006}{695}.

\bibitem{hu10}
\NAME{Hu H., Liu X.-J. \atque Drummond P.}, \IN{Europhys.
  Lett.}{91}{2010}{20005}.

\bibitem{punk09molaron}
\NAME{Punk M., Dumitrescu P.~T. \atque Zwerger W.}, \IN{Physical Review
  A}{80}{2009}{053605}.

\bibitem{wern12}
\NAME{Werner F. \atque Castin Y.}, \IN{Phys. Rev. A}{86}{2012}{013626}.

\bibitem{bart11}
\NAME{Barth M. \atque Zwerger W.}, \IN{Annals of Physics}{326}{2011}{2544}.

\bibitem{hofm13jellium}
\NAME{Hofmann J., Barth M. \atque Zwerger W.}, \IN{Phys. Rev.
  B}{87}{2013}{235125}.

\bibitem{gior08review}
\NAME{Giorgini S., Pitaevskii L.~P. \atque Stringari S.}, \IN{Reviews of Modern
  Physics}{80}{2008}{1215}.

\bibitem{rand14}
\NAME{Randeria M. \atque Taylor E.}, \IN{Annu. Rev. Condens. Matter
  Phys.}{5}{2014}{209}.

\bibitem{zwer12book}
\NAME{Zwerger W.} (Ed.), \TITLE{The BCS-BEC Crossover and the Unitary Fermi
  Gas} (Lecture Notes in Physics, Vol. 836, Springer, Berlin, Heidelberg) 2012.

\bibitem{bert00}
\NAME{Bertsch G.}, \TITLE{Proceedings of the tenth international conference on
  recent progress in many-body theories}, presented at \TITLE{Recent progress
  in many-body theories}, edited by \NAME{Bishop R., Gernoth K.~A., Walet N.~R.
  \atque Xian Y.} (World Scientific, Seattle) 2000.

\bibitem{geze14}
\NAME{Gezerlis A., Pethick C. \atque Schwenk A.},
  \IN{arXiv:1406.6109}{}{2014}{}.

\bibitem{niko07renorm}
\NAME{Nikolic P. \atque Sachdev S.}, \IN{Physical Review A}{75}{2007}{033608}.

\bibitem{nish07CFT}
\NAME{Nishida Y. \atque Son D.~T.}, \IN{Physical Review D}{76}{2007}{086004}.

\bibitem{wern06unitary}
\NAME{Werner F. \atque Castin Y.}, \IN{Phys. Rev. A}{74}{2006}{053604}.

\bibitem{son07bulk}
\NAME{Son D.~T.}, \IN{Phys. Rev. Lett.}{98}{2007}{020604}.

\bibitem{nish06epsilon}
\NAME{Nishida Y. \atque Son D.~T.}, \IN{Phys. Rev. Lett.}{97}{2006}{050403}.

\bibitem{nish07eps}
\NAME{Nishida Y. \atque Son D.~T.}, \IN{Physical Review A}{75}{2007}{063617}.

\bibitem{sach11book}
\NAME{Sachdev S.}, \TITLE{Quantum Phase Transitions, Second Edition} (Cambridge
  University Press, Cambridge, UK) 2011.

\bibitem{enss12QCP}
\NAME{Enss T.}, \IN{Phys. Rev. A}{86}{2012}{013616}.

\bibitem{ku12superfluid}
\NAME{Ku M. J.~H., Sommer A.~T., Cheuk L.~W. \atque Zwierlein M.~W.},
  \IN{Science}{335}{2012}{563}.

\bibitem{ho04uni}
\NAME{Ho T.-L.}, \IN{Phys. Rev. Lett.}{92}{2004}{090402}.

\bibitem{nuss06becbcs}
\NAME{Nussinov Z. \atque Nussinov S.}, \IN{Physical Review
  A}{74}{2006}{053622}.

\bibitem{arno07bertsch}
\NAME{Arnold P., Drut J.~E. \atque Son D.~T.}, \IN{Phys. Rev.
  A}{75}{2007}{043605}.

\bibitem{nish09eps}
\NAME{Nishida Y.}, \IN{Physical Review A}{79}{2009}{013627}.

\bibitem{haus07bcsbec}
\NAME{Haussmann R., Rantner W., Cerrito S. \atque Zwerger W.}, \IN{Phys. Rev.
  A}{75}{2007}{023610}.

\bibitem{legg65}
\NAME{Leggett A.~J.}, \IN{Phys. Rev.}{140}{1965}{A1869}.

\bibitem{gork61}
\NAME{Gor'kov L. \atque Melik-Barkhudarov T.}, \IN{Zh. Eskp. Theor.
  Fiz.}{40}{1961}{1452}.

\bibitem{shee06phase}
\NAME{Sheehy D.~E. \atque Radzihovsky L.}, \IN{Annals of
  Physics}{322}{2007}{1790}.

\bibitem{ohar02science}
\NAME{O'Hara K.~M., Hemmer S.~L., Gehm M.~E., Granade S.~R. \atque Thomas
  J.~E.}, \IN{Science}{298}{2002}{2179}.

\bibitem{bour03}
\NAME{Bourdel T., Cubizolles J., Khaykovich L., Magalh K. M.~F., Kokkelmans S.
  J. J. M.~F., Shlyapnikov G.~V. \atque Salomon C.}, \IN{Phys. Rev.
  Lett.}{91}{2003}{020402}.

\bibitem{bart04}
\NAME{Bartenstein M., Altmeyer A., Riedl S., Jochim S., Chin C.,
  {Hecker-Denschlag} J. \atque Grimm R.}, \IN{Phys. Rev.
  Lett.}{92}{2004}{120401}.

\bibitem{rega04}
\NAME{Regal C.~A., Greiner M. \atque Jin D.~S.}, \IN{Phys. Rev.
  Lett.}{92}{2004}{040403}.

\bibitem{zwie04rescond}
\NAME{Zwierlein M., Stan C., Schunck C., Raupach S., Kerman A. \atque Ketterle
  W.}, \IN{Phys. Rev. Lett.}{92}{2004}{120403}.

\bibitem{kina05heat}
\NAME{Kinast J., Turlapov A., Thomas J.~E., Chen Q., Stajic J. \atque Levin
  K.}, \IN{Science}{307}{2005}{1296}.

\bibitem{part06phase}
\NAME{Partridge G.~B., Li W., Kamar R.~I., Liao Y. \atque Hulet R.~G.},
  \IN{Science}{311}{2006}{503}.

\bibitem{stew06pot}
\NAME{Stewart J.~T., Gaebler J.~P., Regal C.~A. \atque Jin D.~S.}, \IN{Phys.
  Rev. Lett.}{97}{2006}{220406}.

\bibitem{nasc10thermo}
\NAME{Nascimb\`{e}ne S., Navon N., Jiang K.~J., Chevy F. \atque Salomon C.},
  \IN{Nature}{463}{2010}{1057}.

\bibitem{hori10energy}
\NAME{Horikoshi M., Nakajima S., Ueda M. \atque Mukaiyama T.},
  \IN{Science}{327}{2010}{442}.

\bibitem{ku12normal}
\NAME{Van~Houcke K., Werner F., Kozik E., Prokof'ev N.~V., Svistunov B.~V., Ku
  M., Sommer A., Cheuk L.~W., Schirotzek A. \atque Zwierlein M.~W.}, \IN{Nature
  Physics}{8}{2012}{366}.

\bibitem{brac08}
\NAME{Brack M. \atque Bhaduri R.}, \TITLE{Semiclassical Physics}, Frontiers in
  Physics (Westview Press, Boulder, CO (USA)) 2008.

\bibitem{son06symmetry}
\NAME{Son D.~T. \atque Wingate M.}, \IN{Ann. Phys. (NY)}{321}{2006}{197}.

\bibitem{haus08}
\NAME{Haussmann R. \atque Zwerger W.}, \IN{Physical Review
  A}{78}{2008}{063602}.

\bibitem{ashc76}
\NAME{Ashcroft N.~W. \atque Mermin N.~D.}, \TITLE{Solid State Physics} (Holt,
  Rinehardt and Winston, New York) 1976.

\bibitem{ho04virial}
\NAME{Ho T.-L. \atque Mueller E.~J.}, \IN{Phys. Rev. Lett.}{92}{2004}{160404}.

\bibitem{liu09virial}
\NAME{Liu X.-J., Hu H. \atque Drummond P.~D.}, \IN{Phys. Rev.
  Lett.}{102}{2009}{160401}.

\bibitem{kapl11virial}
\NAME{Kaplan D.~B. \atque Sun S.}, \IN{Phys. Rev. Lett.}{107}{2011}{030601}.

\bibitem{carl11bertsch}
\NAME{Carlson J., Gandolfi S., Schmidt K.~E. \atque Zhang S.}, \IN{Phys. Rev.
  A}{84}{2011}{061602}.

\bibitem{buro06TC}
\NAME{Burovski E., Prokof'ev N., Svistunov B. \atque Troyer M.}, \IN{Phys. Rev.
  Lett.}{96}{2006}{160402}.

\bibitem{goul10}
\NAME{Goulko O. \atque Wingate M.}, \IN{Phys. Rev. A}{82}{2010}{053621}.

\bibitem{zinn10}
\NAME{Zinn-Justin J.}, \TITLE{Phase Transitions and Renormalization Group}
  (Oxford Graduate Texts, Oxford University Press) 2010.

\bibitem{lipa03}
\NAME{Lipa J.~A., Nissen J.~A., Stricker D.~A., Swanson D.~R. \atque Chui T.
  C.~P.}, \IN{Phys. Rev. B}{68}{2003}{174518}.

\bibitem{nozi85}
\NAME{Nozi\`{e}res P. \atque Schmitt-Rink S.}, \IN{J. Low Temp.
  Phys.}{59}{1985}{195}.

\bibitem{abri75}
\NAME{Abrikosov A., Gor'kov L. \atque Dzyaloshinski I.}, \TITLE{Methods of
  Quantum Field Theory in Statistical Physics} (Dover Publications, New York)
  1975.

\bibitem{lutt60}
\NAME{Luttinger J.~M. \atque Ward J.~C.}, \IN{Phys. Rev.}{118}{1960}{1417}.

\bibitem{baym61}
\NAME{Baym G. \atque Kadanoff L.~P.}, \IN{Phys. Rev.}{124}{1961}{287}.

\bibitem{haus93}
\NAME{Haussmann R.}, \IN{Z. Phys. B}{91}{1993}{291}.

\bibitem{haus99}
\NAME{Haussmann R.}, \TITLE{Self-consistent Quantum Field Theory and
  Bosonization for Strongly Correlated Electron Systems}, Lecture Notes in
  Physics (Springer Verlag, Berlin) 1999.

\bibitem{he14ward-identity}
\NAME{He Y. \atque Levin K.}, \IN{Phys. Rev. B}{89}{2014}{035106}.

\bibitem{hohe65helium}
\NAME{Hohenberg P.~C. \atque Martin P.~C.}, \IN{Annals of
  Physics}{34}{1965}{291}.

\bibitem{vanh13BDMC}
\NAME{van Houcke K., Werner F., Prokof'ev N. \atque Svistunov B.},
  \IN{arXiv:1305.3901}{}{2013}{}.

\bibitem{dieh07bcsbec}
\NAME{Diehl S., Gies H., Pawlowski J. \atque Wetterich C.}, \IN{Phys. Rev.
  A}{76}{2007}{021602}.

\bibitem{son07graphene}
\NAME{Son D.~T.}, \IN{Phys. Rev. B}{75}{2007}{235423}.

\bibitem{hols93anomalies}
\NAME{Holstein B.}, \IN{American Journal of Physics}{61}{1993}{142}.

\bibitem{fors75}
\NAME{Forster D.}, \TITLE{Hydrodynamic Fluctuations, Broken Symmetry and
  Correlation Functions}, Frontiers in Physics (W. A. Benjamin, Advanced Book
  Program, Reading, Massachusetts) 1975.

\bibitem{zee10book}
\NAME{Zee A.}, \TITLE{Quantum Field Theory in a Nutshell, Second Edition}
  (Princeton University Press, Princeton, NJ) 2010.

\bibitem{petr03three_body}
\NAME{Petrov D.~S.}, \IN{Phys. Rev. A}{67}{2003}{010703}.

\bibitem{pita97symmetry}
\NAME{Pitaevskii L.~P. \atque Rosch A.}, \IN{Phys. Rev. A}{55}{1997}{R853}.

\bibitem{chev02}
\NAME{Chevy F., Bretin V., Rosenbusch P., Madison K.~W. \atque Dalibard J.},
  \IN{Phys. Rev. Lett.}{88}{2002}{250402}.

\bibitem{stoc04}
\NAME{Stock S., Bretin V., Chevy F. \atque Dalibard J.}, \IN{Europhysics
  Letters}{65}{2004}{594}.

\bibitem{adhi86}
\NAME{Adhikari S.}, \IN{Am. J. Phys.}{54}{1986}{362}.

\bibitem{petr01scattering-2D}
\NAME{Petrov D.~S. \atque Shlyapnikov G.~V.}, \IN{Phys. Rev.
  A}{64}{2001}{012706}.

\bibitem{hung11}
\NAME{Hung C.-L., Zhang X., Gemelke N. \atque Chin C.},
  \IN{Nature}{470}{2011}{236}.

\bibitem{desb14}
\NAME{Desbuquois R., Yefsah T., Chomaz L., Weitenberg C., Corman L.,
  Nascimb\`ene S. \atque Dalibard J.}, \IN{Phys. Rev.
  Lett.}{113}{2014}{020404}.

\bibitem{ranc12scaling}
\NAME{Ran\ifmmode~\mbox{\c{c}}\else \c{c}\fi{}on A. \atque Dupuis N.},
  \IN{Phys. Rev. A}{85}{2012}{063607}.

\bibitem{froh11}
\NAME{Fr\"ohlich B., Feld M., Vogt E., Koschorreck M., Zwerger W. \atque K\"ohl
  M.}, \IN{Phys. Rev. Lett.}{106}{2011}{105301}.

\bibitem{somm12}
\NAME{Sommer A.~T., Cheuk L.~W., Ku M. J.~H., Bakr W.~S. \atque Zwierlein
  M.~W.}, \IN{Phys. Rev. Lett.}{108}{2012}{045302}.

\bibitem{hofm12anomaly}
\NAME{Hofmann J.}, \IN{Phys. Rev. Lett.}{108}{2012}{185303}.

\bibitem{vogt12viscosity}
\NAME{Vogt E., Feld M., Fr\"ohlich B., Pertot D., Koschorreck M. \atque K\"ohl
  M.}, \IN{Phys. Rev. Lett.}{108}{2012}{070404}.

\bibitem{tayl12anomaly}
\NAME{Taylor E. \atque Randeria M.}, \IN{Phys. Rev. Lett.}{109}{2012}{135301}.

\bibitem{torm00}
\NAME{Torm\"a P. \atque Zoller P.}, \IN{Phys. Rev. Lett.}{85}{2000}{487}.

\bibitem{chin04gap}
\NAME{Chin C., Bartenstein M., Altmeyer A., Riedl S., Jochim S.,
  Hecker-Denschlag J. \atque Grimm R.}, \IN{Science}{305}{2004}{1128}.

\bibitem{schu08pairsize}
\NAME{Schunck C.~H., Shin Y., Schirotzek A. \atque Ketterle W.},
  \IN{Nature}{454}{2008}{739}.

\bibitem{shin07rf}
\NAME{Shin Y., Schunck C.~H., Schirotzek A. \atque Ketterle W.}, \IN{Phys. Rev.
  Lett.}{99}{2007}{090403}.

\bibitem{stew08arpes}
\NAME{Stewart J.~T., Gaebler J.~P. \atque Jin D.~S.},
  \IN{Nature}{454}{2008}{744}.

\bibitem{dama03arpes}
\NAME{Damascelli A., Hussain Z. \atque Shen Z.-X.}, \IN{Reviews of Modern
  Physics}{75}{2003}{473}.

\bibitem{gaeb10pseudo}
\NAME{Gaebler J.~P., Stewart J.~T., Drake T.~E., Jin D.~S., Perali A., Pieri P.
  \atque Strinati G.~C.}, \IN{Nature Phys.}{6}{2010}{569}.

\bibitem{nish12OPE}
\NAME{Nishida Y.}, \IN{Phys. Rev. A}{85}{2012}{053643}.

\bibitem{carl08gap}
\NAME{Carlson J. \atque Sanjay R.}, \IN{Phys. Rev. Lett.}{100}{2008}{150403}.

\bibitem{schi08gap}
\NAME{Schirotzek A., Shin Y., Schunck C.~H. \atque Ketterle W.}, \IN{Phys. Rev.
  Lett.}{101}{2008}{140403}.

\bibitem{prok08polaron}
\NAME{Prokof'ev N. \atque Svistunov B.}, \IN{Phys. Rev. B}{77}{2008}{020408}.

\bibitem{jo09ferro}
\NAME{Jo G.-B., Lee Y.-R., Choi J.-H., Christensen C.~A., Kim T.~H., Thywissen
  J.~H., Pritchard D.~E. \atque Ketterle W.}, \IN{Science}{325}{2009}{1521}.

\bibitem{bulg09}
\NAME{Magierski P., Wlaz\l{}owski G., Bulgac A. \atque Drut J.~E.}, \IN{Phys.
  Rev. Lett.}{103}{2009}{210403}.

\bibitem{chen09rf}
\NAME{Chen Q. \atque Levin K.}, \IN{Phys. Rev. Lett.}{102}{2009}{190402}.

\bibitem{pier09rf}
\NAME{Pieri P., Perali A. \atque Strinati G.~C.}, \IN{Nat Phys}{5}{2009}{736}.

\bibitem{nasc11fermiliquid}
\NAME{Nascimb\`{e}ne S., Navon N., Pilati S., Chevy F., Giorgini S., Georges A.
  \atque Salomon C.}, \IN{Phys. Rev. Lett.}{106}{2011}{215303}.

\bibitem{somm11}
\NAME{Sommer A., Ku M., Roati G. \atque Zwierlein M.~W.},
  \IN{Nature}{472}{2011}{201}.

\bibitem{baue14}
\NAME{Bauer M., Parish M.~M. \atque Enss T.}, \IN{Phys. Rev.
  Lett.}{112}{2014}{135302}.

\bibitem{feld11}
\NAME{Feld M., Fr\"ohlich B., Vogt E., Koschorreck M. \atque K\"ohl M.},
  \IN{Nature}{480}{2011}{75}.

\bibitem{stew10contact}
\NAME{Stewart J.~T., Gaebler J.~P., Drake T.~E. \atque Jin D.~S.}, \IN{Phys.
  Rev. Lett.}{104}{2010}{235301}.

\bibitem{chin05rf}
\NAME{Chin C. \atque Julienne P.~S.}, \IN{Phys. Rev. A}{71}{2005}{012713}.

\bibitem{baym07}
\NAME{Baym G., Pethick C.~J., Yu Z. \atque Zwierlein M.~W.}, \IN{Phys. Rev.
  Lett.}{99}{2007}{190407}.

\bibitem{zhan08rf}
\NAME{Zhang S. \atque Leggett A.~J.}, \IN{Phys. Rev. A}{77}{2008}{033614}.

\bibitem{schu07pair}
\NAME{Schunck C., Shin Y.-I., Schirotzek A., Zwierlein M. \atque Ketterle W.},
  \IN{Science}{316}{2007}{867}.

\bibitem{lang12}
\NAME{Langmack C., Barth M., Zwerger W. \atque Braaten E.}, \IN{Phys. Rev.
  Lett.}{108}{2012}{060402}.

\bibitem{gelm05viscosity}
\NAME{Gelman B.~A., Shuryak E.~V. \atque Zahed I.}, \IN{Phys. Rev.
  A}{72}{2005}{043601}.

\bibitem{bard14}
\NAME{Bardon A.~B., Beattie S., Luciuk C., Cairncross W., Fine D., Cheng N.~S.,
  Edge G. J.~A., Taylor E., Zhang S., Trotzky S. \atque Thywissen J.~H.},
  \IN{Science}{344}{2014}{722}.

\bibitem{enss13spindiffusion}
\NAME{Enss T.}, \IN{Phys. Rev. A}{88}{2013}{033630}.

\bibitem{kosc13}
\NAME{Koschorreck M., Pertot D., Vogt E. \atque K\"ohl M.}, \IN{Nature
  Physics}{9}{2013}{405}.

\bibitem{turl08perfect}
\NAME{Turlapov A., Kinast J., Clancy B., Luo L., Joseph J. \atque Thomas J.},
  \IN{J. Low Temp. Phys.}{150}{2008}{567}.

\bibitem{cao11viscosity}
\NAME{Cao C., Elliott E., Joseph J., Wu H., Petricka J., Schaefer T. \atque
  Thomas J.~E.}, \IN{Science}{331}{2011}{58}.

\bibitem{elli14conformal}
\NAME{Elliott E., Joseph J.~A. \atque Thomas J.~E.}, \IN{Phys. Rev.
  Lett.}{112}{2014}{040405}.

\bibitem{bruu07viscosity}
\NAME{Bruun G.~M. \atque Smith H.}, \IN{Phys. Rev. A}{75}{2007}{043612}.

\bibitem{jose15viscosity}
\NAME{Joseph J.~A., Elliott E. \atque Thomas J.~E.}, \IN{Phys. Rev.
  Lett.}{115}{2015}{020401}.

\bibitem{enss12spindiffusion}
\NAME{Enss T. \atque Haussmann R.}, \IN{Phys. Rev. Lett.}{109}{2012}{195303}.

\bibitem{poli01viscosity}
\NAME{Policastro G., Son D.~T. \atque Starinets A.~O.}, \IN{Phys. Rev.
  Lett.}{87}{2001}{081601}.

\bibitem{kovt05visc}
\NAME{Kovtun P.~K., Son D.~T. \atque Starinets A.~O.}, \IN{Phys. Rev.
  Lett.}{94}{2005}{111601}.

\bibitem{brig08viscosity}
\NAME{Brigante M., Liu H., Myers R.~C., Shenker S. \atque Yaida S.}, \IN{Phys.
  Rev. Lett.}{100}{2008}{191601}.

\bibitem{scha09perfect}
\NAME{Sch\"afer T. \atque Teaney D.}, \IN{Rep. Prog. Phys.}{72}{2009}{126001}.

\bibitem{sach12review}
\NAME{Sachdev S.}, \IN{Annual Review of Condensed Matter Physics}{3}{2012}{9}.

\bibitem{adam13review}
\NAME{Adams A., Carr L., Sch\"afer T., Steinberg P. \atque Thomas J.}, \IN{New
  J. Phys.}{15}{2013}{045022}.

\bibitem{bali92}
\NAME{Balian R.}, \TITLE{From Microphysics to Macrophysics, Vol. II} (Spinger
  Berlin) 1992.

\bibitem{sido13}
\NAME{Sidorenkov L.~A., Tey M.~K., Grimm R., Hou Y.-H., Pitaevskii L. \atque
  Stringari S.}, \IN{Nature}{498}{2013}{78}.

\bibitem{land49viscosity}
\NAME{Landau L. \atque Khalatnikov I.}, \IN{Sov. Phys. JETP}{19}{1949}{637}.

\bibitem{rupa07viscosity}
\NAME{Rupak G. \atque Sch\"afer T.}, \IN{Physical Review A}{76}{2007}{053607}.

\bibitem{einz87eta}
\NAME{Einzel D. \atque Parpia J.~M.}, \IN{Phys. Rev. Lett.}{58}{1987}{1937}.

\bibitem{heik55}
\NAME{Heikil\"a W. \atque Hollis~Hallet A.}, \IN{Canadian Journal of
  Physics}{33}{1955}{420}.

\bibitem{gold12sumrule}
\NAME{Goldberger W.~D. \atque Khandker Z.~U.}, \IN{Phys. Rev.
  A}{85}{2012}{013624}.

\bibitem{einz91}
\NAME{Einzel D.}, \IN{J. Low Temp. Phys.}{84}{1991}{321}.

\bibitem{fran15}
\NAME{Frank B., Haussmann R. \atque Zwerger W.}, \IN{in preparation}{}{2015}{}.

\end{thebibliography}

\end{document}